\documentclass[onecolumn,aps,prl,showpacs,amsmath,superscriptaddress]{revtex4}
\usepackage{amsthm,amsmath,amssymb}
\usepackage{mathrsfs}

\usepackage{hyperref}
\usepackage{graphicx}
\usepackage{float}

\usepackage{bm}
\usepackage{enumerate}

\usepackage{color}

\newcommand{\e}{\mathrm{e}}
\newcommand{\Li}{\mathrm{Li}}
\renewcommand{\d}{\mathrm{d}}

\begin{document}

\title{Exact results of  one-dimensional repulsive Hubbard model}

\author{Jia-Jia Luo}
\affiliation{Innovation Academy for Precision Measurement Science and Technology, Chinese Academy of Sciences, Wuhan 430071, China}
\affiliation{University of Chinese Academy of Sciences, Beijing 100049, China}

\author{Han Pu}
\affiliation{Department of Physics and Astronomy, and Rice Center for Quantum Materials, Rice University, Houston, Texas 77251-1892, USA}

\author{Xi-Wen Guan}
\email[e-mail:]{xwe105@wipm.ac.cn; xiwen.guan@anu.edu.au}
\affiliation{Innovation Academy for Precision Measurement Science and Technology, Chinese Academy of Sciences, Wuhan 430071, China}
\affiliation{Hefei National Laboratory, University of Science and Technology of China, Hefei 230088, China}
\affiliation{Peng Huanwu Center for Fundamental Theory, Xi'an 710069, China}
\affiliation{Department of Fundamental and Theoretical Physics, Research School of Physics, Australian National University, Canberra ACT 0200, Australia}

\date{\today}

\begin{abstract}
	
We present analytical results of fundamental properties of  one-dimensional (1D) Hubbard model with a repulsive interaction, ranging from fractional excitations to universal thermodynamics, interaction-driven criticality, correlation functions, Contact susceptibilities and quantum cooling. 
Using the exact solutions of the Bethe Ansatz equations of  the Hubbard  model, we first rigorously calculate the gapless spin and charge excitations, exhibiting exotic  features of   fractionalized spinons and holons. 
While we investigate the gapped excitations in terms of the spin string  and the $k-\Lambda$ string bound states  at arbitrary driving fields, showing subtle differences  of spin magnons and   charge $\eta$-pair excitations.    
Based on the analysis on  the fractional charge and spin excitations, the  spin-incoherent Luttinger liquid with only the charge propagation mode is elucidated by the asymptotic of the  two-point correlation functions with the help of the conformal field theory.
For a high density and high spin magnetization region, i.e. near quadruple critical point, we then further analytically  obtain  the thermodynamical properties, dimensionless ratios and scaling functions near quantum phase transitions.
These results of thermodynamics provide rich universal low energy physics  and critical scaling laws for the phase transitions near the quadruple point. 
In particular, we determine additivity rules of spin and charge susceptibilities, and  thermodynamics of spin-incoherent Luttinger liquid.
Finally, in order to capture deeper  insight into  the Mott insulator and interaction driven criticality, we  further study the double occupancy and its associated Contact and Contact susceptibilities  through which an adiabatic  cooling scheme  upon  the quantum criticality  is  proposed  
Our methods  provide  rich  perspectives  of  quantum integrability in  1D  strongly correlated systems  and offer promising  guidance to future experiments with interacting electrons and ultracold atoms with and without a lattice.

\end{abstract}

\author{}
\maketitle

\section{I. Introduction}
The strongly correlated electronic  systems provide us with a powerful platform to investigate  novel and intriguing many-body phenomena, such as high-$T_c$ superconductor \cite{pruschke1995anomalous,dagotto2005complexity,yanase2003theory,miyake2007new}, Mott metal-insualtor transition\cite{mott1968metal,schulz1990correlation,stafford1993scaling,jordens2008mott,kokalj2013thermodynamics}, colossal magnetoresistance\cite{dagotto2005complexity,ramirez1997colossal}, antiferromagnetic correlations \cite{hart2015observation,chiu2018quantum,giamarchi2017firmer,boll2016spin,parsons2016site,schneider2012fermionic,cheuk2016observation, salomon2019direct, mitra2018quantum}  etc., which  cannot be merely explained by compounding of single particle motions. 
Among the continuous and lattice models of interacting electrons and ultracold atoms, one of the prototypical integrable  models is the one-band Hubbard model \cite{Lieb:1968,takahashi1972one,essler2005one}. 
Recently, particular attention has been paid to this model because of  its operability and accessibility in trapping ultracold atoms on lattices \cite{jordens2008mott,koepsell2019imaging,vijayan2020time,scherg2021observing,gall2021competing}.
There have been  numerous publications  focusing on the 1D Hubbard model with on-site interaction as well as its  variants in one or higher  dimensions \cite{hirsch1985two,jarrell1992hubbard,georges1992hubbard,glocke2007half,de2011thermodynamics,ibarra2020thermodynamics,wietek2021mott}, involving the studies of spectral weights and critical exponents for correlation functions\cite{schulz1990correlation,frahm1990critical,frahm1991correlation,stafford1993scaling,kim2020spin,qin2022hubbard}, dynamical transport\cite{qin2022hubbard,ilievski2017ballistic,ilievski2018superdiffusion,guardado2020subdiffusion,bertini2021finite}, interaction quench\cite{moeckel2008interaction,eckstein2009thermalization,schlunzen2017nonequilibrium} etc.
 In this  scenario,  the well-celebrated Landau theory of Fermi liquid \cite{landau1959theory,baym2008landau}  remarkably describes universal low energy physics of interacting fermions and leads to  a wide range of applications in higher dimensional quantum many-body systems.
 
 On the other hand, the low energy excitations of 1D many-body  systems usually lack individual motions of quasiparticles. Instead, they form collective motions of bosons, i.e., Tomonaga-Luttinger liquid (TLL) \cite{tomonaga1950remarks,luttinger1963exactly,haldane1981luttinger}.
 A hallmark of  interacting electrons  in 1D is the splitting of low-lying spin and charge excitations into two separate collective motions \cite{lorenz2002evidence,demler2002fractionalization,kollath2005spin,jompol2009probing,schmidt2010spin,ma2017angle,he2020emergence,senaratne2022spin,vijayan2020time,Scopa2022},  i.e., one solely carrying spin and the other carrying charge. 
  This is one of the most significant features predicted by both the integrable theory and  the TLL theory, namely,  the propagators generated by excitations separate into two massless wave packets with  distinctive velocities $v_c,\, v_s$ in charge and spin degrees of freedom, respectively \cite{voit1993charge,kollath2005spin,Recati:PhysRevLett.90.020401,lorenz2002evidence}. 
  The spin-charge separation phenomenon was experimentally evidenced  in literature \cite{Kim:1996,senaratne2022spin,Kim:2006,auslaender2005spin,jompol2009probing}. 
  This phenomenon was further evidenced  by detecting  spin and charge  propagating  velocities \cite{jompol2009probing,vijayan2020time} from quenching an initial Mott state into   few hole doped states in the 1D Hubbard model. 
  While some high-energy or gapped excitations may contribute to the optical conductivity \cite{veness2016mobile}.  
The spin-charge separation has been recently verified and confirmed   by the  virtue of spin and charge dynamical structure factors in 1D Fermi gas of ultracold atoms \cite{senaratne2022spin}. 
In this research, the team used Bragg beams to excite spin and charge density waves separately and measured the corresponding spin and charge Bragg spectra at different interacting strengths.

By taking into account the curvature correction to the linear charge excitation spectrum and the nonlinear effect of the spin backward scattering in the spin sector, theoretical and experimental results are in good agreement.
Nevertheless, when the temperature $T$ exceeds a characteristic energy  scale of spin, but still less than typical energy of charge  Fermi energy, i.e. $E_s\ll T\ll E_c$,  the spin degree of freedom is fully  disordered, while the charge degree of freedom still remains in propagating mode. 
Such a state  is referred to as spin incoherent Luttinger liquid (SILL)\cite{cheianov2004nonunitary,fiete2004green,fiete2007colloquium,fiete2005theory,fiete2005transport,hew2008spin}. 
 The correlation function of the SILL  shows an  exponential decay of distance in the spin sector, while it remains a power-law decay in the charge sector, giving a characteristic of the emergent SILL \cite{cheianov2004nonunitary,fiete2004green,fiete2007colloquium,fiete2005theory},
 in contrast to  the case of the usual TLL that correlation functions in both spin and charge sectors exhibit power-law decaying in distance \cite{schulz1990correlation,frahm1990critical,meden1999nonuniversality,giamarchi2003quantum,miranda2003introduction}. 
 An evidence for the spin-incoherent Luttinger liquid was recently reported by R Hulet's group at Rice University \cite{Cavazos-Cavazos:2022}. 
 Although there have been extensive study on this model, a comprehensive study of the spin-coherent and incoherent Luttinger liquids, universal thermodynamics,  fractional quasiparticles and Mott phase in the 1D Hubbard model still remains elusive.

Moreover, comparing with  the continuum gas systems, one has a consensus that the Mott insulator phase dominated by antiferromagnetic ordering is of great importance on account of its uniqueness stemming from the umklapp scattering in terms of bosonization \cite{giamarchi1991umklapp}. 
There are ample thoughts and debates on the mechanism for the onset of the Mott insulator for a long time \cite{stafford1993scaling,hubbard1963electron,brinkman1970application,han2016charge}. 
But it is still a big theoretical challenge to capture the essential nature of the insulating phase. 
This massive charge sector with massless spin degree belongs to the category of Luther-Emery model \cite{luther1974backward,voit1998dynamical,orgad2001spectral} whose dynamical spectral function for Mott phase displays one or two singularities without anomalous dimension \cite{voit1998dynamical,orgad2001spectral}. 
A quantity which is experimentally accessible is the double occupancy, embodying charge fluctuations.
It gives insights into the emergence of the Mott phase\cite{glocke2007half,de2011thermodynamics,sciolla2013competition} and 
 conveys  plentiful information apart from Mott phenomenon, for instance, the Pomeranchuk effect \cite{glocke2007half,de2011thermodynamics,wietek2021mott}, antiferromagnetic correlations \cite{wietek2021mott,paiva2010fermions,gorelik2012universal}, adiabatic cooling \cite{werner2005interaction,dare2007interaction,de2011thermodynamics} over a wide  energy  scales and  temperature regimes.
One of the key contents in this paper is  to define the  Contact and Contact susceptibilities through which we further study  adiabatic cooling scheme at  quantum criticality. 

The Bethe Ansatz provides us with a feasible way to solve the 1D Hubbard Hamiltonian \cite{Lieb:1968,Degu2000}. 
The framework of building up the finite-temperature thermodynamics of 1D integrable systems  was attributed to C N Yang and C P Yang's seminal paper \cite{YangCN:1969}. 
Later M Takahashi made an important step forward by treating the thermodynamics of some  more complicated integrable systems  \cite{takahashi1972one}. 
In this scenario, quantum transfer matrix was successfully developed  to the study of the thermodynamics  of integrable models in terms of path integral \cite{klumper1996exact,Batchelor:2007}. 
These two approaches endow us with possibility to acquire the excitation spectra \cite{carmelo1991renormalized,Ess1994a}, universal free Luttinger  liquids information \cite{he2020emergence} and correlation functions \cite{frahm1990critical,qin2022hubbard} for the repulsive Hubbard model and the attractive Hubbard model \cite{DMRG,Penc1995,Krivnov1975,Woyna1983,Bogoliubov1988,Woyna1991,Ess1994a,Ess1994b,Sacra1995,Cheng:2018A,Cheng:2018B,Cheng:2018C} despite big  challenges in actual calculations of physical properties  due to the complexity of the TBA equations at arbitrary temperature.
 For higher dimensionality and various lattice symmetries \cite{hirsch1985two,georges1992hubbard,jarrell1992hubbard,ibarra2020thermodynamics}, there does not seem to have  completely applicable theoretical schemes  and commonly exploited numerical simulation methods including quantum Monte Carlo\cite{jarrell1992hubbard,dare2007interaction,kim2020spin,wietek2021mott}, matrix-decomposition algorithms \cite{white1989numerical}, dynamical mean-field theory \cite{hirsch1985two,ibarra2020thermodynamics,wietek2021mott}, exact diagonalization \cite{caffarel1994exact,wang2020zero}, perturbation theory \cite{timrov2021self,senechal2000spectral}, density-Matrix renormalization group\cite{kollath2005spin,schlunzen2017nonequilibrium} and other generalization \cite{georges1992hubbard,senechal2000spectral}.

The elementary excitations in spin and charge degree and universal thermodynamics of the 1D repulsive Hubbard model are briefly discussed in \cite{LPG:2022}.
In this paper, we develop analytical methods to obtain fundamental properties of  the 1D Hubbard model with repulsive interaction and external fields. %
In particular, using the exact solutions of the Bethe Ansatz equations of  the Hubbard  model, we  rigorously calculate fractional excitation spectra, universal thermodynamics, scaling functions, incoherent  correlation functions, Contact, Contact susceptibilities, quantum cooling and interaction-driven quantum phase transitions in a pedagodical way. 
 Exact results of various excitation spectra reveal exotic features of fractionalized spinons and holons and show subtle differences of spin magnons and charge $\eta$-pair excitations.    
Universal thermodynamical properties of spin and charge Luttigner liquids, dimensionless ratios and scaling functions near quantum phase transitions in terms of chemical potential, magnetic field and interaction are obtained explicitly near quadruple critical point. 
The asymptotic two-point correlation functions for the spin-incoherent Luttinger liquid are derived with  the help of the conformal field theory.
Finally, we study the double occupancy, the Contact, Contact susceptibilities, interaction-driven adiabatic  cooling scheme and criticality and Mott insulator behaviour of the model as well. 

The outline of this paper is the following. In section II, we introduce the basic knowledge of exact Bethe ansatz solution of the repulsive Hubbard model and calculate the excitation spectra of fractional holons and spinons as well as the gapped excitations above the ground state. 
In section III, we initiate our study  on the  thermal and magnetic properties of  TLL, quantum criticality,  non-Fermi liquid behaviour and  the universal laws in the 1D Hubbard model. 
In section IV, we study the SILL from the Bethe ansatz perspective. The asymptotic single particle Green's function and pairing correlation function are  presented.
A depiction of  double occupancy and analytical calculation of Contact and its susceptibilities with respect to the chemical potential, magnetic field and interaction near Mott phase are  given in section V. 
In this section,  we also carry out a  brief analysis on quantum  cooling and interaction driven phase transitions. The last section VI remains for a summary and outlooks.

\section{II. Fractional holons and spinons}

Fundamental low energy physics is intimately related to the elementary  excitations in 1D many-body systems. 
In particular, the low-lying excited states involve many-body correlations and determine the unique dynamics of   the 1D Hubbard model, showing novel features of  fractional holons and spinons. 
In the Mott phase, the excitations show a charge gapped phase with antiferromagnetic correlations in spin sector. 
Spin and charge particle-hole excitations reveal the origins of the spin-coherent TLL and   spin-incoherent liquid in a crossover temperature regime. 
The purpose of this section is aimed to calculate such  typical excitation spectra of the repulsive Hubbard model,  which complement the study in literature, see a review \cite{essler2005one}.

\subsection{II.1 Bethe ansatz equations and root patterns}
In the presence of external potentials, the 1D Hamiltonian of Hubbard model is given by 
\begin{equation}
H=-\sum_{j=1}^{L} \sum_{a=\uparrow \downarrow}\left(c_{j, a}^{+} c_{j+1, a}+c_{j+1, a}^{+} c_{j, a}\right)+u \sum_{j=1}^{L} (1-2n_{j \uparrow}) (1-2n_{j \downarrow})-\mu\hat{N}-2B\hat{S}_z, \label{eq-H}
\end{equation}
where  $c_{j,a}^\dagger$ and $c_{j,a}$ are the creation and annihilation operators of fermions  with spin $a$ ($a=\uparrow$ or $a=\downarrow$) at site $j$ in a  periodic lattice of length $L$, 
$n_{j \sigma}$ is particle density operator of spin $\sigma$ at site $j$, $\hat{N}=\sum_{j=1}^{L}(n_{j \uparrow}+n_{j \downarrow})$ is the total number particle operator, and $\hat{S}_z=\frac{1}{2}\sum_{j=1}^{L}(n_{j \uparrow}-n_{j \downarrow})$ is the magnetization operator. 
The hopping amplitude is set to unity, and $u$ represents an  on-site interaction between particles.
The chemical potential $\mu$ and magnetic field $B$ are  rescaled with respect to the hopping term. 
 $u>0$ for repulsion and $u<0$ for attraction, they can be converted into each other by the Shiba transformation. 
 We choose a unit system by setting  $k_B,\, \hbar$ to $1$ throughout the paper. 

At an absence of the external potential ($\mu=0,B=0$) and even lattice sites, the Hamiltonian \ref{eq-H} possesses SO(4)$\cong$SU(2)$\times$ SU(2)$/\mathbb{Z}_2$ full symmetry composed of spin rotational and $\eta$-pairing invariance\cite{Ess1994a}:
\begin{equation}
{[H,S^{\alpha}]}=0, \qquad 
[H,\eta^{\alpha}]=0, 
\end{equation}
where the spin and $\eta$-pair operators are given by 
\begin{equation}
\begin{aligned}
 S^{\alpha}=&\frac{1}{2}\sum_{j=1}^L\sum_{a,b=\uparrow,\downarrow}c_{j, a}^{+}(\sigma^{\alpha})_b^ac_{j,b},\\
 \eta^x=&-\frac{1}{2}\sum_{j=1}^L(-1)^j\left(c_{j, \uparrow}^{+}c_{j, \downarrow}^{+}+c_{j, \uparrow}c_{j, \downarrow}\right),\\
 \eta^y=&\frac{i}{2}\sum_{j=1}^L(-1)^j\left(c_{j, \uparrow}^{+}c_{j, \downarrow}^{+}-c_{j, \uparrow}c_{j, \downarrow}\right), \\
 \eta^z=&\frac{1}{2}\left(N-L\right), 
\end{aligned}
\end{equation}
which are useful to analyze elementary excitations at zero magnetic field and half-filled band \cite{Ess1994a}.
In the above equation, $\sigma^{\alpha }$ with $\alpha=x,y,z$ denote the Pauli matrices. 

In this paper, we will consider the case where the number of particles $n_c$ is less than one with the magnetization  less than $n_c/2$ for repulsive interaction, i.e.,  it corresponds to a negative chemical potential and positive magnetic field. 
The other argument spaces can be involved in terms of discrete symmetries \cite{essler2005one}.

The eigenvalues of \{$k$\}, \{$\Lambda$\}, respectively the solutions  of charge quasimomentum and spin rapidity  of this model, can be obtained  by solving  the Lieb-Wu equations \cite{Lieb:1968}. 
While Takahashi \cite{takahashi1972one} assumed that the patterns of quasimomenta  \{$k$\}, \{$\Lambda$\} exhibit different patterns,  which determine full states of the model.
The root patterns of repulsive Hubbard model can be categorized into three types of strings: single $k$, length-$n$ $\Lambda$ strings composed of $n$ spin-down electrons, length-$m$ $k-\Lambda$ string containing $m$ spin-down and $m$ spin-up particles. 
Let $M_e,M_n,M^{\prime}_n$ denote the number of $k$, $\Lambda$  strings and  $k-\Lambda$ strings of length $n$, resepctively. 
Therefore, the total particle number $N$ and spin down electron number $M$ read:
\begin{equation}
\begin{aligned}
&M=\sum_{n=1}^{\infty} n\left(M_{n}+M_{n}^{\prime}\right),  \\
&N=\mathcal{M}_{e}+\sum_{n=1}^{\infty} 2 n M_{n}^{\prime}. 
\end{aligned}
\end{equation}

In order to carry easily out our study, we prefer to present the following Bethe ansatz equations and the TBA equations of the 1D Hubbard model with a repulsive interaction. Based on these, we will derive analytically physical properties of the model. 
In terms of the string solutions, the real centers of these roots satisfy the so-called Takahashi's forms of Bethe ansatz equations \cite{takahashi1972one}:
\begin{equation}
\begin{aligned}
&k_{j} L=2 \pi I_{j}-\sum_{n=1}^{\infty} \sum_{\alpha=1}^{M_{n}} \theta\left(\frac{\sin k_{j}-\Lambda_{\alpha}^{n}}{n u}\right)-\sum_{n=1}^{\infty} \sum_{\alpha=1}^{M_{n}^{\prime}} \theta\left(\frac{\sin k_{j}-\Lambda_{\alpha}^{\prime n}}{n u}\right), \\
&\sum_{j=1}^{N-2 M^{\prime}} \theta\left(\frac{\Lambda_{\alpha}^{n}-\sin k_{j}}{n u}\right)=2 \pi J_{\alpha}^{n}+\sum_{m=1}^{\infty} \sum_{\beta=1}^{M_{m}} \Theta_{n m}\left(\frac{\Lambda_{\alpha}^{n}-\Lambda_{\beta}^{m}}{u}\right), \\
&2 L \operatorname{Re}\left[\arcsin \left(\Lambda_{\alpha}^{\prime n}+n \mathrm{i} u\right)\right]=2 \pi J_{\alpha}^{\prime n}+\sum_{j=1}^{N-2 M^{\prime}} \theta\left(\frac{\Lambda_{\alpha}^{\prime n}-\sin k_{j}}{n u}\right)+\sum_{m=1}^{\infty} \sum_{\beta=1}^{M_{m}^{\prime}} \Theta_{n m}\left(\frac{\Lambda_{\alpha}^{n}-\Lambda_{\beta}^{\prime m}}{u}\right), 
\end{aligned}
\end{equation}
where $\theta(x)=2\arctan(x)$ and $\Theta_{n m}$ is defined as
\begin{equation}
\Theta_{nm}(x)=\left\{\begin{array}{l}
\theta\left(\frac{x}{|n-m|}\right)+2 \theta\left(\frac{x}{|n-m|+2}\right)+\cdots+2 \theta\left(\frac{x}{n+m-2}\right)+\theta\left(\frac{x}{n+m}\right), \text { if } n \neq m \\
2 \theta\left(\frac{x}{2}\right)+2 \theta\left(\frac{x}{4}\right)+\cdots+2 \theta\left(\frac{x}{2n-2}\right)+\theta\left(\frac{x}{2n}\right), \text { if } n=m
\end{array}\right..
\end{equation}
The counting numbers $I_j,J_{\alpha}^{n},J_{\alpha}^{\prime n}$ are integer or half-odd integers, which rely on the odevity of string number,
\begin{eqnarray}
	&&I_{j} \text { is } \begin{cases}\text { integer } & \text { if } \sum_{m}\left(M_{m}+M_{m}^{\prime}\right) \text { is even } \\ \text { half-odd integer } & \text { if } \sum_{m}\left(M_{m}+M_{m}^{\prime}\right) \text { is odd }\end{cases},\label{I}\\
&&J_{\alpha}^{n} \text { is } \begin{cases}\text { integer } & \text { if } N-M_{n} \text { is odd } \\ \text { half-odd integer } & \text { if } N-M_{n} \text { is even, }\end{cases}, \label{J}\\
&& J_{\alpha}^{\prime n} \text { is } \begin{cases}\text { integer } & \text { if } L-N+M_{n}^{\prime} \text { is odd } \\ \text { half-odd integer } & \text { if } L-N+M_{n}^{\prime} \text { is even }\end{cases}.\label{J'}
\end{eqnarray}
The classification of these quantum numbers will be needed for characterizing the excitations which will be presented later. 

Every selected set of quantum numbers $I_j,J_{\alpha}^{n},J_{\alpha}^{\prime n}$ uniquely determine the value of quasimomentum $k_j,\,\Lambda_{\alpha}^{n},\, \Lambda_{\alpha}^{\prime n}$, and thus give a specific state, either the ground state or an excited state of various filling of the model. 
These numbers are in the ranges of
\begin{equation}
	\begin{aligned}
	&-\frac{L}{2}<I_{j} \leq \frac{L}{2}, \\
	&\left|J_{\alpha}^{n}\right| \leq \frac{1}{2}\left(N-2 M^{\prime}-\sum_{m=1}^{\infty} t_{n m} M_{m}-1\right), \\
	&\left|J_{\alpha}^{\prime n}\right| \leq \frac{1}{2}\left(L-N+2 M^{\prime}-\sum_{m=1}^{\infty} t_{n m} M_{m}^{\prime}-1\right), \label{limit}
	\end{aligned}
\end{equation}
where $M^{\prime}=\sum_{n=1}^{\infty} n M_{n}^{\prime}$ is the entire number of spin-down particles included in $k-\Lambda$ strings, and $\ t_{m n}=2 \min (m, n)-\delta_{m n}$.

Denote $\rho^p,\sigma_{n}^p,\sigma_{n}^{\prime p}$ ($\rho^h,\sigma_{n}^h,\sigma_{n}^{\prime h}$) as the densities of real quasimomenta $k$, the real part of the length-$n$ spin strings and the real part of length-$n$ $k-\Lambda$ strings for particles (holes), the root distributions of these types of root patterns are given by \cite{takahashi1972one}
\begin{equation}
\begin{aligned}
\rho^p(k)+\rho^h(k)=&\frac{1}{2\pi}+\cos k \sum_{n=1}^{\infty} \int_{-\infty}^{\infty} \d \Lambda a_{n}(\Lambda-\sin k)\left(\sigma_{n}^{\prime p}(\Lambda)+\sigma_{n}^p(\Lambda)\right), \\
\sigma_{n}^p(\Lambda)+\sigma_{n}^h(\Lambda) =& -\sum_{m=1}^{\infty} A_{n m} * \sigma_{m}^p(\Lambda) +\int_{-\pi}^{\pi} \d k a_{n}(\sin k-\Lambda) \rho^p(k),\\
\sigma_{n}^{\prime p}(\Lambda)+\sigma_{n}^{\prime h}(\Lambda) =&\frac{1}{\pi} \operatorname{Re} \frac{1}{\sqrt{1-(\Lambda-\mathrm{i} n u)^{2}}} -\sum_{m=1}^{\infty} A_{n m} * \sigma_{m}^{\prime p}(\Lambda) -\int_{-\pi}^{\pi} \d k a_{n}(\sin k-\Lambda) \rho^p(k), \label{eq-density} 
\end{aligned}
\end{equation}
respectively. In the above equations $a_n(x)=\frac{1}{2\pi}\frac{2nu}{(nu)^2+x^2}$, and the convolution term denoted by $* $ denotes a convolution
\begin{equation}
A_{nm}*f(x)=\int_{-\infty}^{\infty} \frac{\d y}{2 \pi}\frac{\d }{\d x}\Theta_{nm}\left(\frac{x-y}{u}\right)f(y).
\end{equation}
The function $A_{nm}$ denotes a derivative of the function $\Theta_{nm}$, namely 
\begin{equation}
\begin{aligned}
&A_{nm} \left(\frac{x-y}{u}\right) =\frac{1}{2 \pi}\frac{\d }{\d x}\Theta_{nm}\left(\frac{x-y}{u}\right)=\\&\left\{\begin{array}{l}
a_{|n-m|}\left(x-y\right)+2 a_{|n-m|+2}\left(x-y\right)+\cdots+2 a_{n+m-2}\left(x-y\right)+a_{n+m}\left(x-y\right), \text { if } n \neq m \\
2 a_{2}\left(x-y\right)+2 a_{4}\left(x-y\right)+\cdots+2 a_{2n-2}\left(x-y\right)+a_{2n}\left(x-y\right), \text { if } n=m
\end{array}\right..
\end{aligned}
\end{equation}

Following the Yang-Yang method \cite{YangCN:1969}, the true physical equilibrium states at finite temperature can be determined by the minimization of the free energy with respect to the densities. 
The TBA equations for dressed energies $\kappa(k),\varepsilon_n(\Lambda),\varepsilon_n^{\prime}(\Lambda)$, associated with the product of the logarithm of the ratio of hole density to particle density and temperature $\kappa=T\ln\left(\frac{\rho^h}{\rho^p}\right),\varepsilon_n=T\ln\left(\frac{\sigma_n^h}{\sigma_n^p}\right),\varepsilon_n^{\prime}=T\ln\left(\frac{\sigma_n^{\prime h}}{\sigma_n^{\prime p}}\right)$, are given by the following form \cite{takahashi1972one}
\begin{eqnarray}
\kappa(k)&=&-2 \cos k-\mu-2 u-B+\sum_{n=1}^{\infty} \int_{-\infty}^{\infty}\d \Lambda a_{n}(\sin k-\Lambda) T\ln \left(1+\e^{-\frac{\varepsilon_n^{\prime}(\Lambda)}{T}}\right)\nonumber \\
&&-\sum_{n=1}^{\infty} \int_{-\infty}^{\infty} \mathrm{d} \Lambda a_{n}(\sin k-\Lambda) T\ln \left(1+\e^{-\frac{\varepsilon_n(\Lambda)}{T}}\right),\label{eq-TBA-1} \\
\varepsilon_n(\Lambda)&=&2nB-\int_{-\pi}^{\pi}\d k \cos k a_{n}(\sin k-\Lambda) T\ln \left(1+\e^{-\frac{\kappa(k)}{T}}\right)+\sum_{m=1}^{\infty} A_{n m} * T\ln \left(1+\e^{-\frac{\varepsilon_m(\Lambda)}{T}}\right),\label{eq-TBA-2} \\
\varepsilon_n^{\prime}(\Lambda)&=&4 \textmd{Re} \sqrt{1-(\Lambda-\mathrm{i} n u)^{2}}-2 n \mu-4 n u-\int_{-\pi}^{\pi} \d k \cos k a_{n}(\sin k-\Lambda) T\ln \left(1+\e^{-\frac{\kappa(k)}{T}}\right)\nonumber \\
&&+\sum_{m=1}^{\infty} A_{n m} * T\ln \left(1+\e^{-\frac{\varepsilon_m^{\prime}(\Lambda)}{T}}\right).\label{eq-TBA-3}
\end{eqnarray}
It is observed  that the above  dressed energies are even functions with respect to  their variables and admit monotonically increasing behavior in the positive argument space. The Gibbs free energy per site is given by 
\begin{eqnarray}
f&=&-T \int_{-\pi}^{\pi} \frac{\d k}{2 \pi} \ln \left(1+\e^{-\frac{\kappa(k)}{T}}\right)+u\nonumber \\
&&-T \sum_{n=1}^{\infty} \int_{-\infty}^{\infty} \frac{\d \Lambda}{\pi} \operatorname{Re} \frac{1}{\sqrt{1-(\Lambda-\mathrm{i} n u)^{2}}}\ln \left(1+\e^{-\frac{\varepsilon_n^{\prime}(\Lambda)}{T}}\right). \label{eq-f-t} 
\end{eqnarray}
(\ref{eq-f-t}) serves as the equation of state from which one can directly obtain the profiles of ground state and thermodynamic properties of the  Hubbard model at zero and finite temperatures by solving (\ref{eq-TBA-1}), (\ref{eq-TBA-2}) and (\ref{eq-TBA-3}). 
However, finding the solution to the TBA equations~(\ref{eq-TBA-1}), (\ref{eq-TBA-2}) and (\ref{eq-TBA-3}) imposes a big theoretical challenge.
Analytical results of universal thermodynamics, Luttinger liquid properties, magnetic properties, scaling functions in the vicinities of the critical points, as well as behaviour of the Mott phase, quantum cooling are still lacking of a comprehensive study  in literature. 

\subsection{II.2 Ground state}
In the limit of zero temperature, one can see that the dressed energies of the length-$n$ $\Lambda$  strings with $n>1$ and all $k-\Lambda$  are  always positive in the parameter space.
Consequently, they do not contribute to zero temperature thermal and magnetic properties.
The real $k$ and length-$1$ $\Lambda$ string should be cut off at the points where dressed energies switch sign. Thereby, in the ground state, the coupled equations for dressed energies and densities and hole distribution functions \cite{Lieb:1968} are greatly simplified
\begin{eqnarray}
\kappa(k)&=&-2 \cos{k}-\mu-2 u-B+\int_{-A}^{A}\d \Lambda  a_1(\sin{k}-\Lambda) \varepsilon_1(\Lambda), \label{kappa}\\
\varepsilon_1(\Lambda)&=&2 B +\int_{-Q}^{Q} \d k \cos k a_1(\sin{k}-\Lambda) \kappa(k)
-\int_{-A}^{A} \d \Lambda^{\prime} a_{2}\left(\Lambda-\Lambda^{\prime}\right)\varepsilon_1(\Lambda^{\prime}), \label{epsilon}\\
\rho^{p}(k)&=& \theta_{H}(Q-|k|)\left[\frac{1}{2 \pi}+\cos k \int_{-A}^{A} \d \Lambda a_{1}(\sin k-\Lambda) \sigma_{1}^{p}(\Lambda)\right]\nonumber, \\
\sigma_{1}^{p}(\Lambda)&=& \theta_{H}(A-|\Lambda|)\left[-\int_{-A}^{A} \d \Lambda^{\prime} a_{2}\left(\Lambda-\Lambda^{\prime}\right) \sigma_{1}^{p}\left(\Lambda^{\prime}\right)+\int_{-Q}^{Q} \d k a_{1}(\sin k-\Lambda) \rho^{p}(k)\right]\nonumber,  \\
\rho^{h}(k)&=& \theta_{H}(|k|-Q)\left[\frac{1}{2 \pi}+\cos k \int_{-A}^{A} \d \Lambda a_{1}(\sin k-\Lambda) \sigma_{1}^{p}(\Lambda)\right], \nonumber \\
\sigma_{1}^{h}(\Lambda)&=& \theta_{H}(|\Lambda|-A)\left[-\int_{-A}^{A} \d \Lambda^{\prime} a_{2}\left(\Lambda-\Lambda^{\prime}\right) \sigma_{1}^{p}\left(\Lambda^{\prime}\right)+\int_{-Q}^{Q} \d k a_{1}(\sin k-\Lambda) \rho^{p}(k)\right],\nonumber  \\
\sigma_{n}^{h}(\Lambda)& =&-A_{n 1} * \sigma_{1}^{p}(\Lambda)+\int_{-Q}^{Q} \d k a_{n}(\sin k-\Lambda) \rho^{p}(k), \,\, {\rm for}, \, n \geq 2, \nonumber \\
\sigma_{n}^{\prime h}(\Lambda)&=& \frac{1}{\pi} \operatorname{Re} \frac{1}{\sqrt{1-(\Lambda-i n u)^{2}}}-\int_{-Q}^{Q} \d k a_{n}(\sin k-\Lambda) \rho^{p}(k), \nonumber
\end{eqnarray}
where $Q,\,A$ are the Fermi points of charge and spin, respectively, i.e.  $\kappa(Q)=0,\varepsilon_1(A)=0$.
$\theta_{H}$ is the Heaviside step function. 
We observe  that particles only exist within Fermi points, while holons locate outside in the ground state. 
In the above equations, we also present the hole density distribution functions for $\Lambda$ strings and $k-\Lambda$ strings for our later calculation. 
 Defining the total quantities $\rho=\rho^p+\rho^k,\, \sigma_{1}=\sigma_{1}^{p}+\sigma_{1}^{h}$, thus the  density Bethe ansatz equations  can be recast in a more succinct form
\begin{eqnarray}
\rho(k) &=& \frac{1}{2 \pi}+\cos k  \int_{-A}^{A} \d \Lambda a_{1}(\Lambda-\sin k)\sigma_{1}(\Lambda)\label{rho}, \\
\sigma_{1}(\Lambda) &=& - \int_{-A}^{A} \d \Lambda^{\prime} a_{2}(\Lambda-\Lambda^{\prime})\sigma_{1}(\Lambda^{\prime}) +\int_{-Q}^{Q} \d k a_{1}(\sin k-\Lambda) \rho(k). \label{sigma}
\end{eqnarray}
Thus, the particle number and spin-down number can be expressed as in terms of (\ref{rho}) and (\ref{sigma})
\begin{eqnarray}
n_c&=&\int_{-Q}^{Q}\d k\rho(k), \label{n0}\\
n_{\downarrow}&=&\int_{-A}^A\d \Lambda \sigma_1(\Lambda).\label{m0}
\end{eqnarray}
The momenta of the charge and spin quasiparticles for $|k|\le Q, \, |\Lambda|\le A$, and  holons  for $|k|> Q, |\Lambda|> A$ in terms of the quasi-momentum parameters  $k_j,\Lambda_{\alpha}^{n}, \, \Lambda_{\alpha}^{\prime n}$ can be calculated via the following expressions
\begin{eqnarray}
p(k)&=&\frac{2\pi  I_k}{L}=2\pi\int_0^k\d k^{\prime}\rho\left(k^{\prime}\right), \label{q1}\\
p_1(\Lambda)&=&\frac{2\pi J_1}{L}=2\pi\int_0^{\Lambda}\d \Lambda^{\prime}\sigma_1\left(\Lambda^{\prime}\right),\label{q2}\\
p_n(\Lambda)&=&\frac{2\pi J_n}{L}=2\pi\int_0^{\Lambda}\d \Lambda^{\prime}\sigma_n^h\left(\Lambda^{\prime}\right),\ n\geq2,\label{q3}\\
p_n^{\prime}(\Lambda)&=&\frac{2\pi J_n^{\prime}}{L}=-2\pi\int_0^{\Lambda}\d \Lambda^{\prime}\sigma_n^{\prime h}\left(\Lambda^{\prime}\right)+\pi(n+1),\ n\geq1. \label{q4}
\end{eqnarray}

Let us denote $N_{GS}$ be the total particle number, and $M_{GS}$ the spin-down number in the ground state.
 Consider the ground state structure in the presence of magnetic field and $0<n_c<1$ as follows (assume lattice length $L$ is even): 
$
M_e=N_{GS}=2*M_{GS}<L$, $M_1=M_{GS}$ is an odd number  and less than $N_{GS}/2,~ M_n=0$ for $n\geq 2,\, M^{\prime}_n=0$ for $n\geq1$, that is, both charge and spin sectors are partially filled and all particles form charge and spin  Fermi seas, inside of which are referred as antiholon states for charge and spinon states for spin degree of freedom. 
At zero temperature, the system lies in its lowest energy state and particles are symmetrically distributed around  zero momentum.
 Using (\ref{I}) (\ref{J}) and (\ref{q1})-(\ref{q4}), we can evaluate that $\left\{I_j\right\}$ are half-odd integers and fill in the interval $\left[-\frac{N_{GS}-1}{2},\frac{N_{GS}-1}{2}\right]$, whereas  $\left\{ J_{\alpha}^{1}\right\}$ are integer within $\left[-\frac{M_{GS}-1}{2},\frac{M_{GS}-1}{2}\right]$. 
 The momentum distribution $p(k),\, p_1(\Lambda)$ respectively covers the intervals  of  $\left[-\pi n_c,\pi n_c\right]$ and $ \left[-\pi n_{\downarrow},\pi n_{\downarrow}\right]$.
  It is inferred from (\ref{limit}) that 
the values for quantum numbers  $\left\{ J_{\alpha}^{1}\right\}$ satisfiy $\left|J_{\alpha}^{1}\right| \leq \frac{1}{2}\left(N_{GS}-M_{GS}-1\right)$. This implies that there are $N_{GS}-M_{GS}$ vacancies for $J_{\alpha}^{1}$ and thus $N_{GS}-2M_{GS}$ holes are left. 
Building on the quantum numbers of the ground state, elementary excitations can be classified into two types: gapless and gapped excitations. 
Particle-hole excitations as well as their combinations belong to the gapless ones,  which  always occur at sufficient low energy sector.
Whereas the length-$n$ $k-\Lambda$ and $\Lambda$ strings with $n>1$ require an additional energy to be excited. The key points for excitation spectrum  calculation are stated below:
 \begin{itemize} 
	\item dressed energy is an exact excitation energy of the particle with the corresponding momentum, and holon energy corresponds to the negative dressed energy.
	 In other words, energies of particle excitations for different types of quasiparticles are  $\kappa(k)$, $\varepsilon_n(\Lambda)$ and $\varepsilon^{\prime}_n(\Lambda)$,
	 while their corresponding holon excitations are $-\kappa(k)$, $ -\varepsilon_n(\Lambda)$ and $-\varepsilon^{\prime}_n(\Lambda)$, respectively.
	
	\item by changing quantum numbers  $M_e$, $M_n$ and $M^{\prime}_n$, one can  analyze the excited structure with respect to ground state for capturing the holon or particle excitations of each type strings. 
	
	\item if  there is no extra hole or particle appearing over  the ground state, the  particle-hole excitation can exist.  This type of excitation does not alter the total particle number or parity.
	
	\item  individual particle and holon excitations can be transformed  into each other, 	\textbf{e.g.}, in figure~\ref{fig-ph}, a holon excitation can be created by conducting two particle-hole excitations over a particle excitation.

\begin{figure}[H] 
	\centering    
	\includegraphics[scale=0.45]{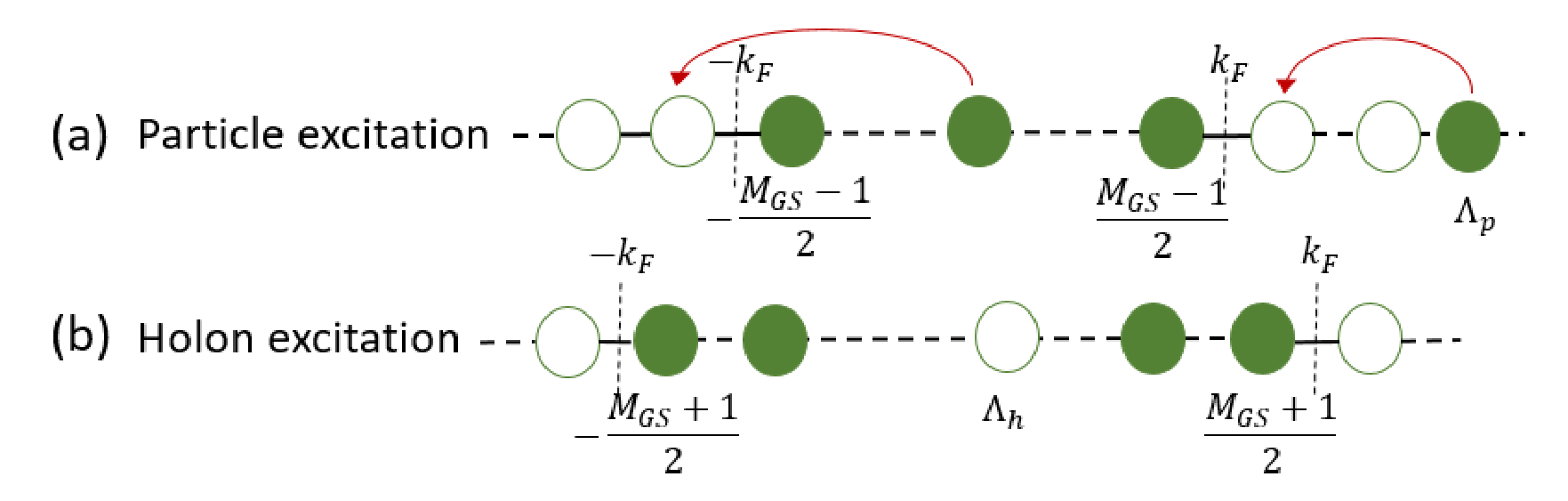}  	
	\caption{Illustration of two equivalent excitations: a transformation between particle excitation and holon excitation.}    
	\label{fig-ph}    
\end{figure}
	
	\item when $I_j$ converts from half-odd (interger) to interger (half-odd), extra constant momentum $\pm\pi n_c$ needs  to be added to total momentum with a change of  parity (In the later study, we only consider the case of $\pi n_c$ which can find common in continuous model. The case for $-\pi n_c$ is symmetric with the former.),  see, for example, figure~\ref{fig-g-ex}.\\

\begin{figure}[H] 
	\centering    
	\includegraphics[scale=0.45]{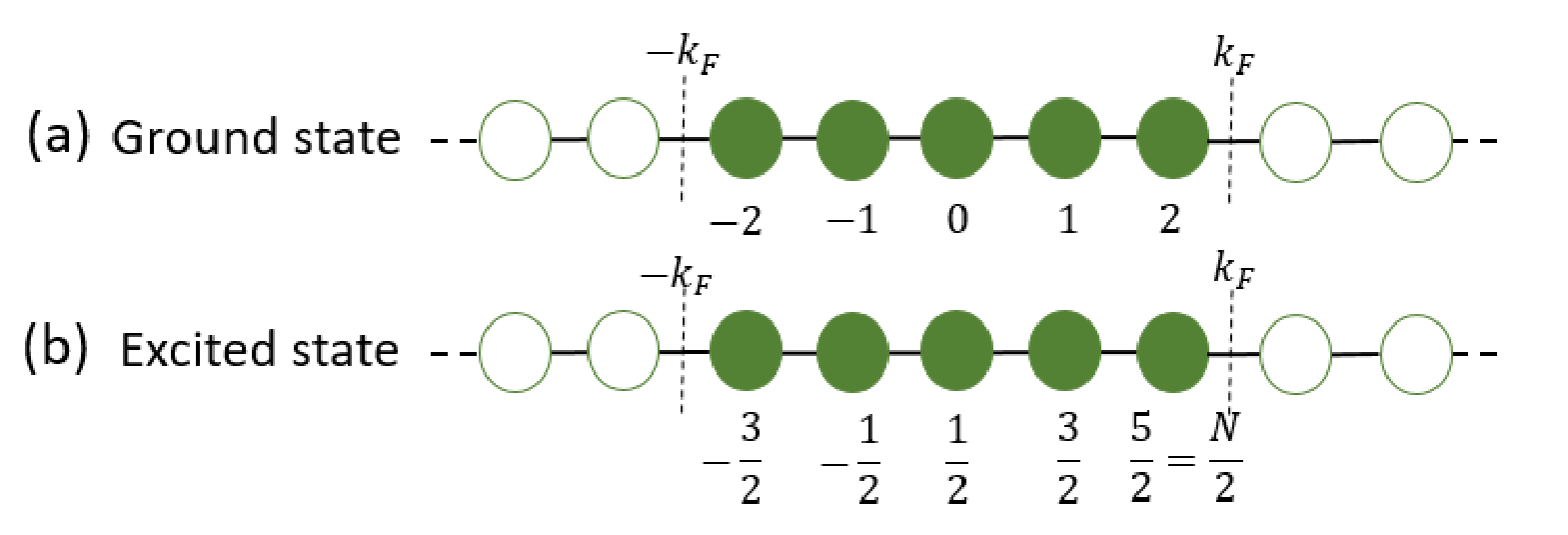}  	
	\caption{Illustration of the variation of parity in excitation. }    
	\label{fig-g-ex}     
\end{figure}
	
	\item a multi-parametric excitation can be decomposed into few-parametric elementary excitations. Once few particle excitations are determined, analytical expressions of multiple particle excitations are created by few elementary excitations. We elucidate this approach in later, \textbf{e.g.}, $M_e=N_{GS}+2$ is compose of double excitations with   $M_e=N_{GS}+1$.
\end{itemize}

The techniques mentioned above enable us to classify the processes of complicated excitations, based on which we can conceive  unique collective nature of fractional excitations. In what follows we will present various spectrum patterns of elementary excitations, which are helpful for latter analysis. 

\subsection{II.3 Fractional charge and spin excitations}

\begin{figure}[H] 
	\centering    
	\includegraphics[scale=0.73]{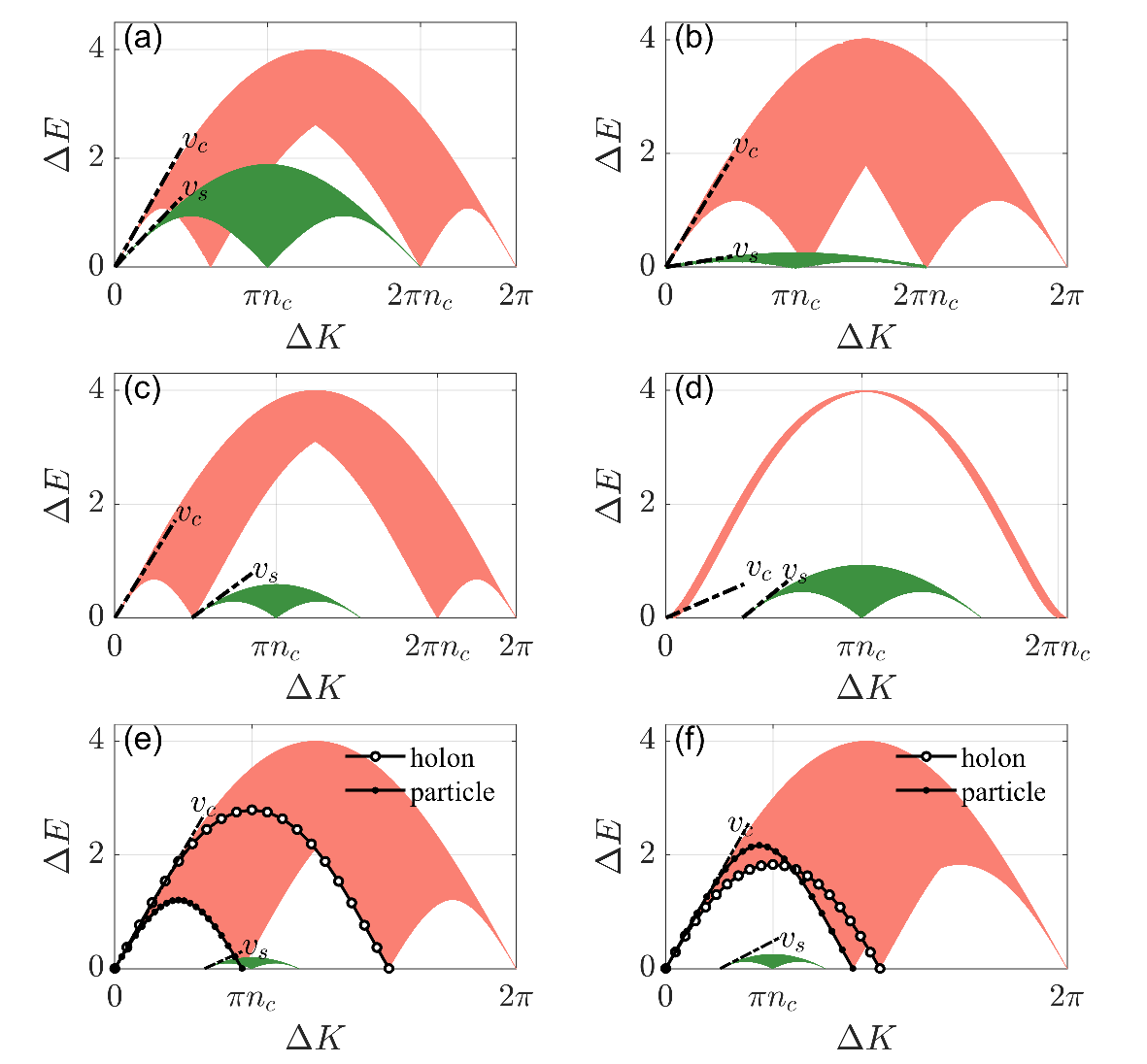}  
	\caption{Elementary excitations in spin and charge. The shaded area in orange represents a particle-hole excitation of charge, where  one particle  inside  the Fermi sea is excited to outside  at different fillings.  Whereas the shaded area in green  presents a two-holon excitation in spin sector. All figures  are drawn in the first Brillouin zone. Interaction $u$ is fixed to unity for all figures except (b):  (a) For the case away from half-filled band at zero magnetic field with $B=0,\mu=-1.5765$, we have charge and spin sound  velocities  $v_c=2.0907,\, v_s=1.2259$, respectivly; (b)  For strong coupling at zero magnetic field with $B=0,\mu=-19.22,u=10$,  we have charge and spin sound  velocities  $v_c=1.8533,v_s=0.1824$, respectively. (c) presents a general case with a presence of magnetization, $B=0.555,\mu=-1.32$;
		(d) presents charge and spin excitations near half filled phase with a finite magnetization, $B=0.555,\mu=-0.722$; (e) shows excitations near a spin-polarized band, $B=0.555,\mu=-1.77$.  At a low energy scale, holon excition energy (empty-circles) is higher than particle excitation energy (filled-circles); (f) shows excitations at $B=0.3,\mu=-2.5$ for a comparison  with the figure  (e). }            
	\label{Cph-Shh}      
\end{figure}

 Elementary excitations of the Hubbard model were studied in literature, see review \cite{essler2005one}. Here  we present a comprehensive understanding of fractional excitations and various gapped excitations which provide deep insight into understanding Luttinger liquid, spin incoherent liquid and  low energy physics.

\subsubsection{Two holon excitations}
We first study the elementary excitation of flipping one spin-down particle, for which we have a change $M_e=L,\, M_1=M_{GS}-1$.
Here we take $M_{GS}$ an odd number. 
 From (\ref{I}), (\ref{J}), we can determine that the quantum number $\left\{I_j\right\} $ are integers and $J_{\alpha}^{1}$ are half-odd integers within ranges $-\frac{L}{2}<I_{j} \leq \frac{L}{2}$ and $\left|J_{\alpha}^{1}\right| \leq \frac{1}{2}\left(N_{GS}-M_{GS}\right)$, respectively. 
 Therefore the vacancies for spin are $N_{GS}-M_{GS}+1$. 
 In contrast to the ground state, one more vacancy arises with a diminished particle  in the spin sector, and thus two holes $\Lambda_{h1},\, \Lambda_{h2}$ carrying separately spin $\frac{1}{2}$  are created in the spinon band for $\Lambda$. 
 For this spin flipping excitation, $\eta$ pairing and spin magnetization $\left(\Delta\eta^z,\Delta S^z\right)$ are determined through $\Delta \eta^z=\frac{\Delta N}{2},\, \Delta S^z=\frac{\Delta N-2\Delta M}{2}$ with $N,\, M$ total particle number and spin-down number in the excited state, i.e., $\left(\Delta\eta^z,\Delta S^z\right)=(0,1)$ in this sample. 
 
 On the other hand, the particle-hole excitation can be constructed in the charge sector  with a particle taken out from $|k|<Q$ to $|k|>Q$, leaving a holon $k_h$ inside the antiholon band and a particle $k_p$ in the holon states. 
 The energies and momenta of the particle-hole excitation in the charge sector and the two-spinon excitation in the spin sector are 
\begin{eqnarray}
E_{ph}&=&\kappa(k_p)-\kappa(k_h), \qquad P_{ph}=p(k_p)-p(k_h), \\
E_{ss}&=&-\varepsilon_1(\Lambda_{h1})-\varepsilon_1(\Lambda_{h2}), \qquad P_{ss}=-p_1(\Lambda_{h1})-p_1(\Lambda_{h2})+\pi n_c.
\end{eqnarray}

In figure~\ref{Cph-Shh}, we demonstrate the results of the particle-hole and the two-spinon excitations for various parameters. 
Obviously, one universal nature of low-energy excitation spectra is the existence of spin-charge separation behavior \cite{voit1993charge,kollath2005spin,lorenz2002evidence}, visualizing  the theory of the spin-charge separated  TLL. 
In the absence of the  magnetic field, the system has spin rotation symmetry. 
Figure~\ref{Cph-Shh} (a) shows the particle-hole excitation in the charge sector when away from half-filled lattice and the two-spinon  excitation in the spin sector which  fully accounts for the first Brillouin zone. 

In contrast, Figure~\ref{Cph-Shh} (b) shows the  particle-hole excitation  and the two-spinon  excitation for strong coupling limit, reminiscent of the spin-charge separated  spectra in the 1D Yang-Gaudin model \cite{he2020emergence}.
The slopes of the dispersion are characterised by velocities $v_c,\,v_s$, displaying $v_s\ll v_c$, which were  recently measured experimentally in a quasi-1D trapped  repulsive Fermi gas \cite{senaratne2022spin}.
This regime offers a remarkable possibility to study the spin-incoherent Luttinger liquid (SILL), which occurs when the  temperature exceeds the spin characteristic energy, but is still less than the charge energy. 
In the SILL, the spin degree of freedom is frozen. However, the charge degree of freedom propagates as a sound mode, behaving like spinless fermions \cite{cheianov2004nonunitary,fiete2004green,fiete2007colloquium}.

\begin{figure}[H] 
	\centering    
	\includegraphics[scale=0.4]{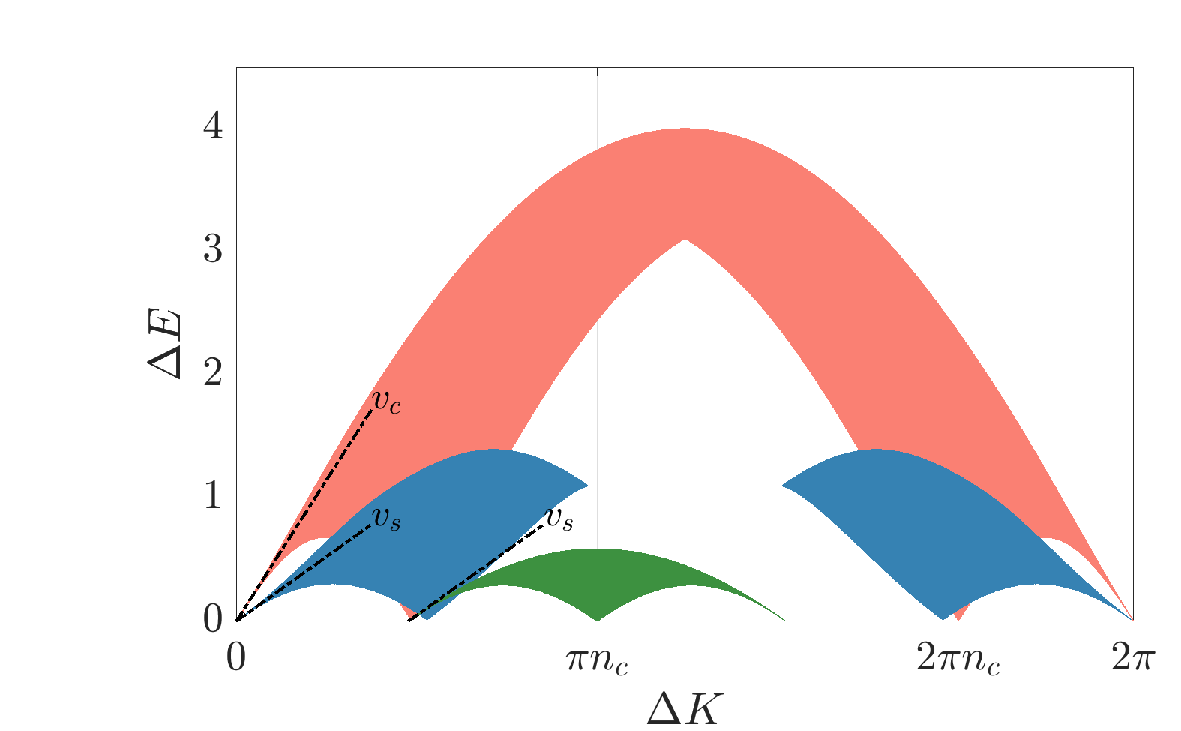}  
	\caption{The shaded area in orange represents a particle-hole excitation of charge. The blue shaded area  denotes spin particle-hole excitation. The green  color denotes the two-spinon excitation, same as  figure~\ref{Cph-Shh} (c). Parameter setting: $B=0.555,\mu=-1.32,u=1$.}       
	\label{Cph-Shh-Sph}     
\end{figure}

Figure~\ref{Cph-Shh} (c), (d) presents the  particle-hole excitation  and the two-spinon  excitation  for an arbitrary  magnetic field,  away from half-filling and near half-filling, respectively.
In these two cases, the spin degree is highly suppressed in the presence of the magnetic field.
The spin and charge  velocities are significantly different, and the spectra of spin and charge display a large separation. 
Exspecially for figure~\ref{Cph-Shh} (d),
the charge band is very narrow, behaving like a single particle, which was experimentally studied in \cite{vijayan2020time}.
The spin velocity exceeds the sound velocity of charge. 
Figure~\ref{Cph-Shh} (e), (f) shows that in the charge sector the energy of holon excitation at small momentum can exceed that of particle  excitation in the Hubbard model  due to the cosine term presented in the charge dispersion, which is prohibited in continuous model\cite{giamarchi2003quantum,he2020emergence}. 
Nevertheless, when the system approaches the Mott phase, charge excitation gradually shrinks to the cases of the figure~\ref{Cph-Shh} (d), indicating an emergence of single particle excitation in the charge sector \cite{vijayan2020time}. In general, the spin sector exhibits similar behaviour as that of  Heisenberg chain\cite{ovchinnikov1969excitation,nocera2016magnetic}.
When magnetic field is applied, the two-spinon excitation is no longer the  low-energy one in the sense that it does not fill the entire Brillouin area. 
The role played by the low-energy excitation in spin degree of freedom is the spin particle-hole spectrum as depicted in figure~\ref{Cph-Shh-Sph}.

\subsubsection{Antiholon- and holon-spinon excitations}
\label{me_m1}
Let's further consider adding or removing a particle with spin up or down over  the ground state in the Hubbard model.
There are four configurations with the following quantum numbers: $(a) M_e=N_{GS}+1, M_1=M_{GS}+1$; (b)  $M_e=N_{GS}-1, M_1=M_{GS}$; (c) $M_e=N_{GS}+1, M_1=M_{GS}$; (d) $M_e=N_{GS}-1, M_1=M_{GS}-1$.
These four excitations can be classified by the charge and spin magnetization values $\left(\Delta\eta^z,\Delta S^z\right)$:
correspondingly, 
(a) antiholon-spinon particle $(\frac{1}{2},-\frac{1}{2})$, (b) holon-spinon particle $(-\frac{1}{2},-\frac{1}{2})$, (c) antiholon-spinon hole $(\frac{1}{2},\frac{1}{2})$, and (d) holon-spinon hole $(-\frac{1}{2},\frac{1}{2})$.
Consequently, (a) has one antiholon particle in the charge and one spinon particle in the spin sector; 
 (b) has one  holon in the charge and one spinon particle in the spin sector; 
  (c) has an antiholon particle in the charge and one spinon hole in the spin sector; 
   (d) has  one holon in the charge and one spinon hole  in the spin sector. 
 Analogous to previous discussion, we obtain the following results of excitation spectra 
\begin{eqnarray}
\begin{array}{ll} 
(a) & \left\{\begin{array}{l}
E_{\overline{h}\overline{s}}=\kappa(k_p)+\varepsilon_1(\Lambda_{p})\\P_{\overline{h}\overline{s}}=p(k_p)+p_1(\Lambda_p)+\pi n_c
\end{array},\right.\\
(b)&  \left\{\begin{array}{l}
E_{h\overline{s}}=-\kappa(k_h)+\varepsilon_1(\Lambda_{p})\\P_{h\overline{s}}=-p(k_h)+p_1(\Lambda_p)
\end{array},\right.\\
(c)&  \left\{\begin{array}{l}
E_{\overline{h}s}=\kappa(k_p)-\varepsilon_1(\Lambda_{h})\\P_{\overline{h}s}=p(k_p)-p_1(\Lambda_h)
\end{array},\right.\\
(d) & \left\{\begin{array}{l}
E_{hs}=-\kappa(k_h)-\varepsilon_1(\Lambda_{h})\\P_{hs}=-p(k_h)-p_1(\Lambda_h)+\pi n_c
\end{array}.\right.
\end{array}
\end{eqnarray}

\begin{figure}[H] 
	\centering    
	\includegraphics[scale=0.4]{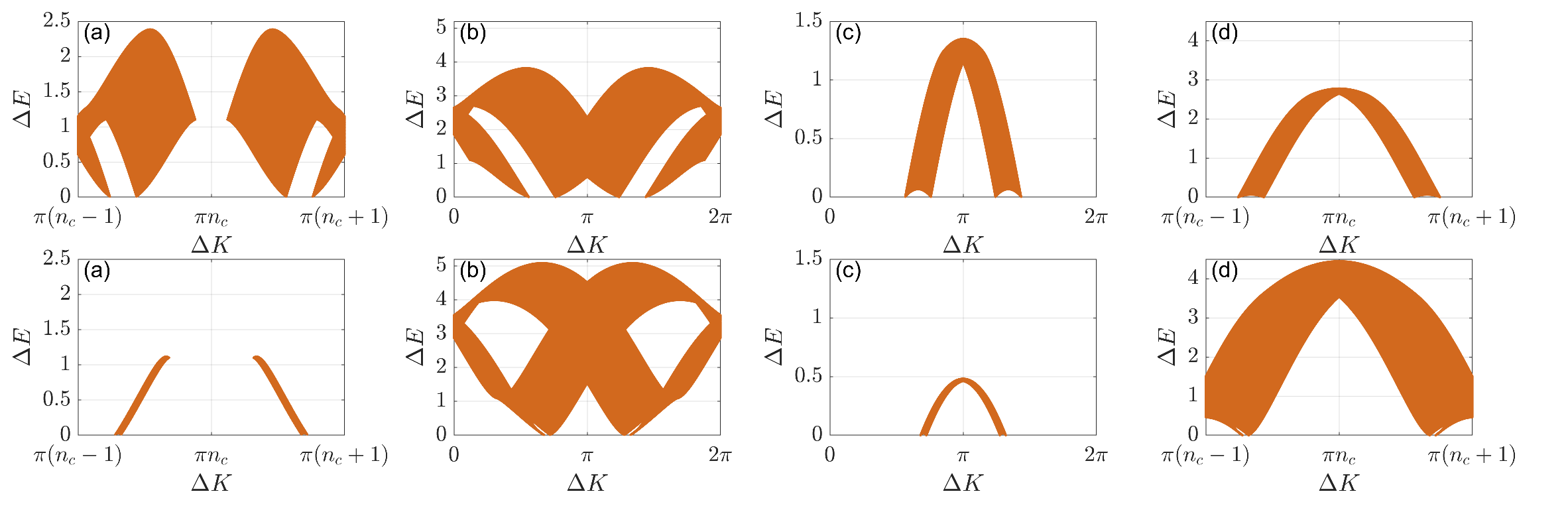}  
	\caption{From left to right: (a) antiholon-spinon particle excitation $(\frac{1}{2},-\frac{1}{2})$, (b)  holon-spinon particle excitation$(-\frac{1}{2},-\frac{1}{2})$, (c) antiholon-spinon hole excitation $(\frac{1}{2},\frac{1}{2})$ and (d)  holon-spinon hole excitation $(-\frac{1}{2},\frac{1}{2})$. Upper panel: excitation spectra in the low spin-down density area with the potentials  $B=0.555,\mu=-1.85,u=1$ and density $n_c=0.6609,n_{\downarrow}=0.0962$;  Lower  panel: excitation spectra near  half-filling with the potentials  $B=0.555,\mu=-0.722,u=1$ and density $n_c=0.9775,n_{\downarrow}=0.298$. In each panel, the excitation bands in  (a) and (d)  are centered around the $\pi n_c$, whereas the excitation bands in  (b) and (c) are centered around  $\pi$.} 
	\label{Me-1-M1-1}    
\end{figure}

In the light of the above discussion, these four excitation patterns can be converted to each other through  certain numbers of particle-hole excitations. Their continuum spectra are depicted in figure~\ref{Me-1-M1-1}.
  Comparing upper (low spin-down density) and  lower (near half-filling) panels of the figure~\ref{Me-1-M1-1}, we observe that the cases (a)(c)(d), which are related to the antiholon excitations in charge sector or spinon hole excitations in spin part, are severely affected by the filling and magnetization. 
When spin sector tends to vanish or charge sector becomes gapped, the profiles of excitations behave like single particle behaviours.

In general, we observe that the gapless excitations occur only at a sufficiently small energy scale. 
The charge and spin excitations are well decoupled only in certain configurations demonstrated in  figure~\ref{Cph-Shh} and figure~\ref{Cph-Shh-Sph}.
This spin-charge separation phenomenon is unique in 1D strongly correlated electron systems and can be described by the spin-coherent and spin-incoherent TLL theory.  
We will further discuss the spin-coherent and spin-incoherent TLLs in the 1D Hubbard model below.

\subsection{II.4 $k-\Lambda$ string excitations}
\label{kL}
\begin{figure}[H] 
	\centering    
	\includegraphics[scale=0.4]{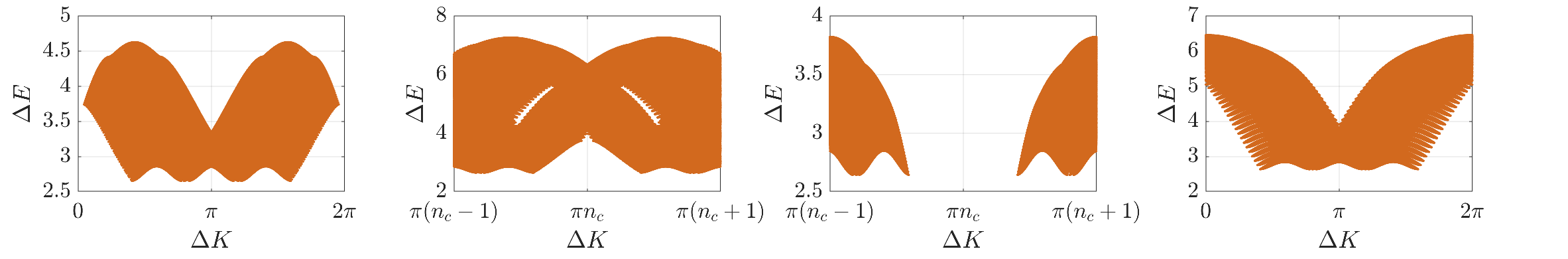}  
	\caption{From left to right: length-1 $k-\Lambda$ string is added to the excitations given in case 'Antiholon- and holon-spinon excitations': (a) antiholon-spinon particle excitation, (b) holon-spinon particle excitation, (c) antiholon-spinon hole excitation and (d) holon-spinon hole excitation. The parameters are $B=0.555,\mu=-1.32,u=1$.}       
	\label{Me-1-M1-1-M1p-1}     
\end{figure}

In comparison with the continuous systems, one peculiar excitation of 1D Hubbard model is the gapped bound states called $k-\Lambda$ string, with length-1 formed by two electrons with opposite spins due to lattice effect.
 Such states  make important contributions to optical conductivity close to half-filling \cite{veness2016mobile}. 
 The continuum spectra  are shaped like a parabola with a convex opening, see the bottom of the excitation spectra in figure~\ref{Me-1-M1-1-M1p-1}.
The length-$1$ $k-\Lambda$ string shows to be the lowest charge gapped excitation among the other   $k-\Lambda$ strings, leading to the appearance of quasiparticle $\Lambda_{1}$ in the distribution of  quantum number $J_{\alpha}^{\prime1}$.
Such a gapped excitation also leads to the change of the parity of $I_j$. 
 It can be seen  that the excited energy of this length-$1$ $k-\Lambda$  has an energy and momentum  $E=\varepsilon^{\prime}_1(\Lambda_{1}), \, P=p^{\prime}_1(\Lambda_{1})+\pi n_c$.  
For a more general consideration of $k-\Lambda$ excitations, one $k-\Lambda$ string is added to the occupation numbers as discussed in section Antiholon- and holon-spinon excitations, i.e., for antiholon-spinon $M_e=N_{GS}+1,M_1=M_{GS}+1,M^{\prime}_1=1$, of which
we decompose the multi-parameter excitation into elementary ones. 
Similarly, we obtain the information of such gapped excited states.  
The energy and momentum of this compound case read 

\begin{eqnarray}
&&(a) \, \left\{\begin{array}{l}
E_{\overline{h}\overline{s}}=\kappa(k_p)+\varepsilon_1(\Lambda_{p})+\varepsilon^{\prime}_1(\Lambda_{1})\\P_{\overline{h}\overline{s}}=p(k_p)+p_1(\Lambda_p)+p^{\prime}_1(\Lambda_{1})
\end{array},\right.\\
&& (b)\,  \left\{\begin{array}{l}
E_{h\overline{s}}=-\kappa(k_h)+\varepsilon_1(\Lambda_{p})+\varepsilon^{\prime}_1(\Lambda_{1})\\P_{h\overline{s}}=-p(k_h)+p_1(\Lambda_p)+p^{\prime}_1(\Lambda_{1})+\pi n_c
\end{array},\right.\\
&& (c)\,  \left\{\begin{array}{l}
E_{\overline{h}s}=\kappa(k_p)-\varepsilon_1(\Lambda_{h})+\varepsilon^{\prime}_1(\Lambda_{1})\\P_{\overline{h}s}=p(k_p)-p_1(\Lambda_h)+p^{\prime}_1(\Lambda_{1})+\pi n_c
\end{array},\right.\\
&& (d)\, \left\{\begin{array}{l}
E_{hs}=-\kappa(k_h)-\varepsilon_1(\Lambda_{h})+\varepsilon^{\prime}_1(\Lambda_{1})\\P_{hs}=-p(k_h)-p_1(\Lambda_h)+p^{\prime}_1(\Lambda_{1})
\end{array}.\right.
\end{eqnarray}
The corresponding excitation spectra are plotted in figure~\ref{Me-1-M1-1-M1p-1}. The length-$1$ $k-\Lambda$ string leads to gapped continuum bands over the states (a) antiholon-spinon excitation, (b) holon-spinon excitation, (c) antiholon-spinon excitation and (d) holon-spinon excitation.

\subsection{II.5 High spin string excitations}
\begin{figure}[H] 
	\centering    
	\includegraphics[scale=0.45]{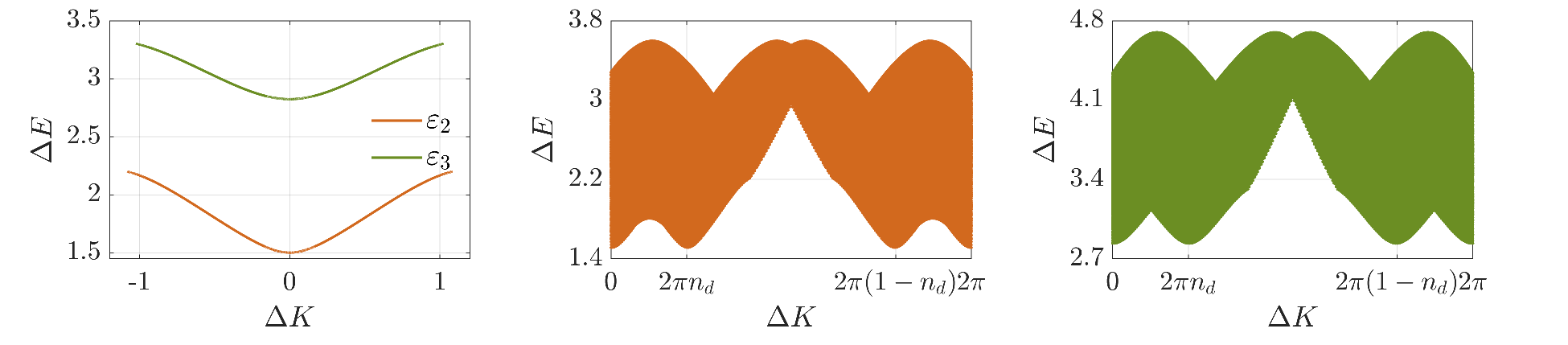}  
	\caption{Excitation spectra of length-$n$ spinon strings. From left to right: one-parametric excitation of  length-2 $\Lambda$ string (orange line) and length-3 $\Lambda$ string (green line); the length-$1$  $\Lambda$ particle-hole excitation with a length-$2$ $\Lambda$ particle excitation; length-$1$ $\Lambda$ particle-hole excitation with a  length-$3$ $\Lambda$ particle excitation. The parameters are $B=0.555,\mu=-1.32,u=1$.}       
	\label{M1-m1-M23-1}     
\end{figure}

Now we discuss the excitations of high spin string. 
Due to the antiferromagnetic ordering, the spin $\Lambda$ string bound states in the 1D Heisenberg chain 
have received a great deal of interest \cite{essler2005one,WangZ:2018,molter2014bound}. 
Like the spin root patterns in 1D repulsive Fermi gas, the $\Lambda$ strings with length greater than $1$  are gapped too. 
We consider the case of the quantum  numbers $M_1=M_{GS}-1, \, M_n=1$ with $n>1$, leading to an extra particle $\Lambda_n$ created in the parameter space of $J_{\alpha}^{n}$.
 This configuration corresponds to the process of flipping $n-1$ spin-up and simultaneously removing a spin-down from the channel of $J_{\alpha}^{1}$ to constitute a length-$n$ $\Lambda$ string, and thus one has the  energy $\varepsilon_n(\Lambda_n)$ and momentum $p_n(\Lambda_n)$ for the newly generated quasiparticle. 
 This situation is different from a multi-spinon-magnon excitation, where the quantum number of down-spin $M$ is fixed \cite{essler2005one}.  
 Meanwhile we can apply a particle-hole excitation in spin sector together with the  length-$n$ $\Lambda$ string. 
 In general, its energy and momentum are given by
\begin{eqnarray}
E&=&-\varepsilon_1(\Lambda_h)+\varepsilon_1(\Lambda_p)+\varepsilon_n(\Lambda_n),\\
P&=& -p_1(\Lambda_h)+p_1(\Lambda_p)+p_n(\Lambda_n).
\end{eqnarray}
In figure~\ref{M1-m1-M23-1}, we present the continuum excitation spectra for high spin strings of length-$2$ and -$3$. 
 The magnitudes of the energy gaps depend on the length of the spinon bound states. 

\section{III. Universal thermodynamics}

The 1D Hubbard model lays out profound many-body physics at zero and finite temperatures.
 It exhibits rich quantum phases and universal thermodynamics when the temperature is much less than the Fermi energy. 
 A comprehensive understanding of the universal low temperature behaviour still remains challenging due to complexity of the Bethe ansatz equations and its $N!$-many terms of the wave function.
 Most study on thermal and magnetic properties of the 1D Hubbard model  has been carried out at zero temperature and half-filling or in dilute limit due to the complexity and intricacy of spin and charge root string patterns in the Bethe ansatz equations and the TBA equations. 
Even in the low-temperature regime, there is no well established understanding of universal behaviour, such as TLL and quantum scaling functions etc.
Likely, this can be resorted to numerical simulation by iterative means \cite{kawakami1989thermodynamic,usuki1990thermodynamic,mott1968metal,qin2022hubbard,ilievski2017ballistic}.
However,  analytical results of the TLL and quantum criticality are still elusive. 
 In this section,  we will present our analytical and numerical study of universal thermodynamics and quantum criticality of the 1D Hubbard model by means of the TBA equations near quadruple critical point. In this regime, the model is rarely studied.
\subsection{III.1 Dimensionless ratios and phase diagram }
 In this subsection, we  study the dimensionless ratios and their application in the 1D Hubbard model. 
Before doing so, let us first give a brief view of the ground state phase diagram determined by equations~(\ref{kappa}) and (\ref{epsilon}). 
In the limit of zero temperature, the dressed energies $\kappa(k)$ and $\varepsilon_1(\Lambda)$ can be used to analyze the phase transitions and  thermodynamic quantities. 
In the grand canonical ensemble, the magnetic field and chemical potential  determine the integration boundaries $Q,\, A$ through the conditions  $\kappa(Q)=0,\varepsilon_1(A)=0$.
 The particle density $n_c=\int_{-Q}^Q\d k\rho(k)$,  the spin-down electrons per site $n_{\downarrow}=\int_{-A}^A\d \Lambda\sigma_1(\Lambda)$ and the magnetization  $m=n_c/2-n_{\downarrow}$ can be obtained from the root densities equations~(\ref{rho}) and (\ref{sigma}).
The magnetic field and chemical potential drive the system from one phase to another at zero temperature.
Phase transitions usually occur with the conditions $\kappa(0)=0$ or $\varepsilon_1(0)=0$, or with the presence of an energy gap $\kappa(\pi)=0$ in part.
Using these conditions we may give five phases of states \cite{essler2005one,takahashi1974low}: (I)  vacuum with $Q=0,\, A=0$ or $n_c=m=0$; (II) partially filled and spin fully polarized state  with $0<Q<\pi,\, A=0$ or $0<n_c<1,\, m=n_c/2$; the boundary condition is given  by $Q=0$ and $Q=\pi$; (III) half-filled and spin fully polarized state  with $Q=\pi,\, A=0$ or $n_c=1,m=1/2$; (IV) partially filled and magnetized band with $0<Q<\pi,\, 0<A\leq\infty$ or $0<n_c<1,0\leq m<n_c/2$; the phase boundary  between phase (II) and (IV) is given by $\kappa(0)<0<\kappa(\pi),\, \varepsilon_1(0)=0$; (V) half-filled, magnetized band, also referred to as Mott insulator with $Q=\pi,\, 0<A\leq \infty$ or $n_c=1,0\leq m<n_c/2$, the phase boundary between (V) and (IV) is given  by $\kappa(\pi)=0,\varepsilon_1(0)<0$. 
Substituting each phase boundary conditions into the dressed energy equations (\ref{kappa}) and (\ref{epsilon}), we can analytically determine all critical fields in the $B-\mu$ plane, also see \cite{essler2005one}.
As a matter of fact, thermal properties feature dramatic quantum  fluctuations around a quantum critical point (QCP),
whereas the dimensionless ratios, the ratios between the fluctuations of two types of sources, can become comparable to temperature, for instance the susceptibility Wilson ratio $R^{\chi_s}_w$ \cite{Wilson:1975} and the compressibility Wilson ratio $R^{\chi_c}_w$ \cite{guan2013wilson,yu2016dimensionless} 
\begin{equation}
R^{\chi_s}_w=\frac{4}{3}\left(\frac{\pi k_B}{\mu_Bg}\right)^2\frac{\chi_s}{C_v/T}, \qquad R^{\chi_c}_w=\frac{\pi^2k^2_B}{3}\frac{\chi_c}{C_v/T}, 
\end{equation}
describing  the competition between magnetic fluctuation or particle number fluctuation and thermal fluctuations, respectively.  
In the above expressions, $\chi_s$ ($\chi_c$) is the magnetic susceptibility (compressibility), respectively. 

These ratios can be constants independent of temperature in the TLL phase. 
Therefore the ground state phase diagram can be characterized by the dimensionless ratios.  

Moreover, the Gr{\"u}neisen ratio \cite{kuchler2003divergence,boehler1980experimental}, which was introduced by Eduard Gr{\"u}neisen \cite{Gruneisen-AdP-1908}  in the beginning of 20th Century in the study of the effect of volume change of a crystal lattice on its vibrational frequencies, 
\begin{equation}
\Gamma=\frac{V\frac{\d p}{\d T}|_{V,N}}{\frac{\d E}{\d T}|_{V,N}}=\frac{1}{T}\frac{\frac{\partial^2p}{\partial\mu^2}\frac{\partial p}{\partial T}-\frac{\partial^2p}{\partial\mu\partial T}\frac{\partial p}{\partial \mu}}{\frac{\partial^2p}{\partial\mu^2}\frac{\partial^2p}{\partial T^2}-\left(\frac{\partial^2p}{\partial\mu\partial T}\right)^2},
\end{equation}
has been widely used to  study the caloric effect of solids and phase transitions associated with the changes of volume, chemical potential, interaction and magnetic field.   
Similar to the magnetic Gr{\"u}neisen ratio, which quantifies the magnetocaloric effect in the  refrigeration with the variation of magnetic field, the interaction driven  Gr{\"u}neisen ratio will be studied later, quantifying  the caloric effect in the  refrigeration driven by the variation of the interaction strength.  

\begin{figure}[H] 
	\centering        
	\includegraphics[scale=0.4]{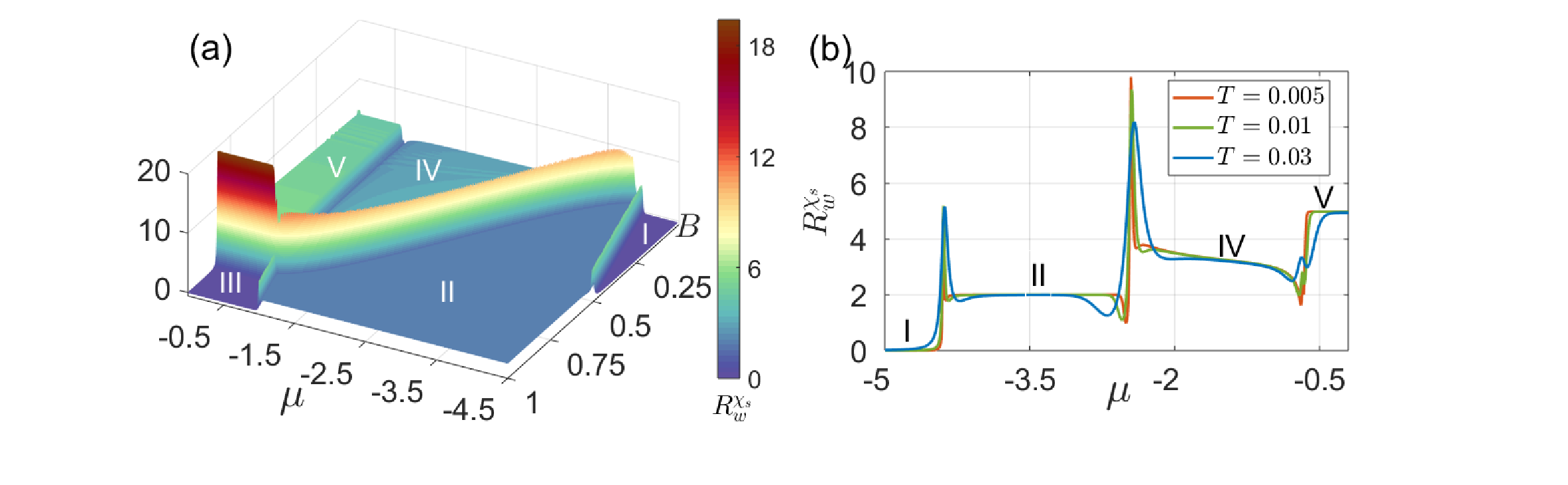}     
	\caption{(a) The 3D plot of the Wilson ratio $R^{\chi_s}_w$ maps out  phase diagram of the 1D Hubbard model  at low temperatures, which significantly marks different phases of states.    The parameter setting reads: $T=0.005$ and $u=1$. (b) Plot of Wilson ratio $R^{\chi_s}_w$ at fixed $B=0.4,u=1$.  As chemical potantial varies, four phases I II IV V can be reached.}            
	\label{fig-R}         
\end{figure}

A significant aspect of these ratios is their characterization of the TLL and their scaling behaviour  near critical points \cite{guan2013wilson,yu2016dimensionless,yu2020gruneisen,Cheng:2018A}. In figure~\ref{fig-R} (a), we present the three-dimensional plot of the magnetic Wilson ratio in the $B-\mu-R^{\chi_s}_w$ parameter space, while in figure~\ref{fig-R} (b) we show the Wilson ratio as a function of $\mu$ for fixed $B$. 
It is remarkable that the magnetic Wilson ratio is suddenly enhanced  near the QCP, distinguishing  different phases.
Moreover, in terms of bosonization results of the magnetic susceptibility, specific heat, Luttinger parameter and velocity, the Wilson ratios are given by 
\begin{eqnarray}
\text{II:}\quad R^{\chi_s}_w&\approx& 2,\nonumber\\\text{IV:}\quad R^{\chi_s}_w&\approx&4v_cK_s/(v_s+v_c),\nonumber\\\text{V:}\quad R^{\chi_s}_w&\approx& 8K_s,\nonumber\\\text{I and III:}\quad R^{\chi_s}_w&=& 0,\label{Rs}
\end{eqnarray}
where $K_s$ is the Luttinger parameter for spin, and  $v_{c,s}$ are sound velocities for charge and spin, respectively. These relations (\ref{Rs}) are in excellent agreement with figure~\ref{fig-R}.

The ground state of the 1D repulsive Hubbard model consists of five phases: 
empty lattice I, partially filled and fully-polarized phase II, fully-filled and -polarized phase III, partially-filled and -polarized phase IV, and fully-filled and partially-polarized phase V (Mott insulator) in figure~\ref{fig-R}. 
Among these, the phase IV shows the  most abundant  physics 
In this phase, charge and spin degrees of freedom coexist and dramatically make up the phase of spin-charge separated TLLs \cite{tomonaga1950remarks,luttinger1963exactly,haldane1981luttinger}, showing non-Landau Fermi liquid behaviour in one dimension. 
 In the TLL phase, all quasiparticles form collective motions and thus  decouple into two propagating modes with different different velocities $v_c,\, v_s$\cite{voit1993charge,kollath2005spin,lorenz2002evidence}.
  which rely on the value of the coupling and can be effectively estimated from TBA equations. The separated phenomenon can also be observed from their excitation spectra. In APPENDIX A, we show that the phase IV can be seen as two-component free fluids characterised by additivity of free energy at low-energy level, while phase II and V are representatives of single TLL:
\begin{eqnarray}
f&=&f_0-\frac{\pi T^2}{6}\left(\frac{1}{v_c}+\frac{1}{v_s}\right) \quad  \text {phase IV},\label{f-TLL2}\\
f&=&f_0-\frac{\pi T^2}{6}\frac{1}{v_c}\quad \text {phase II},\label{f-TLLc}\\
f&=&f_0-\frac{\pi T^2}{6}\frac{1}{v_s}\quad \text {phase V},\label{f-TLLs}
\end{eqnarray}
where $v_c$ and $v_s$ are defined through $v_c=\kappa^{'}(Q)/(2\pi \rho(Q)) $, $v_s=\varepsilon_1^{'}(A)/(2\pi \sigma_1(A))$ with cut-off points $Q,A$ respectively. 
The relation between free energy and velocity is a universal common nature enjoyed by a large family of systems\cite{he2020emergence,guan2013fermi,Cheng:2018A}.
 Taking derivative of free energy with respect to temperature,
  specific heat is directly derived $C_v=\pi T /3\left(1/v_c+1/v_s\right),C_v=\pi T/(3v_c),C_v=\pi T/(3v_s)$ for IV,II and V, respectively.
The above results (\ref{f-TLL2})-(\ref{f-TLLs}) provide the universal leading order correction of the temperature at low energy of  interacting many-body systems, i.e., characteristic of thermodynamics in the TLLs. 
%

\subsection{III.2 Contributions from $k-\Lambda$ string in phase III}
In the low-temperature limit $\mu,B>T$, the low energy behaviours of the system are immune to the gapped string excitations
 when the states are away from the QCPs in figure~\ref{fig-R}. 
By analysing the TBA equations (\ref{eq-TBA-1}), (\ref{eq-TBA-2}) and (\ref{eq-TBA-3}),  we observe that it is reasonable to ignore the gapped string states and retain only the gapless string states in both $\Lambda$ and $k-\Lambda$ strings at low energy. 
The driving terms in these equations significantly determine the contributions to the low energy states.
The length-$n$ $\Lambda$ strings and  $k-\Lambda$ string have lesser  contributions to the low energy states when $n$ is larger. 
Here we just calculate the contributions from the length-$1$ $\Lambda$ and $k-\Lambda$ strings in the phase III in which both are gapped. 
The TBA equations suggest that the greater the value of the chemical potential $\mu$ and magnetic field $B$ relative to the temperature $T$, the smaller the contributions of the $k-\Lambda$ and $\Lambda$  strings.
In the phase III,  the absolute value of chemical potential is small, while the magnetic field $B$ is large. 
Therefore, without losing generality, we here consider the length-$1$ $k-\Lambda$ and $\Lambda$  strings  at low temperature $|\mu|\sim T$ and  $B>T$ in the phase III. 
After tedious iterations on free energy shown in APPENDIX B, we obtain a close form of the free energy 
\begin{equation}
f=-\mu-u-B+g_{\frac{3}{2}}T^{\frac{3}{2}}+g_2T^{2}+g_{\frac{5}{2}}T^{\frac{5}{2}}+O(T^3,\e^{-\frac{4B}{T}},\e^{\frac{4\mu}{T}}), \label{f}
\end{equation}
where the coefficients are given explicitly by
\begin{eqnarray}
g_{\frac{3}{2}}&=&-\frac{\lambda_1\pi^{\frac{1}{2}}}{\eta^{\frac{1}{2}}_1}\e^{-\frac{2B+\eta_0}{T}}+\frac{f_{\frac{1}{2}}}{2\pi^{\frac{1}{2}}}\e^{\frac{2\mu}{T}}+\frac{f_{\frac{3}{2}}}{2\pi^{\frac{1}{2}}},\label{g} \\
g_{2}&=&-\left(\frac{\lambda_1f_{\frac{3}{2}}}{\eta^{\frac{1}{2}}_1u}+\frac{f_{\frac{1}{2}}}{2\eta^{\frac{1}{2}}_1\pi u}\right)\e^{-\frac{2B+\eta_0}{T}}+\frac{f_{\frac{1}{2}}f_{\frac{3}{2}}}{4\pi u}\e^{\frac{2\mu}{T}}, \nonumber\\
g_{\frac{5}{2}}&=&-\left(\frac{\lambda_1f^2_{\frac{3}{2}}}{2\eta^{\frac{1}{2}}_1\pi^{\frac{1}{2}}u^2}+\frac{\lambda_2\pi^{\frac{1}{2}}}{4\eta^{\frac{3}{2}}_1}+\frac{f_{\frac{1}{2}}f_{\frac{3}{2}}}{2\eta^{\frac{1}{2}}_1\pi^{\frac{3}{2}}u^2}\right)\e^{-\frac{2B+\eta_0}{T}}+\left(\frac{3f_{\frac{1}{2}}f^2_{\frac{3}{2}}}{32\pi^{\frac{3}{2}}u^2}+\frac{f_{\frac{3}{2}}}{32\pi^{\frac{1}{2}}}\right)\e^{\frac{2\mu}{T}}+\frac{f_{\frac{5}{2}}}{32\pi^{\frac{1}{2}}}, \nonumber
\end{eqnarray}
where we denote the function $f_n=\Li_n\left(-\e^{\frac{2-\mu-2u-B}{T}}\right)$, and the parameters $\lambda_1=\frac{1}{\pi\sqrt{1+u^2}}$, $\lambda_2=\frac{1-2u^2}{\pi(1+u^2)^{5/2}}$, $\eta_0=4(u-\sqrt{1+u^2})$, and $\eta_1=\frac{2}{(1+u^2)^{3/2}}$.
 By analysing (\ref{g}), we observe that in the functions $g_{\frac{3}{2},2,\frac{5}{2}}$, the terms involving exponents $\e^{-\frac{2B+\eta_0}{T}}$ and $ \e^{\frac{2\mu}{T}}$ come from gapped length-$1$ $\Lambda$ and $k-\Lambda$ strings, respectively, while the terms that do not contain any exponential terms stem from the contributions of gapless charge $k$. 
 In the above results, we have omitted the contributions from the strings with the lengths more than  length-$2$.

\begin{figure}[t] 
	\centering    
	\includegraphics[scale=0.5]{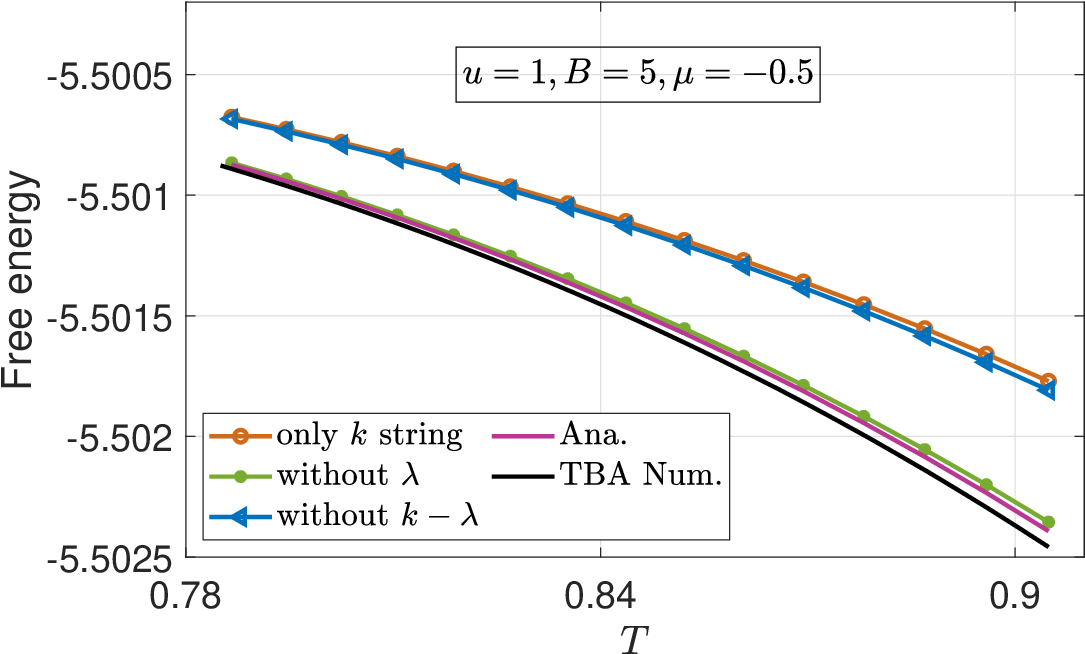}  
	\caption{Free energy for gapped length-$1$  $\Lambda$ and $k-\Lambda$ strings. The black and purple solid curves denote the numerical result and analytical results (\ref{f})  of the free energy, respectively.  The red empty-circle-solid line denotes the only contribution from the charge $k$ degree of freedom. The green filled-circle-solid line denotes the free energy without $\Lambda-\Lambda$ strings, while the blue triangular-solid line denotes the free energy without $k-\Lambda$ strings. It is obvious that the $\Lambda$ string contribution is negligible. 
}       
	\label{fig-fT}     
\end{figure}

From figure~\ref{fig-fT}, we observe that our analytic expression of the free energy is consistent with the numerical result obtained from the TBA equations 
(\ref{eq-TBA-1}), (\ref{eq-TBA-2}) and (\ref{eq-TBA-3}). 
The contributions from the $k-\Lambda$ string are greater than that from the $\Lambda$ strings due to the choice of parameters $B>T$ and $|\mu|\sim T$.  
Moreover, from the absolute value of free energy, we observe that the order of the $k-\Lambda$ string contributions is $O(10^{-4})$ as temperature tends to $T=1$. 
In general, it can be safely ignored in the low temperature behaviour. 
In the phase III, the contributions from the gapped $\Lambda$ strings are very small and negligible too. 
But the length-$1$ $\Lambda$ string plays an important role in the phase IV and V.  
More detailed calculations of the free energy for the phase III are given in APPENDIX B.

\subsection{III.3 Magnetic properties at zero temperature near quadruple critical point}

The magnetic properties of the 1D Hubbard model at zero temperature provide a benchmark for understanding universal low energy physics. 
In order to get insightful analytical results of magnetic properties of the model, here we try to analytically  calculate of the magnetic properties under the conditions of high density (Fermi point $Q$ tends to $\pi$) and high magnetization limits ($A$ tends to $0$), namely, the region in figure~\ref{fig-R} around the quadruple point.
The physics in this region has not been studied thoroughly in the literature. 
The main idea  is to expand the TBA equations in terms of small quantities $\delta=\pi-Q$ and $A$. 
To this end, we  need to calculate the integral interval of charge from the range $[0,Q]$ to $[0,\pi-Q]$ and thus the second term of (\ref{epsilon}) can be re-expressed by substituting $k$ by $\pi-k$, namely
\begin{eqnarray}
&&\int_{-Q}^{Q} \d k \cos k a_1(\sin{k}-\Lambda) \kappa(k) \label{kappa-1}\\
&=&\int_{-Q}^0\d k \cos k (a_1(\sin{k}+\Lambda)+a_1(\sin{k}-\Lambda))\kappa(k)\nonumber\\
&=&-4\operatorname{Re}\sqrt{1-(\Lambda-\mathrm{i}u)^2}+4u+\int_{0}^{\pi-Q} \d k \cos k (a_1(\sin{k}+\Lambda)+a_1(\sin{k}-\Lambda)) \kappa(\pi-k).\nonumber
\end{eqnarray}
Using the above expression, the spin dressed energy is rewritten in terms of small integral boundaries
\begin{eqnarray}
\varepsilon_1(\Lambda)&=& \varepsilon^0(\Lambda) +\int_{0}^{\pi-Q} \d k \cos k (a_1(\sin{k}+\Lambda)+a_1(\sin{k}-\Lambda)) \kappa(\pi-k)\nonumber\\
&&
-\int_{-A}^{A} \d \Lambda^{\prime} a_{2}\left(\Lambda-\Lambda^{\prime}\right) \varepsilon_1(\Lambda^{\prime}),\label{varepsilon1-1}
\end{eqnarray}
where $\varepsilon^0(\Lambda)=2B-4\operatorname{Re}\sqrt{1-(\Lambda-\mathrm{i}u)^2}+4u$ is the leading term. 
We further write the dressed energies $\kappa(\pi-k)=\sum_{i=0}^{\infty}\kappa^i(\pi-k)\delta^i,\varepsilon_1(\Lambda)=\sum_{i=0}^{\infty}\varepsilon^i(\Lambda)A^i$,  where the coefficients $\kappa^i,\, \varepsilon^i$ can be determined by iterations of the TBA equations. 
Here, we present $\kappa(k)$ in terms of $\delta =\pi-Q$
\begin{eqnarray}
&&\int_{0}^{\pi-Q} \d k \cos k (a_1(\sin{k}+\Lambda)+a_1(\sin{k}-\Lambda)) \kappa(\pi-k)\nonumber \\
&=&\delta \,2a_1(\Lambda)\kappa^0(\pi)+\delta^2\, (2a_1(\Lambda)\kappa^1(\pi)-a_1(\Lambda)\kappa^{0'}(\pi))\nonumber \\
&&+\delta^3\,(2a_1(\Lambda)\kappa^2(\pi)-a_1(\Lambda)\kappa^{1'}(\pi)+\frac{1}{3}(a_1^{''}(\Lambda)-a_1(\Lambda))\kappa^0(\pi)+\frac{1}{3}a_1(\Lambda)\kappa^{0''}(\pi)),
\end{eqnarray}
where the prime $'$ ($''$) denotes first (second) derivative with respect to $k$.  In order to derive above expansion, the following derivative formula   are used 
\begin{eqnarray}
F(n)&=&\int_{0}^{a(n)} \d xf(x,n),\nonumber\\
F^{'}(n)&=&\int_{0}^{a(n)} \d x\frac{\partial{f(x,n)}}{\partial{n}}+f(a(n),n)\frac{\d a(n)}{\d n},
\end{eqnarray}
here $n$ stands for a small quantity. 
Using the above expressions, we may calculate (\ref{varepsilon1-1}) 
\begin{equation}
\varepsilon_1(\Lambda)=\varepsilon^0(\Lambda)-\int_{-A}^{A} \d \Lambda^{\prime} a_{2}\left(\Lambda-\Lambda^{\prime}\right) \varepsilon_1(\Lambda^{\prime})\label{e1},
\end{equation}
where the leading term $\varepsilon^0(\Lambda)$ incorporating the role of charge is modified as:
\begin{eqnarray}
\varepsilon^0(\Lambda)&=&2B-4\operatorname{Re}\sqrt{1-(\Lambda-\mathrm{i}u)^2}+4u\\
&&+\delta(2a_1(\Lambda)\kappa^0(\pi))\nonumber\\
& &+\delta^2(2a_1(\Lambda)\kappa^1(\pi)-a_1(\Lambda)\kappa^{0'}(\pi))\nonumber \\
&&+\delta^3(2a_1(\Lambda)\kappa^2(\pi)-a_1(\Lambda)\kappa^{1'}(\pi)+\frac{1}{3}(a_1^{''}(\Lambda)-a_1(\Lambda))\kappa^0(\pi)+\frac{1}{3}a_1(\Lambda)\kappa^{0''}(\pi)). \nonumber 
\end{eqnarray}

Taking an expansion on equation (\ref{e1}) and comparing  order by order with respect to the small $A$ on both sides, we obtain the coefficients of the first three orders in $\varepsilon_1(\Lambda)$ 
\begin{eqnarray}
\varepsilon_1(\Lambda) &=&\varepsilon^0(\Lambda)+A(-2a_2(\Lambda)\varepsilon^0(0))\nonumber\\
& &+A^2(-2a_2(\Lambda)\varepsilon^1(0)-a_2(\Lambda)\varepsilon^{0'}(0))\\
&&+A^3(-a_2(\Lambda)\varepsilon^{1'}(0)-2a_2(\Lambda)\varepsilon^2(0)-\frac{1}{3}(a_2^{''}(\Lambda)\varepsilon^0(0)+a_2(\Lambda)\varepsilon^{0''}(0))), \nonumber
\end{eqnarray}
with 
\begin{eqnarray}
\varepsilon^0(0)&=&2B+4(u-\sqrt{1+u^2})+\frac{2}{\pi u}\kappa^0(\pi)\delta+\frac{1}{\pi u}(2\kappa^1(\pi)-\kappa^{0'}(\pi))\delta^2,\\
\varepsilon^{0'}(0)&=&0,\quad \varepsilon^{0''}(0)=\frac{4}{(1+u^2)^{\frac{3}{2}}}.
\end{eqnarray}
 The quantities associated with the charge degrees of freedom can be calculated using the same manner. 
 Consequently,  expressions of charge and spin equations can be written in terms of the orders of $\delta$ and $A$
\begin{eqnarray}
\kappa(k)&=&\bar{\delta}\bm{M_c}\bar{A}^T,\\
\varepsilon_1(\Lambda)&=&\bar{\delta}\bm{M_s}\bar{A}^T,
\end{eqnarray}
where $\bar{\delta}=\left(\delta^0,\delta^1,\delta^2,\delta^3\right),\bar{A}=\left(A^0,A^1,A^2,A^3\right)$, $T$ denotes transpose and $\bm{M_c},\bm{M_s}$ are two matrices which take the following forms
\begin{equation}
\begin{aligned}
\bm{M_c}=&\left(\begin{matrix}
-2\cos k-\mu-2u-B&0&0&0\\
2a_1(\sin{k})\beta_2&\frac{4}{\pi u}a_1(\sin{k})\beta_1&0&0\\-\frac{2}{\pi u}a_1(\sin{k})\beta_2&-\frac{4}{\pi^2 u^2}a_1(\sin{k})\beta_1+\frac{8}{\pi^2 u^2}a_1(\sin{k})\beta_2&0&0\\\left(\begin{matrix} \frac{2}{\pi^2 u^2}a_1(\sin{k})\beta_2\\+\frac{1}{3}a_1^{''}(\sin{k})\beta_2\\+\frac{4}{3(1+u^2)^{\frac{3}{2}}}a_1(\sin{k})\end{matrix}\right)&0&0&0	
\end{matrix}\right), 
\end{aligned}
\end{equation}
\begin{equation}
\begin{aligned}
\bm{M_s}=
&\begin{pmatrix}
\left(\begin{matrix}2B+4u\\-4\operatorname{Re}\sqrt{1-(\Lambda-\mathrm{i}u)^2}\end{matrix}\right)&2a_1(\Lambda)\beta_1&0&\left(\begin{matrix}\frac{1}{3}(a_1^{''}(\Lambda)\\-a_1(\Lambda))\beta_1\\-\frac{2}{3}a_1(\Lambda)\end{matrix}\right)\\
-2a_2(\Lambda)\beta_2&\left(\begin{matrix}\frac{4}{\pi u}(a_1(\Lambda)\beta_2\\-a_2(\Lambda)\beta_1)\end{matrix}\right)&\frac{8}{\pi^2 u^2}a_1(\Lambda)\beta_1&0\\\frac{2}{\pi u}a_2(\Lambda)\beta_2&\left(\begin{matrix}-\frac{4}{\pi^2 u^2}a_1(\Lambda)\beta_2\\-\frac{8}{\pi^2 u^2}a_2(\Lambda)\beta_2\\+\frac{4}{\pi^2 u^2}a_2(\Lambda)\beta_1\end{matrix}\right)&0&0\\\left(\begin{matrix} -\frac{2}{\pi^2 u^2}a_2(\Lambda)\beta_2\\-\frac{1}{3}a_2^{''}(\Lambda)\beta_2\\-\frac{4}{3(1+u^2)^{\frac{3}{2}}}a_2(\Lambda)\end{matrix}\right)&0&0&0\end{pmatrix}.
\end{aligned}
\end{equation}

To simplify our notations in the following analysis, we define 
\begin{equation}
\beta_1=2-\mu-2u-B, \qquad \beta_2=2B+4(u-\sqrt{1+u^2}),\label{beta-1-2}
\end{equation}
 which characterize the leading contributions in the charge and spin dressed energies, respectively. 
By analysing  the construction of the matrices $\bm{M_c},\bm{M_s}$, it can be deduced that $\beta_1$ and $\beta_2$ are at least the second orders of $A$ or $\delta$, namely we have 
\begin{eqnarray}
\kappa(k)&=&-2\cos k-\mu-2u-B+2a_1(\sin{k})\beta_2A+\frac{4}{3(1+u^2)^{\frac{3}{2}}}a_1(\sin{k})A^3, \label{kappa0}\\
\varepsilon_1(\Lambda)&=&2B+4(u-\operatorname{Re}\sqrt{1-(\Lambda-\mathrm{i}u)^2})\nonumber \\
&&+2a_1(\Lambda)\beta_1\delta-\frac{2}{3}a_1(\Lambda)\delta^3-2a_2(\Lambda)\beta_2A-\frac{4}{3(1+u^2)^{\frac{3}{2}}}a_2(\Lambda)A^3.
\end{eqnarray}
In the above results, we only retain terms up to the third order. 
 By applying the Fermi point conditions $\kappa(Q)=0$ and $\varepsilon_1(A)=0$, we obtain $\beta_1$ and $\beta_2$ as  
\begin{eqnarray}
\beta_1&=&\delta^2+\frac{4\eta_1}{3\pi u}A^3, \label{eq-beta1}\\
\beta_2&=&-\frac{4}{3\pi u}\delta^3-\eta_1\left(A^2+\frac{2}{3\pi u}A^3\right) \label{eq-beta2}
\end{eqnarray}
with  $\eta_1=\frac{2}{(1+u^2)^{3/2}}$. 
Utilizing iteration repeatedly, the Fermi point conditions give rise to (up to the order $O(\beta^2_1,\beta^2_2)$)
\begin{eqnarray}
\delta^2&=&\beta_1-\frac{4}{3\pi u\eta^{\frac{1}{2}}_1}(-\beta_2)^{\frac{3}{2}}\left\{1-\frac{1}{\pi u}\left[(\frac{-\beta_2}{\eta_1})^{\frac{1}{2}}+\frac{2\beta^{\frac{3}{2}}_1}{(-\beta_2)}\right]\right\},\\
\eta_1 A^2&=&-\beta_2-\frac{2}{3\pi u}\left[\frac{(-\beta_2)^{\frac{3}{2}}}{\eta^{\frac{1}{2}}_1}+2\beta_1^{\frac{3}{2}}\right]+\frac{2}{3\pi^2 u^2}\frac{(-\beta_2)^2}{\eta_1}\nonumber\\
&&+\frac{8}{3\pi^2 u^2\eta^{\frac{1}{2}}_1}\left[\frac{1}{2}(-\beta_2)^{\frac{1}{2}}\beta_1^{\frac{3}{2}}+(-\beta_2)^{\frac{3}{2}}\beta_1^{\frac{1}{2}}\right].\label{d-A}
\end{eqnarray}

Similarly, we apply the above technique to deal with the density Bethe ansatz equations (\ref{rho}) and (\ref{sigma}) and find 
the matrix form:
\begin{eqnarray}
\rho(k)&=&\bar{\delta}\bm{M_{dc}}\bar{A}^T,\\
\sigma_1(\Lambda)&=& \bar{\delta}\bm{M_{ds}}\bar{A}^T,
\end{eqnarray}
where the matrices $\bm{M_{dc}},\bm{M_{ds}}$ are given by 
\begin{eqnarray}
\bm{M_{dc}}&= &\begin{pmatrix}
\frac{1}{2\pi}&0&0&0\\
2\cos ka_1(\sin{k})\lambda_1&-\frac{2}{\pi^2 u}\cos ka_1(\sin{k})&0&0\\-\frac{2}{\pi u}\cos ka_1(\sin{k})\lambda_1&\frac{4}{\pi u}\cos ka_1(\sin{k})\left(\frac{2}{\pi u}\lambda_1+\frac{1}{2\pi^2 u}\right)&0&0\\ \left(\begin{matrix}\frac{2}{\pi^2 u^2}\cos ka_1(\sin{k})\lambda_1\\+\frac{1}{3}a_1^{''}(\sin{k})\cos k\lambda_1\\+\frac{1}{3}\cos ka_1(\sin{k})\lambda_2\end{matrix}\right)&0&0&0
\end{pmatrix}, \\
\bm{M_{ds}}&= &\begin{pmatrix}
\frac{1}{\pi}\operatorname{Re}\frac{1}{\sqrt{1-(\Lambda+\mathrm{i}u)^2}}&-\frac{1}{\pi}a_1(\Lambda)&0&-\frac{1}{6\pi}a_1^{''}(\Lambda)\\
-2a_2(\Lambda)\lambda_1&\frac{4}{\pi u}(a_1(\Lambda)\lambda_1+\frac{1}{2\pi}a_2(\Lambda))&-\frac{4}{\pi^3 u^2}a_1(\Lambda)&0\\\frac{2}{\pi u}a_2(\Lambda)\lambda_1&\left(\begin{matrix}-\frac{4}{\pi^2 u^2}a_1(\Lambda)\lambda_1\\-\frac{8}{\pi^2 u^2}a_2(\Lambda)\lambda_1\\-\frac{2}{\pi^3 u^2}a_2(\Lambda)\end{matrix}\right)&0&0\\\left(\begin{matrix} -\frac{2}{\pi^2 u^2}a_2(\Lambda)\lambda_1\\-\frac{1}{3}a_2^{''}(\Lambda)\lambda_1\\-\frac{1}{3}a_2(\Lambda)\lambda_2\end{matrix}\right)&0&0&0	
\end{pmatrix},
\end{eqnarray}
where the constant values $\lambda_1=\frac{1}{\pi\sqrt{1+u^2}}$, $\lambda_2=\frac{1-2u^2}{\pi(1+u^2)^{5/2}}$ related  to the interaction scale. 
By integrating the densities within the Fermi points (\ref{n0}), (\ref{m0}), the particle number $n_c$ and down-spin number $n_{\downarrow}$ are given by 
\begin{eqnarray}
n_c&=&1-2\int_{0}^{\pi-Q}\d k \rho(\pi-k)=1-2\delta\left(\frac{1}{2\pi}-\frac{2\lambda_1}{\pi  u} A+\frac{2\lambda_1}{\pi^2 u^2} A^2\right)-\frac{4}{\pi^3 u^2}\delta^2 A, \label{nc-c}\\
n_{\downarrow}&=&2\left[A\left(\lambda_1-\frac{2}{\pi  u}\delta\left(\frac{1}{2\pi}-\frac{2\lambda_1}{\pi  u} A\right)\right)-A^2\frac{1}{\pi u}\left(\lambda_1-\frac{1}{\pi^2 u}\delta\right)+A^3\left(\frac{\lambda_1}{\pi^2 u^2}+\frac{\lambda_2}{6}\right)\right].\label{nd-s}
\end{eqnarray} 

 It is more intuitive to express the  Fermi points in terms of particle numbers. 
 Based on the above relations, we have the small parameters near the  Fermi points $\delta=\pi-Q$ and $A$ in terms of doping  parameter $\hat{n}_c=1-n_c$ and $n_{\downarrow}$, 
\begin{eqnarray}
\delta&=&\pi\hat{n}_c\left(1+\frac{2}{u}n_{\downarrow}+\frac{4}{u^2}n_{\downarrow}^2\right), \label{eq-delta}\\
A&=&\frac{\pi u}{2}\zeta_1 n_{\downarrow}\left[1+\zeta_1\left(\hat{n}_c+\frac{n_{\downarrow}}{2}\right)+\zeta^2_1\left(\hat{n}_c+\frac{n_{\downarrow}}{2}\right)^2-\zeta_2n_{\downarrow}^2\right]\label{eq-A}
\end{eqnarray}
with $\zeta_1=\frac{1}{\pi  u\lambda_1}$,  and $\zeta_2=\frac{\lambda_2}{24\lambda^3_1}$. 
Using the free energy (\ref{eq-f-t}) with $T=0$ and the above expression (\ref{kappa0}) for $\kappa(k)$,  we may obtain
\begin{equation}
 f=-\mu-u-B-\frac{8\lambda_1}{3(1+u^2)^{\frac{3}{2}}}A^3-\frac{2}{3\pi}\delta^3.
\end{equation}

On the other hand, in terms of the discrete symmetries of the repulsive Hubbard model \cite{essler2005one}, the free energy of the repulsive system can be transformed to that of the attractive case via the relation $f_r(\mu,B,T,u)=f_a(-B,-\mu,T,-u)-\mu-B$, where subscript $a(r)$ means attractive (repulsive) interaction.
It turns out that the results obtained not only  cover that of the attractive model \cite{Cheng:2018A} in the low density and strong coupling limits, but also are valid for the arbitrary interaction strength away from the strong coupling limit. 
In general, it is straightforward to show that there is a one-to-one correspondence from the high density area for repulsive case to low density regime for attractive case \cite{Cheng:2018A}. 
 Take the derivative of free energy, one can obtain the relationship between their thermodynamic quantities:
\begin{eqnarray}
\frac{\partial f_r}{\partial \mu}&=&-\frac{\partial f_a}{\partial B}-1  \longrightarrow n_{r,\mu}=1-2m_{a,B},\\
\frac{\partial f_r}{\partial B}&=&-\frac{\partial f_a}{\partial \mu}-1 \longrightarrow 2m_{r,B}=1-n_{a,\mu}.
\end{eqnarray}

\begin{figure}[t] 
	\centering        
	\includegraphics[scale=0.47]{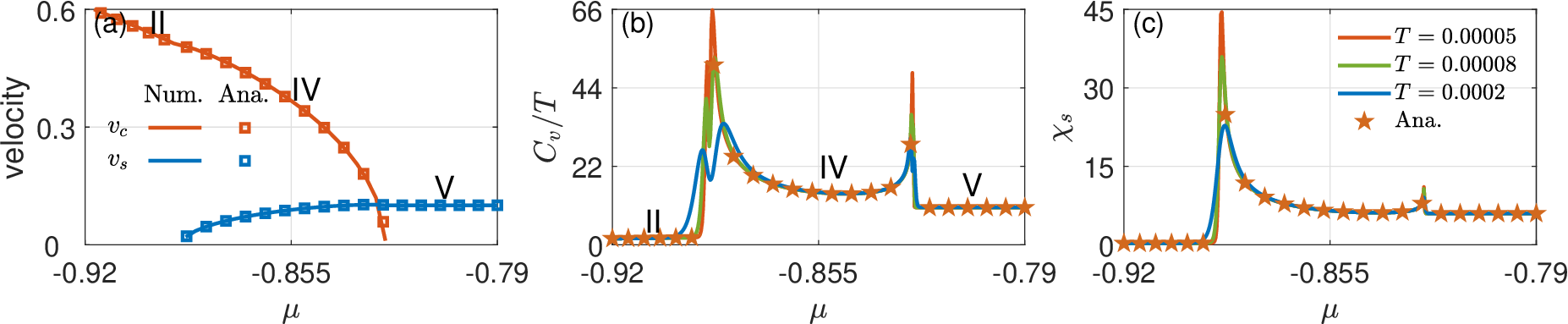}     
	\caption{(a) Spin and charge velocities as function of chemical potential in phase II, IV and V, in  which analytic results are given by the equations (\ref{kappa-Q})-(\ref{sigma-A}) for IV,  equation (\ref{II}) for II and equation (\ref{V-vs}) for V, respectively.
(b) shows the analytical and numerical results of the specific heat in terms of chemical potential. The curves of $C_v/T$ at different temperatures  collapse to a single curve of the analytical result in each phase. 
(c) shows the magnetic susceptibility: agreement between the numerical and analytical results which  are given by equations (\ref{chi-gs}), (\ref{II}), (\ref{V-chis}).
In the TLL phases, the susceptibility  does not depend  on  temperature. 
 The whole numerical setting reads $B=0.825,\,u=1$.}            
	\label{fig-v}         
\end{figure}

Using these results, and by the definitions of spin and charge characteristic velocities $v_c=\frac{\kappa^{'}(k)}{2\pi \rho(k)}|_{k=Q}$, $v_s=\frac{\varepsilon_1^{'}(\Lambda)}{2\pi \sigma(\Lambda)}|_{\Lambda=A}$, we obtain the following quantities 
\begin{eqnarray}
\kappa^{'}(Q)&=&2\delta-\frac{1}{3}\delta^3, \label{kappa-Q}\\
\rho(Q)&=&\frac{1}{2\pi}-\frac{2\lambda_1}{\pi u}A+\frac{2\lambda_1}{\pi^2 u^2}A^2-\frac{2\lambda_1}{\pi^3 u^3}A^3+\frac{2\lambda_1}{\pi u^3}\delta^2 A+\frac{2\lambda_1}{3\pi u^3}A^3\nonumber\\
&&-\frac{\lambda_2}{3\pi u}A^3+\frac{\lambda_1}{\pi u}\delta^2A+\frac{2}{\pi^3 u^2}\delta A-\frac{2}{\pi^4 u^3}\left(1+4\pi\lambda_1\right)\delta A^2, \label{rho-Q}\\
\varepsilon_1^{'}(A)&=&2\eta_1A+\frac{\eta_2}{6}A^3, \label{varepsilon-A}\\
\sigma_1(A)&=&\lambda_1+\frac{\lambda_2}{2}A^2-\frac{1}{\pi^2 u}\delta+\frac{4\lambda_1}{\pi^2 u^2}\delta A-\frac{8\lambda_1}{\pi^3 u^3}\delta A^2\nonumber\\&&+\frac{1}{\pi^2 u^3}\delta A^2-\frac{4}{\pi^4 u^3}\delta^2 A+\frac{1}{3\pi^2 u^3}\delta^3-\frac{\lambda_1}{\pi u} A+\frac{1}{\pi^3 u^2}\delta A\nonumber\\
&&+\frac{\lambda_1}{3\pi u^3}A^3+\frac{\lambda_1}{\pi^2 u^2}A^2-\frac{1}{\pi^4 u^3}\delta A^2-\frac{\lambda_1}{\pi^3 u^3}A^3-\frac{\lambda_2}{6\pi u}A^3\label{sigma-A}
\end{eqnarray}
with $\eta_2=12\frac{1-4u^2}{\pi(1+u^2)^{7/2}}$. 
The analytic results and numerical simulation of sound velocities and thermodynamic properties by the solution of TBA equations are showed in figure~\ref{fig-v}. 
From the figure~\ref{fig-v} (a), we observe that $v_c$ approaches zero at the critical point  for the phase transition from magnetized phase IV to the Mott phase V.
On the other hand, $v_s$ goes to zero that signifies an approach to the spin fully polarized phase II. 
 Thus the velocities can serve as a signature of quantum phase trasnition. 
 In figure~\ref{fig-v} (b),  we show the specific heat in terms of velocities at various temperatures: demonstrating the analytical results in comparing with numerical calculations. 
 The analytical results of the specific heat are given by 
 \begin{eqnarray}
 	C_v&=&\frac{\pi T}{3}\left(\frac{1}{v_c}+\frac{1}{v_s}\right) \quad  \text {phase IV},\\
 	C_v&=&\frac{\pi T}{3}\frac{1}{v_c}\quad \text {phase II},\\
 	C_v&=&\frac{\pi T}{3}\frac{1}{v_s}\quad \text {phase V}.
 \end{eqnarray} 
The quantity $C_v/T$ becomes temperature independent for the TLL states. 

\subsection{III.4 Additivity rules of charge and spin susceptibilities}
Now we calculate the charge and spin susceptibilities $\chi_c=\frac{\partial n_c}{\partial \mu}|_{B(m)}$ and $\chi_s=\frac{\partial m}{\partial B}|_{\mu(n_c)}$ with $m=\frac{n_c}{2}-n_{\downarrow}$. 
In the grand canonical ensemble driven by external potentials $B$ and $\mu$, we have the following expressions of  spin and charge susceptibilities 
\begin{eqnarray}
\chi_c&=&\frac{\partial n}{\partial \mu}|_B=\left(-\frac{1}{\pi}+\frac{4 \lambda_1}{\pi u}A-\frac{4 \lambda_1}{\pi^2 u^2}A^2-\frac{8}{\pi^3 u^2}\delta A \right)\frac{\partial \delta}{\partial \mu}\nonumber\\
&& +\left(\frac{4 \lambda_1}{\pi u}\delta-\frac{8\lambda_1}{\pi^2 u^2}\delta A-\frac{4}{\pi^3 u^2}\delta^2\right)\frac{\partial A}{\partial \mu}=\chi_c^{(1)}+\chi_c^{(2)}, \label{chi-gc}\\
\chi_s&=&\frac{\partial m}{\partial B}|_\mu=\left[-\frac{1}{2\pi}-\frac{4}{\pi^3 u^2}\delta A+\frac{2}{\pi^2 u}\left(1+\pi \lambda_1\right)A-\frac{2}{\pi^3 u^2}\left(1+5\pi \lambda_1\right)A^2 \right]\frac{\partial \delta}{\partial B}\nonumber\\&&+\left[-2\lambda_1\left(1-\frac{2}{\pi u}A+\frac{3}{\pi^2 u^2}A^2\right)-\frac{2}{\pi^3 u^2}\delta^2\right.\nonumber\\&&+\left.\frac{2}{\pi^2 u}\left(1+\pi \lambda_1\right)\delta-\frac{4}{\pi^3 u^2}\left(1+5\pi \lambda_1\right)\delta A-\lambda_2 A^2\right]\frac{\partial A}{\partial B}\nonumber\\&=&\chi_s^{(1)}+\chi_s^{(2)}. \label{chi-gs}
\end{eqnarray}
We observe that the  charge susceptibilities $\chi_c^{(1)},\, \chi_c^{(2)}$ (the spin susceptibility $\chi_s^{(1)},\, \chi_s^{(2)}$)  denote the contributions from different resources $\delta $ and $A$, respectively, see  (\ref{chi-gc}) and (\ref{chi-gs}). 
 The susceptibilities can be  split into two parts which are reminiscent of the additivity rules found  in the FFLO state of attractive situation \cite{Cheng:2018A}.
  It should be noted that the decomposition terms stem from two sources, i.e., the changes of charge and spin with respect to the chemical potential and magnetic field. 
  In the (\ref{chi-gc}) and (\ref{chi-gs}), $\frac{\partial \delta}{\partial \mu}$, $\frac{\partial A}{\partial \mu}$, $\frac{\partial \delta}{\partial B}$ and $\frac{\partial A}{\partial B}$ can be determined by taking the derivative of both sides of (\ref{eq-beta1}) and  (\ref{eq-beta2}) (note: $\beta_1,\, \beta_2$ depend on $\mu$ and $ B$):
\begin{eqnarray}
\frac{\partial A}{\partial \mu}&=&\frac{1}{\eta_1\left(A+\frac{A^2}{\pi u}\right)\frac{\pi u}{\delta}-\frac{4\eta_1}{\pi u}A^2}, \label{gd-1} \\
\frac{\partial \delta}{\partial \mu}&=&-\frac{\left(A+\frac{A^2}{\pi u}\right)}{2\delta \left(A+\frac{A^2}{\pi u}\right)-\frac{8}{\pi^2 u^2}\delta^2 A^2}, \label{gd-2}\\
\frac{\partial A}{\partial B}&=&-\frac{1-\frac{1}{\pi u}\delta}{ \eta_1\left(A+\frac{A^2}{\pi u}\right)-\frac{4\eta_1}{\pi^2 u^2}\delta A^2},\label{gd-3} \\
\frac{\partial \delta}{\partial B}&=&-\frac{1-\frac{3A}{\pi u}}{2\delta \left(1+\frac{A}{\pi u}\right)-\frac{8}{\pi^2 u^2}\delta^2 A}. \label{gd-4}
\end{eqnarray}
We plot the magnetic susceptibility in figure~(\ref{fig-v}) (c) and observe that the susceptibility does not depend on temperature in the TLL regime, but becomes divergent in the vicinities of the critical points.

In the canonical ensemble with fixed density and magnetization, we take the total derivative of (\ref{eq-delta}) and (\ref{eq-A}) with respect to $n_c$ and $m$ under the condition $\d n_{\downarrow}=\d n_c/2-\d m$
\begin{eqnarray}
\d\delta&=&\left[\pi\left(1+\frac{2}{u}n_{\downarrow}+\frac{4}{u^2}n_{\downarrow}^2\right)-\pi\hat{n}_c\left(\frac{1}{u}+\frac{4}{u^2}n_{\downarrow}\right)\right]\d\hat{n}_c-\pi\hat{n}_c\left(\frac{2}{u}+\frac{8}{u^2}n_{\downarrow}\right)\d m, \label{dedelta}\\
\d A&=&\d\hat{n}_c\left\{-\frac{\pi u\zeta_1}{4}\left[1+\zeta_1\left(\hat{n}_c+\frac{n_{\downarrow}}{2}\right)+\zeta^2_1\left(\hat{n}_c+\frac{n_{\downarrow}}{2}\right)^2-\zeta_2n_{\downarrow}^2\right]\right.\nonumber\\&&\left.+\frac{\pi u\zeta_1}{2}n_{\downarrow}\left[\frac{3\zeta_1}{4}+\frac{3\zeta^2_1}{2}\left(\hat{n}_c+\frac{n_{\downarrow}}{2}\right)+\zeta_2n_{\downarrow}\right]\right\}\nonumber\\&&+\d m\left\{-\frac{\pi u\zeta_1}{2}\left[1+\zeta_1\left(\hat{n}_c+\frac{n_{\downarrow}}{2}\right)+\zeta^2_1\left(\hat{n}_c+\frac{n_{\downarrow}}{2}\right)^2-\zeta_2n_{\downarrow}^2\right]\right.\nonumber\\&&\left.+\frac{\pi u\zeta_1}{2}n_{\downarrow}\left[-\frac{\zeta_1}{2}-\zeta^2_1\left(\hat{n}_c+\frac{n_{\downarrow}}{2}\right)+2\zeta_2n_{\downarrow}\right]\right\}. \label{dA}
\end{eqnarray}
Here $\frac{\partial \mu}{\partial n_c}|_m,\frac{\partial B}{\partial m}|_{n_c}$ can be similarly extracted from (\ref{eq-beta1}) (\ref{eq-beta2}).
 Similar to the (\ref{gd-1})-(\ref{gd-4}) with the difference in the quantity being differentiated, we have 
\begin{eqnarray}
 &&-\frac{\partial \mu}{\partial \hat{n}_c(m)}|_{m(\hat{n}_c)}-\frac{\partial B}{\partial \hat{n}_c(m)}|_{m(\hat{n}_c)}=2\delta \frac{\partial \delta}{\partial \hat{n}_c(m)}|_{m(\hat{n}_c)}+\frac{4\eta_1}{\pi u}A^2\frac{\partial A}{\partial \hat{n}_c(m)}|_{m(\hat{n}_c)}, \\
 &&\frac{\partial B}{\partial \hat{n}_c(m)}|_{m(\hat{n}_c)}=-\frac{2}{\pi u}\delta^2\frac{\partial \delta}{\partial \hat{n}_c(m)}|_{m(\hat{n}_c)}-\eta_1\left(A+\frac{A^2}{\pi u}\right)\frac{\partial A}{\partial \hat{n}_c(m)}|_{m(\hat{n}_c)}.
\end{eqnarray}
Therefore, by solving this set of equations together with total differential expressions (\ref{dedelta}) and (\ref{dA}), we obtain the charge and spin susceptibilities in terms of $n_c,\, m$ in the canonical ensemble 
\begin{eqnarray}
\frac{1}{\chi_c}&=&\frac{\partial \mu}{\partial n_c}|_m=2\delta\left(1-\frac{\delta}{\pi u}\right)\frac{\partial \delta}{\partial \hat{n}_c}|_m-\eta_1A\left(1-\frac{3A}{\pi u}\right)\frac{\partial A}{\partial \hat{n}_c}|_{m}\nonumber\\
&=& \frac{1}{\chi_c^{(1)}}+\frac{1}{\chi_c^{(2)}}, \label{chi-c}\\
\frac{1}{\chi_s}&=&
\frac{\partial B}{\partial m}|_{\hat{n}_c}=-\frac{2}{\pi u}\delta^2\frac{\partial \delta}{\partial m}|_{\hat{n}_c}-\eta_1A\left(1+\frac{A}{\pi u}\right)\frac{\partial A}{\partial m}|_{\hat{n}_c}\nonumber\\
&=&\frac{1}{\chi_s^{(1)}}+\frac{1}{\chi_s^{(2)}}.\label{chi-s}
\end{eqnarray}
In contrast to the the case for fixed external potentials, these formulas expose reciprocal additivity relation at a fixed magnetization and density, see (\ref{chi-c}) and (\ref{chi-s}). 
The two ensembles are related to each other by the Jacobian determinant evaluated by total differential of $\mu,\,B$ with respect to $n,\,m$. 
Apart from the numerical arithmetic and approximation method as we have conducted above, we would like to discuss dressed charge matrix $\mathrm{Z}$\cite{frahm1990critical,frahm1991correlation,essler2005one}
\begin{equation}
\mathrm{Z}=\left(\begin{matrix}\xi_{cc}(Q)&\xi_{cs}(A)\\\xi_{sc}(Q)&\xi_{ss}(A)\end{matrix}\right),\label{Z}
\end{equation} 
whose elements are determined by 
\begin{equation} \xi_{a b}\left(x_{b}\right)=\delta_{a b}+\sum_{d} \int_{-X_{d}}^{X_{d}} \d x_{d} \xi_{a d}\left(x_{d}\right) K_{d b}\left(x_{d}, x_{b}\right).
\end{equation} 
Here the  kernels are given by 
\begin{eqnarray}
K_{c c}\left(x_{c}, y_{c}\right)&=&0,\nonumber\\
K_{s c}\left(x_{c}, x_{s}\right)&=&a_{1}\left(\sin \left(x_{c}\right)-x_{s}\right),\nonumber\\
K_{c s}\left(x_{c}, x_{s}\right)&=&\cos \left(x_{c}\right) a_{1}\left(\sin \left(x_{c}\right)-x_{s}\right),\nonumber\\
K_{s s}\left(x_{s}, y_{s}\right)&=&-a_{2}\left(x_{s}-y_{s}\right).
\end{eqnarray}
These dressed charges can be used to  calculate  rigorous solutions for susceptibilities and conformal dimensions in asymptotics of correlation functions
\cite{frahm1990critical,frahm1991correlation}.  
We will discuss about this study later.

One advantage of this method is that the iteration process is greatly simplified both analytically and numerically, and the final result is clear and unambiguous. 
In the grand canonical ensemble, some exact relations are relevant to the dressed charge matrix (\ref{Z}), for example \cite{essler2005one}, 
\begin{eqnarray}
	\chi_c|_B&=&\frac{\partial n_c}{\partial \mu}|_B=\frac{\mathrm{Z}^2_{cc}}{\pi v_c}+\frac{\mathrm{Z}^2_{cs}}{\pi v_s}=\chi_c^{(1)}+\chi_c^{(2)},\label{chi-g-1} \\
	\chi_s|_{\mu}&=&\frac{\partial m}{\partial B}|_{\mu} =\frac{(\mathrm{Z}_{cc}-2\mathrm{Z}_{sc})^2}{2\pi v_c}+\frac{(\mathrm{Z}_{cs}-2\mathrm{Z}_{ss})^2}{2\pi v_s}=\chi_s^{(1)}+\chi_s^{(2)}. \label{chi-g-2}
\end{eqnarray}
Regarding the circumstance of variable changes, we can work out the Jacobian determinant from the equations (\ref{chi-g-1}) and (\ref{chi-g-2}), namely, $J=2(\mathrm{det}\mathrm{Z})^2/(\pi^2v_cv_s)$.
As a result, we obtain \footnote{The second equality in the equation (6.79) of the book \cite{essler2005one} misses a factor of 2.} for canonical ensemble
\begin{eqnarray}
	\frac{1}{\chi_c}|_{m}&=&\frac{\partial \mu}{\partial n_c}|_m=\frac{\pi}{4}\frac{v_c(\mathrm{Z}_{cs}-2\mathrm{Z}_{ss})^2+v_s(\mathrm{Z}_{cc}-2\mathrm{Z}_{sc})^2}{(\mathrm{det}\mathrm{Z})^2}=\frac{1}{\chi_c^{(1)}}+\frac{1}{\chi_c^{(2)}}, \label{chi-1}\\
	\frac{1}{\chi_s}|_{n_c}&=&\frac{\partial B}{\partial m}|_{n_c} =\frac{\pi}{2}\frac{v_c\mathrm{Z}_{cs}^2+v_s\mathrm{Z}_{cc}^2}{(\mathrm{det}\mathrm{Z})^2}=\frac{1}{\chi_s^{(1)}}+\frac{1}{\chi_s^{(2)}}. 
	\label{chi-2}
\end{eqnarray}
Comparing our analytic results (\ref{chi-gc}) and (\ref{chi-gs}) (or (\ref{chi-c}), (\ref{chi-s})) with the dressed charge formula (\ref{chi-g-1}) and (\ref{chi-g-2}) (or (\ref{chi-1}),  (\ref{chi-2})), we observe that the derivative term with respect to  $\delta$ in the  (\ref{chi-gc}) is equivalent to the first term associated with the charge velocity $v_c$ in (\ref{chi-g-1}), whereas the derivative term  with respect to  $A$ in  (\ref{chi-gc}) is equivalent to the second  term associated with the spin velocity $v_s$ in (\ref{chi-g-1}), respectively. 
Similar correspondences can be found from (\ref{chi-gs}) and (\ref{chi-g-2}) in the grand canonical ensemble, as well as in these equations (\ref{chi-c}), (\ref{chi-s}) and  (\ref{chi-1}),  (\ref{chi-2}), respectively. 

 We remark that the results we obtained can also be applied to the phase II and the Mott phase V, i.e., omitting the terms related to $A$ in II or $\delta$ in phase V, respectively.
 Explicitly,  the corresponding thermodynamic quantities and velocities are given by\\
\textbf{phase II:}
\begin{equation}
n_c=1-\frac{\delta}{\pi},
n_{\downarrow}=0,
\chi_c=\frac{1}{2\pi \delta},
\chi_s=\frac{1}{4\pi \delta},
v_c=2\delta -\frac{1}{3}\delta^3, \label{II}
\end{equation}
\textbf{phase V:}
\begin{eqnarray}
n_c&=&1,\label{V-nc}\\
n_{\downarrow}&=&\int_{-A}^{A}\d\Lambda \sigma_1(\Lambda)=2\left[\lambda_1A-\frac{\lambda_1}{\pi u}A^2+A^3\left(\frac{\lambda_1}{\pi^2 u^2}+\frac{\lambda_2}{6}\right)\right], \label{V-ndown} \\
\chi_c&=&0, \label{V-chic} \\
\chi_s&=&\left[2\lambda_1\left(1-\frac{2}{\pi u}A+\frac{3}{\pi^2 u^2}A^2\right)+\lambda_2 A^2\right]\frac{1}{ \eta_1A\left(1+\frac{A}{\pi u}\right)}, \label{V-chis} \\
v_s&=&\frac{\eta_1A+\frac{\eta_2}{12}A^3}{\pi\left(\lambda_1+\frac{\lambda_2}{2}A^2-\frac{\lambda_1}{\pi u} A+\frac{\lambda_1}{3\pi u^3}A^3+\frac{\lambda_1}{\pi^2 u^2}A^2-\frac{\lambda_1}{\pi^3 u^3}A^3-\frac{\lambda_2}{6\pi u}A^3\right)}. \label{V-vs}
\end{eqnarray}

\begin{figure}[H] 
	\centering        
	\includegraphics[scale=0.52]{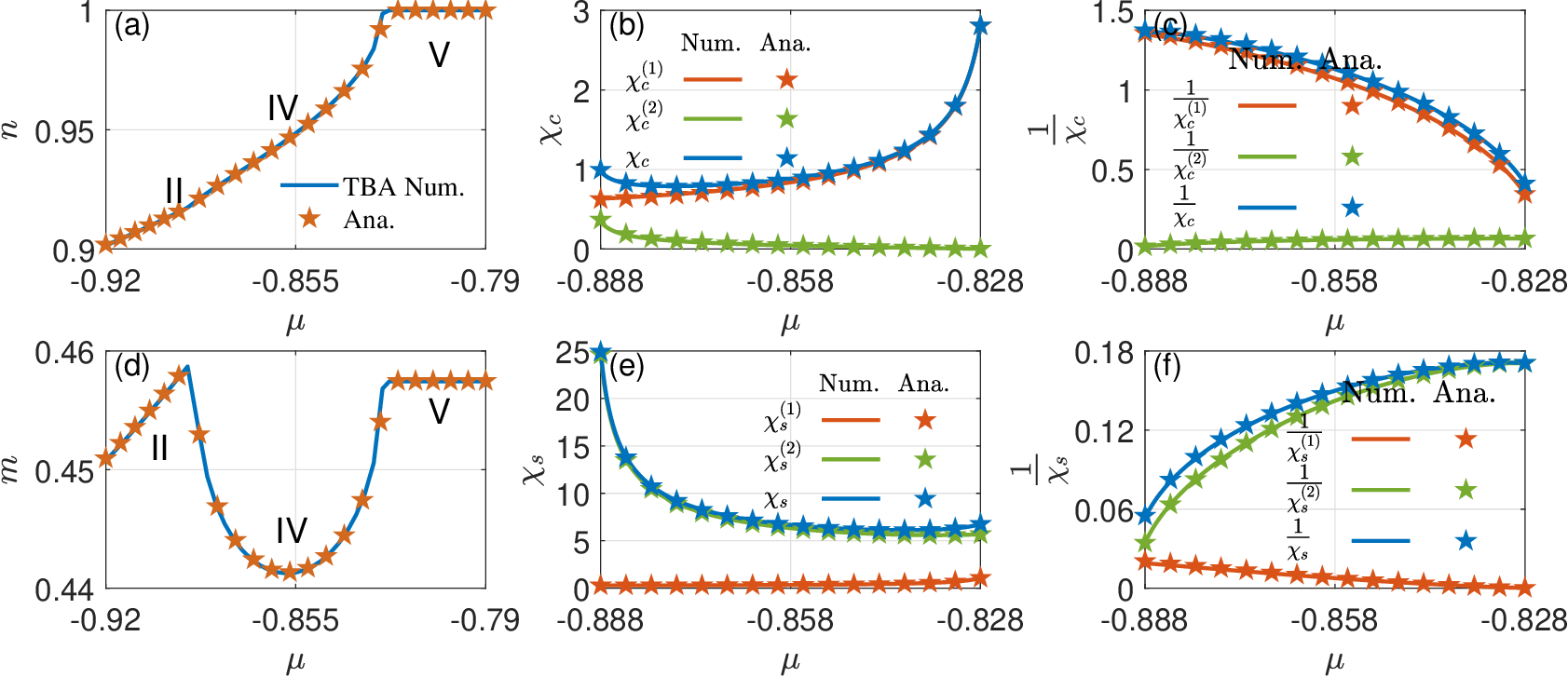}     
	\caption{ Charge density (a) and (d) magnetization vs chemical potential in phases  II, IV and V: agreement between  analytic results given by (\ref{II}) for phase II, (\ref{nc-c}) (\ref{nd-s}) for phase IV,  (\ref{V-nc})(\ref{V-ndown}) for phase V and the numerical calculations from the TBA equations. 
	The compressibility (b) and susceptibility (e) in grand canonical ensemble show agreement between the analytic results  given by (\ref{chi-gc}) (\ref{chi-gs}) and numerical simulations based on (\ref{chi-g-1}) (\ref{chi-g-2}). While compressibility (c) and susceptibility (f) in canonical ensemble show agreement between  the analytic results  given by \ref{chi-c} (\ref{chi-s}) and numerical simulations  based on (\ref{chi-1}) (\ref{chi-2}). All figures  are set to $B=0.825,\, u=1$.}            
	\label{fig-n0-m0-x0}         
\end{figure}

Finally, we make direct comparison between our analytical results and numerical results from the TBA equations.
 In figure~\ref{fig-n0-m0-x0}, the first row shows density (a), compressibility (b) in the grand canonical ensemble and (c) in the  canonical ensemble.
 Whereas  the second row shows the  magnetization (d), susceptibility (e) in the grand canonical ensemble and (f) in the canonical ensemble.
  The density and magnetization in figure~\ref{fig-n0-m0-x0} (a)(d) cross quantum phases from II, to IV  and V as chemical potential varies.
   In figure~\ref{fig-n0-m0-x0} (b), (c), (e), (f), we observe additivity rules in both the grand canonical and the canonical ensemble. 
   All analytic results agree well with the corresponding numerical results in this figure. 

\subsection{III.5 Quantum criticality near quadruple critical point}
\label{sec_cri}

The results obtained in the previous subsections present an elementary understanding of the ground state properties and the behaviour of the TLLs.
Although the criticality induced by the variation of some external potentials such as magnetic field and chemical potential has been well studied in literature, this is not the case for the interaction-driven quantum critical behaviour, even though interaction plays a central role in many-body systems.
In this subsection, we study quantum phase transitions and universal scaling functions of properties of the 1D Hubbard model in terms of external fields. And in the last section we focus on the interaction driven quantum  transitions. 
From the phase diagram figure~(\ref{fig-R}), it can be observed  that a phase transition occurs in  the repulsive Hubbard model when a degree of freedom appears, disappears or become gapped. 
Although phase transition  occurs  at zero temperature, thermal and quantum fluctuations can reach the same level of the energy scale in the V-shaped critical regions at finite temperatures. 
A natural question is whether quantum critical region has its own set of universal laws similar to the TLL.  
The obvious answer, of course, is that the quantum criticality for the same universality class of models is insensitive to the microscopic details of the systems and shares general and universal critical phenomena characterised by the critical exponents \cite{sachdev2011quantum,hazzard2011techniques}. 
For example, the universal scaling laws of magnetization and susceptibility satisfy \cite{sachdev2011quantum,hazzard2011techniques}
\begin{eqnarray}
m&=&m_0+T^{d/z+1-1/(vz)}O_1\left[\frac{\mu-\mu_c}{T^{1/vz}},\frac{B-B_c}{T^{1/vz}},\frac{u-u_c}{T^{1/vz}}\right], \label{QC-m}\\
\chi_s&=&\chi_{s0}+T^{d/z+1-2/(vz)}O_2\left[\frac{\mu-\mu_c}{T^{1/vz}},\frac{B-B_c}{T^{1/vz}},\frac{u-u_c}{T^{1/vz}}\right], \label{QC-chi}
\end{eqnarray}
respectively. 
Where $m,\, \chi_s$ represent the first and second orders of thermodynamic quantities respectively, $\mu_c,\, B_c,\,u_c$ denotes the critical fields at the critical point and $z,\, v$ stand for the dynamical and the correlation critical exponent, respectively. 
The scaling functions (\ref{QC-m}) and (\ref{QC-chi}) consist of two parts: the first term denotes the  background contributions stemming from the unchanged degrees of freedom, and the second term accounts for the singular part stemming from the sudden change of the density of state of one degree of freedom. 
 Another salient feature of critical region is that the characteristic length scale diverges, providing us with a feasible opportunity to capture system information by fractional exclusive statistics (FES) \cite{sachdev2011quantum,fukui1995haldane,vitoriano2009fractional,zhang2022interaction}. 

 In what follows, we embark on the analytical derivation of the singular behaviours of thermodynamical properties involved solely charge quasimomenta $k$ and length-$1$ $\Lambda$ strings. 
 Without losing generality, we first consider the dressed energy equations for the magnetized  phase IV
\begin{eqnarray}
\kappa(k)&=&-2 \cos{k}-\mu-2 u-B-T \int_{-\infty}^{\infty}\d \Lambda  a_1(\sin{k}-\Lambda) \ln(1+e^{-\frac{\varepsilon_1(\Lambda)}{T}}),\label{ku}\\
\varepsilon_1(\Lambda)&=&2 B -T\int_{-\pi}^{\pi} \d k \cos k a_1(\sin{k}-\Lambda) \ln(1+e^{-\frac{\kappa(k)}{T}})\nonumber \\
&&+T \int_{-\infty}^{\infty} \d \Lambda^{\prime} a_{2}\left(\Lambda-\Lambda^{\prime}\right)  \ln(1+e^{-\frac{\varepsilon_1(\Lambda^{\prime})}{T}}),\label{eu}
\end{eqnarray}
where $a_n(x)=\frac{1}{2 \pi}\frac{2 n u}{\left(n u\right)^2 + x^2}$. 
There are five phase transitions in total, I-II, II-III, III-V, II-IV and IV-V, among which the four cases can be treated uniformly around the quadruple point. 

{\bf I-II phase transition:} To this end, we first study the phase transition from an empty lattice phase I to partially filled II with the absence of down-spin.
In this transition, the phase transition boundary  is simply expressed as $2+2u+B_c+\mu_c=0$ with subscript $B_c$ and $\mu_c$ stand for the critical fields. 
Thus the charge dressed energy is given by $\kappa(k)=-2\cos k-\mu-2u-B$, showing free fermions on a 1D lattice. 
This gives the free energy $f=u+\frac{T^{3/2}}{2\pi^{1/2}}\Li_{\frac{3}{2}}\left(-\e^{\frac{\Delta B+\Delta \mu}{T}}\right)$. 
Using fundamental thermodynamic relations, we directly obtain the first and second order thermodynamic quantities  
\begin{eqnarray}
m&=&\frac{n}{2}=-\frac{T^{\frac{1}{2}}}{4\pi^{\frac{1}{2}}}f_{\frac{1}{2}}, \\
\chi_s&=&\frac{\chi_c}{2}=-\frac{T^{-\frac{1}{2}}}{4\pi^{\frac{1}{2}}}f_{-\frac{1}{2}},\\
\frac{c_v}{T}&=&-\frac{3T^{-\frac{1}{2}}}{8\pi^{\frac{1}{2}}}f_{\frac{3}{2}}+\frac{T^{-\frac{3}{2}}\left(\Delta B+\Delta \mu\right)}{2\pi^{\frac{1}{2}}}f_{\frac{1}{2}}-\frac{T^{-\frac{5}{2}}\left(\Delta B+\Delta \mu\right)^2}{2\pi^{\frac{1}{2}}}f_{-\frac{1}{2}} 
\end{eqnarray}
with $f_n=\Li_n\left(-\e^{\frac{\Delta B+\Delta \mu}{T}}\right)$.

For the remaining four phase transitions, what we need to prepare before obtaining the scaling forms  is to express the dressed energy equations and the free energy in terms of polylog functions. 
Under the assumption of $n\rightarrow1,\, n_{\downarrow}\rightarrow0$, we expand the kernal function $a_n$ around $k=\pi$  and $\Lambda =0$,  the coupled equations become as 
\begin{eqnarray}
\kappa(k)&=&-2 \cos{k}+2\left(\frac{1}{\pi u^3}I_1-\frac{6}{\pi u^5}I_2 \right)\sin^2k+\left( -\mu-2u-B-\frac{2}{\pi u}I_1+\frac{2}{\pi u^3}I_2\right)\nonumber\\
&=&-2\cos{k}+2C_1\sin^2k+C_2, \label{QC-kappa}\\
\varepsilon_1(\Lambda)&=&\Lambda^2\left(\eta_1+\frac{2J_1}{\pi u^3}-\frac{12J_2}{\pi u^5}-\frac{I_1}{4\pi u^3}+\frac{3I_2}{8\pi u^5}\right)\nonumber\\&&+\left(2B+4(u-\sqrt{1+u^2})-\frac{2J_1}{\pi u}+\frac{2J_2}{\pi u^3}+\frac{I_1}{\pi u}-\frac{I_2}{4\pi u^3}\right)\nonumber\\&=&D_1\Lambda^2+D_2 ,\label{QC-lambda}
\end{eqnarray}
where the functions $C_1,\, C_2,\, D_1,\, D_2$  denote the corresponding  coefficients
\begin{eqnarray}
C_1&=& \frac{1}{\pi u^3}I_1-\frac{6}{\pi u^5}I_2,\label{C1}\\
C_2&=&-\mu-2u-B-\frac{2}{\pi u}I_1+\frac{2}{\pi u^3}I_2,\label{C2}\\
D_1 &=&\eta_1+\frac{2J_1}{\pi u^3}-\frac{12J_2}{\pi u^5}-\frac{I_1}{4\pi u^3}+\frac{3I_2}{8\pi u^5},\label{D1}\\
D_2&=&2B+4(u-\sqrt{1+u^2})-\frac{2J_1}{\pi u}+\frac{2J_2}{\pi u^3}+\frac{I_1}{\pi u}-\frac{I_2}{4\pi u^3}.\label{D2}
\end{eqnarray}
while the integrals $(I_1,I_2)$ and $(J_1,J_2)$ are related to spin and charge degrees of freedom and  make contributions only near zero point. 
Integrating by parts, the formal solutions in terms of polylog functions read
\begin{eqnarray}
I_1&=&\int_{0}^{\infty} \d \Lambda  T \ln(1+\e^{-\frac{\varepsilon_1(\Lambda)}{T}})=-\frac{T^{\frac{3}{2}}\pi^{\frac{1}{2}}}{2D^{\frac{1}{2}}_1}\Li_{\frac{3}{2}}\left(-\e^{\frac{-D_2}{T}}\right), \label{QC-I1}\\
I_2&=&\int_{0}^{\infty} \d \Lambda \Lambda^2 T \ln(1+\e^{-\frac{\varepsilon_1(\Lambda)}{T}})=-\frac{T^{\frac{5}{2}}\pi^{\frac{1}{2}}}{4D^{\frac{3}{2}}_1}\Li_{\frac{5}{2}}\left(-\e^{\frac{-D_2}{T}}\right),\label{QC-I2} \\
J_1&=&T\int_{0}^{\pi} \d k \cos k \ln(1+\e^{\frac{\kappa(k)}{T}})\nonumber\\&=&\frac{T^{\frac{3}{2}}}{\sqrt{1-2C_1}}\Gamma(\frac{3}{2})\Li_{\frac{3}{2}}\left(-\e^{\frac{2+C_2}{T}}\right)-\frac{T^{\frac{5}{2}}}{8(1-2C_1)^{\frac{5}{2}}}\Gamma(\frac{5}{2})\Li_{\frac{5}{2}}\left(-\e^{\frac{2+C_2}{T}}\right),\label{QC-J1} \\
J_2&=&T\int_{0}^{\pi} \d k \cos k \sin^2 k \ln(1+\e^{\frac{\kappa(k)}{T}})=\frac{T^{\frac{5}{2}}}{3(1-2C_1)^{\frac{3}{2}}}\Gamma(\frac{5}{2})\Li_{\frac{5}{2}}\left(-\e^{\frac{2+C_2}{T}}\right). \label{QC-J2}
\end{eqnarray}
With the help of the presentations of these four integrals, the Gibbs free energy is given  by 
\begin{equation}
f=-\mu-u-B-2\lambda_1I_1-\lambda_2I_2+\frac{1}{\pi}J_1+\frac{1}{2\pi}J_2.  \label{free-energy-QC}
\end{equation}

Although we have greatly simplified the dressed equations, the four integrals $(I_1,I_2)$ and $(J_1,J_2)$ are still coupled to each other.
 These four quantities are intertwined and are needed to carefully distinguish the primary and secondary contributions according to the critical field conditions. 
  Based on the above equations (\ref{QC-kappa})-(\ref{QC-J2}), we proceed to evaluate critical behaviours from three aspects: critical fields, polylog functions and free energy. 
The polylog functions contain  the sources of criticality, from which  we can derive the scaling function of the free energy for quantum critical region. 
For the part that causes criticality, we take expansions under the limit of $T\gg\Delta B$ or $T\gg\Delta \mu$.

{\bf II-III  phase transition:} For the phase transition II-III, the critical field is determined by $2-2u-B_c-\mu_c=0$ and $\varepsilon_1(\Lambda)$ is gapped. 
 The free energy is given by 
 \begin{eqnarray}
  f=-\mu-u-B+\frac{T^{\frac{3}{2}}}{2\pi^{\frac{1}{2}}}\Li_{\frac{3}{2}}\left(-\e^{\frac{-\Delta B-\Delta \mu}{T}}\right), 
  \end{eqnarray}
  leading to the scattering forms of thermodynamics 
  \begin{eqnarray}
  m&=&\frac{n}{2}=\frac{1}{2}+\frac{T^{\frac{1}{2}}}{4\pi^{\frac{1}{2}}}f_{\frac{1}{2}},\\
  \chi_s&=&\frac{\chi_c}{2}=-\frac{T^{-\frac{1}{2}}}{4\pi^{\frac{1}{2}}}f_{-\frac{1}{2}},\\
  \frac{C_v}{T}&=&-\frac{3T^{-\frac{1}{2}}}{8\pi^{\frac{1}{2}}}f_{\frac{3}{2}}-\frac{T^{-\frac{3}{2}}\left(\Delta B+\Delta \mu\right)}{2\pi^{\frac{1}{2}}}f_{\frac{1}{2}}-\frac{T^{-\frac{5}{2}}\left(\Delta B+\Delta \mu\right)^2}{2\pi^{\frac{1}{2}}}f_{-\frac{1}{2}}
  \end{eqnarray}
  with $f_n=\Li_n\left(-\e^{\frac{-\Delta B-\Delta \mu}{T}}\right)$. 
 These present universal scaling behaviour for the charge gapped phase transition in a lattice.  
 The situation is very similar to the case of I-II transition where  only charge sector exists.

{\bf III-V  phase transition:} For the phase transition III-V, the emergence of spin degree of freedom on a half-filled lattice  generates the criticality with respect to the critical field  $B_c=2\sqrt{1+u^2}-2u$, independent of chemical potential.
This  gives the integrable $I_1\approx-\frac{T^{3/2}\pi^{1/2}}{2\eta^{1/2}_1}\Li_{\frac{3}{2}}\left(-\e^{-\frac{2\Delta B}{T}}\right)$. 
In this case $\kappa(k)$ is always less than zero in charge momentum space.
For the charge degree, $|2+C_2|\gg T$, thus the $J_1$ with respect to temperature magnitude can be evaluated by 
\begin{equation}
J_1=\frac{1}{2}\pi^{\frac{1}{2}}T^{\frac{3}{2}}\Li_{\frac{3}{2}}\left(-\e^{\frac{2+C_2}{T}}\right)\approx\frac{1}{2}\pi^{\frac{1}{2}}T^{\frac{3}{2}}\left[-\e^{\frac{2+C_2}{T}}+\frac{1}{2^{\frac{3}{2}}}\e^{\frac{2(2+C_2)}{T}}\cdots\right]\ll\frac{1}{2}\pi^{\frac{1}{2}}T^{\frac{3}{2}}\approx I_1.
\end{equation}
For this reason, $J_1,J_2$ can be safely neglected, equivalent to the case of XXX spin chain. 
Whereas for spin degrees of freedom, the term $I_1$ is relevant to the low temperature criticality, in contrast the integral $I_2$  has higher order of power of the  temperature than that of $I_1$.
Under this circumstance, we determine the free energy near the phase transition III-V as
\begin{eqnarray}
  f=-\mu-u-B+\frac{T^{\frac{3}{2}}\pi^{\frac{1}{2}}\lambda_1}{\eta^{\frac{1}{2}}_1}\Li_{\frac{3}{2}}\left(-\e^{-\frac{2\Delta B}{T}}\right).
  \end{eqnarray}
  Using the standard thermodynamic relation, 
  we give the following forms of scaling functions for density, magnetization, compressibility, susceptibility and specific heat
\begin{eqnarray}
 n&=&1,\quad m=\frac{1}{2}+\frac{T^{\frac{1}{2}}\pi^{\frac{1}{2}}\lambda_1}{\eta^{\frac{1}{2}}_1}f_{\frac{1}{2}},\\
 \chi_c&=&0, \quad \chi_s=-\frac{2T^{-\frac{1}{2}}\pi^{\frac{1}{2}}\lambda_1}{\eta^{\frac{1}{2}}_1}f_{-\frac{1}{2}},\\
\frac{C_v}{T}&=&-\frac{3T^{-\frac{1}{2}}\pi^{\frac{1}{2}}\lambda_1}{4\eta^{\frac{1}{2}}_1}f_{\frac{3}{2}}-\frac{2T^{-\frac{3}{2}}\pi^{\frac{1}{2}}\lambda_1\Delta B}{\eta^{\frac{1}{2}}_1}f_{\frac{1}{2}}-\frac{4T^{-\frac{5}{2}}\pi^{\frac{1}{2}}\lambda_1\Delta B^2}{\eta^{\frac{1}{2}}_1}f_{-\frac{1}{2}}
\end{eqnarray}
with $f_n=\Li_n\left(-\e^{-\frac{2\Delta B}{T}}\right)$.

{\bf II-IV phase transition:} For quantum phase transition II-IV,  the phase transition occurs at $\varepsilon_1(0)=0$ with charge dispersion  $\kappa(k)=-2\cos k-\mu-2u-B$. 
Let $Q$ denote Fermi point of $\kappa(k)$, which satisfies $2\cos Q=-\mu-2u-B$. 
Via (\ref{eq-beta1}) corresponding to $\kappa(Q)=0$, the critical field is obtained
\begin{eqnarray}
B_c&=&2\sqrt{1+u^2}-2u-\frac{2}{3\pi u}(\pi-Q)^3,\nonumber\\
\mu_c&=&-2\cos Q-2u-B_c=2-2\sqrt{1+u^2}-(\pi-Q)^2+\frac{2}{3\pi u}(\pi-Q)^3. 
\end{eqnarray}
From (\ref{QC-kappa}) and (\ref{QC-lambda}) at  criticality, we observe that the term $C_1$ is a small quantity, while $\frac{2+C_2}{T}=\frac{\kappa(\pi)}{T}$ in polylog function in  $J_1,\, J_2$  is  a large quantity.
Using the property of the polylog function
\begin{eqnarray}
\Li_p(-\e^w)\stackrel{w>>1}{\longrightarrow}-2\sum_{k=0}^{\infty}\eta(2k)\frac{w^{p-2k}}{\Gamma(p-2k+1)}+O(-\e^w), \label{polylog-function}
\end{eqnarray}
we have 
\begin{eqnarray}
J_1&\approx&\frac{1}{2}(1+C_1)\left[-\frac{4}{3}\left(\kappa(\pi)\right)^{\frac{3}{2}}-\frac{\pi^2T^2}{6}\left(\frac{1}{\kappa(\pi)}\right)^{\frac{1}{2}}\right]\nonumber\\&&+\frac{3}{32}(1+5C_1)\left[\frac{8}{15}\left(\kappa(\pi)\right)^{\frac{5}{2}}+\frac{\pi^2T^2}{3}\left(\kappa(\pi)\right)^{\frac{1}{2}}\right], \label{J1-Func} \\
J_2&\approx&-\frac{1}{4}(1+3C_1+\frac{15}{2}C^2_1)\left[\frac{8}{15}\left(\kappa(\pi)\right)^{\frac{5}{2}}+\frac{\pi^2T^2}{3}\left(\kappa(\pi)\right)^{\frac{1}{2}}\right].\label{J2-Func} 
\end{eqnarray}
While $I_1,\, I_2$ indicate that they are the orders of $O(T^{\frac{3}{2}})$ and $O(T^{\frac{5}{2}})$,  respectively.
Therefore we reasonably ignore the effect of $I_2$.
The leading term $I_1$ is given by 
\begin{eqnarray}
I_1\approx-\frac{T^{\frac{3}{2}}\pi^{\frac{1}{2}}}{2D^{\frac{1}{2}}_1}\Li_{\frac{3}{2}}\left(-\e^{-\frac{D_2}{T}}\right).
\end{eqnarray}
In the above equations (\ref{J1-Func}) and (\ref{J2-Func}), we have an approximation $C_1\approx\frac{1}{\pi u^3}I_1,C_2\approx-\mu-2u-B-\frac{2}{\pi u}I_1$. 
It follows that 
\begin{eqnarray}
\kappa(\pi)&=&2+C_2=2-\mu-2u-B-\frac{2}{\pi u}I_1=2+2\cos Q-\frac{2}{\pi u}I_1=\beta_1-\frac{2}{\pi u}I_1,
\end{eqnarray}
where $\beta_1=2-\mu-2u-B$ which is defined previously.
Follow these approximations, we rewrite the functions $J_1$ and $J_2$ 
\begin{eqnarray}
J_1&\approx&-\frac{2}{3}\beta_1^{\frac{3}{2}}+\frac{2\beta_1^{\frac{1}{2}}}{\pi u}I_1-\frac{\pi^2T^2}{12\beta_1^{\frac{1}{2}}}-\frac{2\beta_1^{\frac{3}{2}}}{3\pi u^3}I_1+\frac{1}{20}\beta_1^{\frac{5}{2}}-\frac{\beta_1^{\frac{3}{2}}}{4\pi u}I_1+\frac{\pi^2T^2\beta_1^{\frac{1}{2}}}{32}, \label{eq-J1}\\
J_2&\approx&-\frac{2}{15}\beta_1^{\frac{5}{2}}+\frac{2\beta_1^{\frac{3}{2}}}{3\pi u}I_1-\frac{\pi^2T^2\beta_1^{\frac{1}{2}}}{12}. \label{eq-J2}
\end{eqnarray}
Substituting (\ref{eq-J1}) to the definition of $D_2$ in (\ref{D2}) which reflects the distance away from QCP,
 $D_2$ can then be represented with $\Delta B,\, \Delta \mu$ as 
\begin{eqnarray}
\Delta t\equiv:D_2&\approx&2B+4(u-\sqrt{1+u^2})-\frac{2J_1}{\pi u}\approx2B+4(u-\sqrt{1+u^2})+\frac{4}{3\pi u}\beta_1^{\frac{3}{2}}\nonumber\\&=&2B+4(u-\sqrt{1+u^2})+\frac{4}{3\pi u}\left(\beta_{1c}^{\frac{3}{2}}-\frac{3\beta_{1c}^{\frac{3}{2}}}{2}(\Delta B+\Delta \mu)\right)-2B_c+2B_c\nonumber\\&=&2\left(1-\frac{\beta_{1c}^{\frac{1}{2}}}{\pi u}\right)\Delta B-\frac{2\beta_{1c}^{\frac{1}{2}}}{\pi u}\Delta \mu, 
\end{eqnarray}
where $\beta_{1c}=2-\mu_c-2u-B_c$.
In the above equation,  we only inserted  the leading term $-\frac{2}{3}\beta_1^{\frac{3}{2}}$ of  $J_1$ that is irrelevant of $T$.
While $J_2$ has been neglected and the identity $2-\beta_1-2u-B-\mu=0$ has been used. 
Substituting the results of $J_1,\,J_2,\, I_1$ into the free energy (\ref{free-energy-QC}), we have 
\begin{equation}
f\approx-\mu-u-B-\frac{2}{3\pi}\beta_1^{\frac{3}{2}}-\frac{1}{60\pi}\beta_1^{\frac{5}{2}}-\frac{\pi T^2}{12\beta_1^{\frac{1}{2}}}-\frac{\pi T^2\beta_1^{\frac{1}{2}}}{96}+\frac{T^{\frac{3}{2}}\pi^{\frac{1}{2}}a_0}{\eta^{\frac{1}{2}}_1}\Li_{\frac{3}{2}}\left(-\e^{-\frac{\Delta t}{T}}\right),\label{f-24}
\end{equation}
here  $a_0=\lambda_1-\frac{\beta_1^{\frac{1}{2}}}{\pi^2 u}+\beta_1^{\frac{3}{2}}\left(\frac{2\lambda_1}{3\pi u^3\eta_1}-\frac{1}{24\pi^2 u}+\frac{1}{3\pi^2 u^3}\right)$ and $\Delta  t=2\left(1-\frac{\beta_{1c}^{\frac{1}{2}}}{\pi u}\right)\Delta B-\frac{2\beta_{1c}^{\frac{1}{2}}}{\pi u}\Delta \mu$.
Taking  derivatives of free energy with respect to $B,\, \mu$ and $T$,  we obtain the following scaling forms of the  thermal and magnetic properties 
\begin{eqnarray}
n&=&1-\frac{\beta_1^{\frac{1}{2}}}{\pi}-\frac{\beta_1^{\frac{3}{2}}}{24\pi}-\frac{2 T^{\frac{1}{2}}\beta_{1c}^{\frac{1}{2}} a_0}{\pi^{\frac{1}{2}}u\eta^{\frac{1}{2}}_1}f_{\frac{1}{2}},\label{QC-II-IV-1}\\
m&=&\frac{1}{2}-\frac{\beta_1^{\frac{1}{2}}}{2\pi}-\frac{\beta_1^{\frac{3}{2}}}{48\pi}+\frac{ T^{\frac{1}{2}}\left(1-\frac{\beta_{1c}^{\frac{1}{2}}}{\pi u}\right)\pi^{\frac{1}{2}}a_0}{\eta^{\frac{1}{2}}_1}f_{\frac{1}{2}}, \label{QC-II-IV-2}\\
\chi_c&=&\frac{1}{2\pi\beta_1^{\frac{1}{2}}}+\frac{\beta_1^{\frac{1}{2}}}{16\pi}-\frac{4 T^{-\frac{1}{2}}\beta_{1c}a_0}{\pi^{\frac{3}{2}}u^2\eta^{\frac{1}{2}}_1}f_{-\frac{1}{2}}, \label{QC-II-IV-3}\\
\chi_s&=&\frac{1}{4\pi\beta_1^{\frac{1}{2}}}+\frac{\beta_1^{\frac{1}{2}}}{32\pi}-\frac{2 T^{-\frac{1}{2}}\left(1-\frac{\beta_{1c}^{\frac{1}{2}}}{\pi u}\right)^2\pi^{\frac{1}{2}}a_0}{\eta^{\frac{1}{2}}_1}f_{-\frac{1}{2}},\label{QC-II-IV-4}\\
\frac{C_v}{T}&=&\frac{\pi}{6\beta_1^{\frac{1}{2}}}+\frac{\pi \beta_1^{\frac{1}{2}}}{48}-\frac{3 T^{-\frac{1}{2}}\pi^{\frac{1}{2}}a_0}{4\eta^{\frac{1}{2}}_1}f_{\frac{3}{2}}-\frac{2 T^{-\frac{3}{2}}\pi^{\frac{1}{2}}a_0}{\eta^{\frac{1}{2}}_1}\left[\left(1-\frac{\beta_{1c}^{\frac{1}{2}}}{\pi u}\right)\Delta B-\frac{\beta_{1c}^{\frac{1}{2}}}{\pi u}\Delta \mu\right]f_{\frac{1}{2}}\nonumber 
\\&&-\frac{ 4T^{-\frac{5}{2}}\pi^{\frac{1}{2}}a_0}{\eta^{\frac{1}{2}}_1}\left[\left(1-\frac{\beta_{1c}^{\frac{1}{2}}}{\pi u}\right)\Delta  B-\frac{\beta_{1c}^{\frac{1}{2}}}{\pi u}\Delta \mu\right]^2f_{-\frac{1}{2}} \label{QC-II-IV-5}
\end{eqnarray}
with $f_n=\Li_n\left(-\e^{-\frac{\Delta t}{T}}\right)$.

It is further noticed from (\ref{f-24}) that the background of free energy involves two types of terms, the temperature-dependent term $-\frac{\pi T^2}{12\beta_1^{\frac{1}{2}}}-\frac{\pi T^2\beta_1^{\frac{1}{2}}}{96}$ comes from the contribution of TLL, whereas the remining terms $-\mu-u-B-\frac{2}{3\pi}\beta_1^{\frac{3}{2}}-\frac{1}{60\pi}\beta_1^{\frac{5}{2}}$ is equivalent to the energy of non-interacting electrons on a lattice.
 For simplicity and without losing generality, we only keep the order of $\beta_1^{\frac{3}{2}}$ in the free energy. 
 For the 1D repulsive Hubbard model, zero-temperature background is given by 
 \begin{eqnarray}
 f_{\mathrm{H}}\approx-\mu-u-B-\frac{2}{3\pi}\beta_1^{\frac{3}{2}}\approx \beta_1-2-\frac{2}{3\pi}\beta_1^{\frac{3}{2}}. \label{II-IV-free-E-Back}
 \end{eqnarray}
  While the ground state free energy of a  non-interacting lattice system is given by 
 \begin{eqnarray}
 f_{\mathrm{NI}}=-\frac{1}{\pi}\left[\sum_{\uparrow,\downarrow}2\sin(k_{F,\sigma})+\mu_{\sigma}\sin(k_{F,\sigma})\right], \label{II-IV-free-E-NI}
 \end{eqnarray}
 in which $k_{F,\sigma}=\arccos(-\mu_{\sigma}/2)=\pi n_{\sigma}, \mu_{\uparrow}=\mu+B, \mu_{\downarrow}=\mu-B$. 
 In the limit of high density at spin-polarized phase, these two free energies (\ref{II-IV-free-E-Back}) and  (\ref{II-IV-free-E-NI}) are equivalent 
\begin{equation}
f_{\mathrm{NI}}\approx-\frac{2}{\pi}\left(\beta_1^{\frac{1}{2}}-\frac{1}{6}\beta_1^{\frac{3}{2}}\right)-(\mu+B)n \approx-\frac{2}{\pi}\left(\beta_1^{\frac{1}{2}}-\frac{1}{6}\beta_1^{\frac{3}{2}}\right)-(2-\beta_1)\left(1-\frac{\beta_1^{\frac{1}{2}}}{\pi}\right)=f_{\mathrm{H}}. 
\end{equation}
In light of quantum criticality, we refer to the phase II as the background.
 Thus the terms that do not contribute to the singular part  in thermodynamic quantities can be readily recognised, namely,
 \begin{eqnarray}
 \chi^{\mathrm{II}}_c&=&\frac{1}{2\pi\beta_1^{\frac{1}{2}}}+\frac{\beta_1^{\frac{1}{2}}}{16\pi},\\\chi^{\mathrm{II}}_s&=&\frac{1}{4\pi\beta_1^{\frac{1}{2}}}+\frac{\beta_1^{\frac{1}{2}}}{32\pi},\\\frac{C^{\mathrm{II}}_v}{T}&=&\frac{\pi}{6\beta_1^{\frac{1}{2}}}+\frac{\pi\beta_1^{\frac{1}{2}}}{48}.
 \end{eqnarray}
  These background results are in agreement with (\ref{II}) with $\delta=\beta_1^{\frac{1}{2}}\left(1+\frac{1}{24}\beta_1\right)$, here $\beta_1=2-\mu-2u-B \approx(\pi-Q)^2-\frac{1}{12}(\pi-Q)^4=\delta^2-\frac{1}{12}\delta^4$.  
  Furthermore the Wilson ratios for the phase II are given by 
\begin{equation}
R^{\chi_s}_w=\frac{4\pi^2}{3}\frac{\chi_s}{C_v/T}\approx2,\qquad R^{\chi_c}_w=\frac{\pi^2}{3}\frac{\chi_c}{C_v/T}\approx 1.
\end{equation}

{\bf IV-V phase transition:} The phase transition from  IV to V displays a novel subtlety of quantum  criticality with charge.  
At this phase transition, the charge degree of freedom  becomes gapped and  the dressed energy $\kappa(\pi)$ approaches zero. 
We can directly use the dressed energy equations with the help of (\ref{eq-beta2}) and $\kappa(\pi)=0$ under the condition $\delta=0$, giving the critical fields 
\begin{eqnarray}
\mu_c&=&2-2\sqrt{1+u^2}+\frac{1}{(1+u^2)^{\frac{3}{2}}}A^2_c\left(1-\frac{2}{\pi u}A_c\right),  \label{QC-IV-V-1}\\
B_c&=&-2u+2\sqrt{1+u^2}-\frac{1}{(1+u^2)^{\frac{3}{2}}}A^2_c\left(1+\frac{2}{3\pi u}A_c\right). \label{QC-IV-V-2}
\end{eqnarray}
From (\ref{beta-1-2}) and (\ref{d-A}), it  is  convenient to represent $A_c$ in term of $B_c$ 
\begin{equation}
A_c^2=-\frac{\beta_2}{\eta_1}-\frac{2}{3\pi u}\frac{(-\beta_2)^{\frac{3}{2}}}{\eta^{\frac{3}{2}}_1}+\frac{2}{3\pi^2 u^2}\frac{(-\beta_2)^2}{\eta^2_1}, \label{A-c}
\end{equation}
with  $\eta_1=\frac{2}{(1+u^2)^{3/2}}$. Due to the criticality induced from charge degree, precisely opposite to the transition from phase II to IV, we rewrite the (\ref{QC-IV-V-2}) as 
\begin{eqnarray}
2B+4(u-\sqrt{1+u^2}) +\frac{2\eta_1 }{3\pi u}A^3  +\eta_1 A^2=0. \label{Boundary}
\end{eqnarray}
This equation represents the boundary condition. Using the relation $-D_2$ at zero temperature is approximate by $\eta_1A^2$, we get the main contribution $I_1=2\eta_1A^3/3$ of the first iteration. Comparing this equation with the boundary condition and (\ref{D1})(\ref{D2}), up to the leading contribution,
we have $-D_2 \approx  \eta_1A^2+2J_1/(\pi u)$, which is the leading order  in the arguments of the  polylog functions in $I_1$ and $I_2$.
The integral $I_1$ and $I_2$ serve as the background contributions resulting from of the spin degrees of freedom of the TLL. 
Up to the leading order of the temperature contributions and using the approximation (\ref{polylog-function}), we obtain the function for the second iteration
\begin{eqnarray}
I_1&\approx&\frac{2}{3}\eta_1A^3+\frac{\pi^2T^2}{12\eta_1A}+\frac{2A}{\pi u}J_1-\frac{2A^3}{3\pi u^3}J_1+\frac{\pi T^2}{144u^3\eta_1}A^2, \\
I_2&\approx&\frac{2}{15}\eta_1A^5+\frac{\pi^2T^2}{12\eta_1}A+\frac{2A^3}{3\pi u}J_1. 
\end{eqnarray}
The term $2+C_2$ in the charge dressed energy determines the criticality in the charge degrees of freedom.
Substituting the above expressions of $I_1$ and $I_2$ into  the $C_2$ in (\ref{C2}) and using the critical fields (\ref{QC-IV-V-1}) and (\ref{QC-IV-V-2}), we give 
\begin{equation}
\Delta t\equiv: 2+C_2=-\Delta \mu-\left(1-\frac{4A_c}{\pi u+A_c}\right)\Delta B.
\end{equation}
Subsequently, the scaling function $J_1$ is given by 
\begin{equation}
J_1=\frac{1}{2}T^{\frac{3}{2}}\pi^{\frac{1}{2}}\left(1+\frac{2\eta_1}{3\pi u^3}A^3\right)\Li_{\frac{3}{2}}\left(-\e^{\frac{\Delta t}{T}}\right).
\end{equation}
Using the leading order contribution in $J_1$ and $I_1,\,I_2$, we finally obtain the following scaling form of the Gibbs free energy 
\begin{equation}
f\approx-\mu-u-B-\frac{4}{3}\lambda_1\eta_1A^3-\frac{\pi^2 T^2\lambda_1}{6\eta_1A}-\frac{2}{15}\lambda_2\eta_1A^5-\frac{\pi^2 T^2\lambda_2}{12\eta_1}A+\frac{T^{\frac{3}{2}}b_0}{2\pi^{\frac{1}{2}}}\Li_{\frac{3}{2}}\left(-\e^{\frac{\Delta t}{T}}\right) \label{f-45}
\end{equation}
with $b_0=1-\frac{4\lambda_1}{u}A+\left(\frac{\eta_1}{\pi}+2\lambda_1-u^2\lambda_2\right)\frac{2A^3}{3u^3}$. Using the  thermodynamic relations, the scaling forms of thermal and magnetic properties are  given explicitly by
\begin{eqnarray}
n&=&1+\frac{ T^{\frac{1}{2}}b_0}{2\pi^{\frac{1}{2}}}f_{\frac{1}{2}},\label{QC-IV-V-n} \\
m&=&\frac{1}{2}-\frac{2\lambda_1 A}{1+\frac{A}{\pi u}}-\frac{1}{3}\lambda_2A^3+\frac{ T^{\frac{1}{2}}\left(1-\frac{4A_c}{\pi u+A_c}\right)b_0}{4\pi^{\frac{1}{2}}}f_{\frac{1}{2}},\label{QC-IV-V-m} \\
\chi_c&=&-\frac{ T^{-\frac{1}{2}}b_0}{2\pi^{\frac{1}{2}}}f_{-\frac{1}{2}}, \label{QC-IV-V-chic}\\
\chi_s&=&\frac{2\lambda_1 }{\eta_1A\left(1+\frac{A}{\pi u}\right)^3}+\frac{\lambda_2}{\eta_1}\frac{A}{1+\frac{A}{\pi u}}-\frac{ T^{-\frac{1}{2}}\left(1-\frac{4A_c}{\pi u+A_c}\right)^2b_0}{4\pi^{\frac{1}{2}}}f_{-\frac{1}{2}}, \label{QC-IV-V-chis}\\
\frac{C_v}{T}&=&\frac{\pi^2 \lambda_1}{3\eta_1A}+\frac{\pi^2 \lambda_2A}{6\eta_1}-\frac{3 T^{-\frac{1}{2}}b_0}{8\pi^{\frac{1}{2}}}f_{\frac{3}{2}}-\frac{ T^{-\frac{3}{2}}b_0}{2\pi^{\frac{1}{2}}}\left[\Delta \mu+\left(1-\frac{4A_c}{\pi u+A_c}\right)\Delta B\right]f_{\frac{1}{2}}\nonumber\\&&-\frac{ T^{-\frac{5}{2}}b_0}{2\pi^{\frac{1}{2}}}\left[\Delta \mu+\left(1-\frac{4A_c}{\pi u+A_c}\right)\Delta B\right]^2f_{-\frac{1}{2}}, \label{QC-IV-V-heat}
\end{eqnarray}
where we denoted
$f_n=\Li_n\left(-\e^{\frac{\Delta t }{T}}\right)$. 

Moreover, we observe that the phase V represents the source of the background for the spin degrees of freedom.
Thus we may easily identify the contributions from Mott phase V in the above scaling functions, namely 
\begin{eqnarray}
 \chi^{\mathrm{V}}_c&=&0,\\
 \quad \chi^{\mathrm{V}}_s&=&\frac{2\lambda_1 }{\eta_1A\left(1+\frac{A}{\pi u}\right)^3}+\frac{\lambda_2}{\eta_1}\frac{A}{1+\frac{A}{\pi u}},\\
 \frac{C^{\mathrm{V}}_v}{T}&=&\frac{\pi^2 \lambda_1}{3\eta_1A}+\frac{\pi^2 \lambda_2A}{6\eta_1}.
 \end{eqnarray}
 Comparing above results with equations (\ref{V-nc}) to (\ref{V-vs}), we find these results consistent with each other. 
 Consequently, the Wilson ratios are given by 
\begin{equation}
R^{\chi_s}_w=\frac{4\pi^2}{3}\frac{\chi_s}{C_v/T}\approx\frac{8}{\left(1+\frac{A}{\pi u}\right)^3}+\frac{8\lambda_2A^3}{\pi u\lambda_1},\quad R^{\chi_c}_w=\frac{\pi^2}{3}\frac{\chi_c}{C_v/T}=0,
\end{equation}
showing the nature of the TLL at low temperature. 

\begin{figure}[H] 
	\centering    
	\includegraphics[scale=0.45]{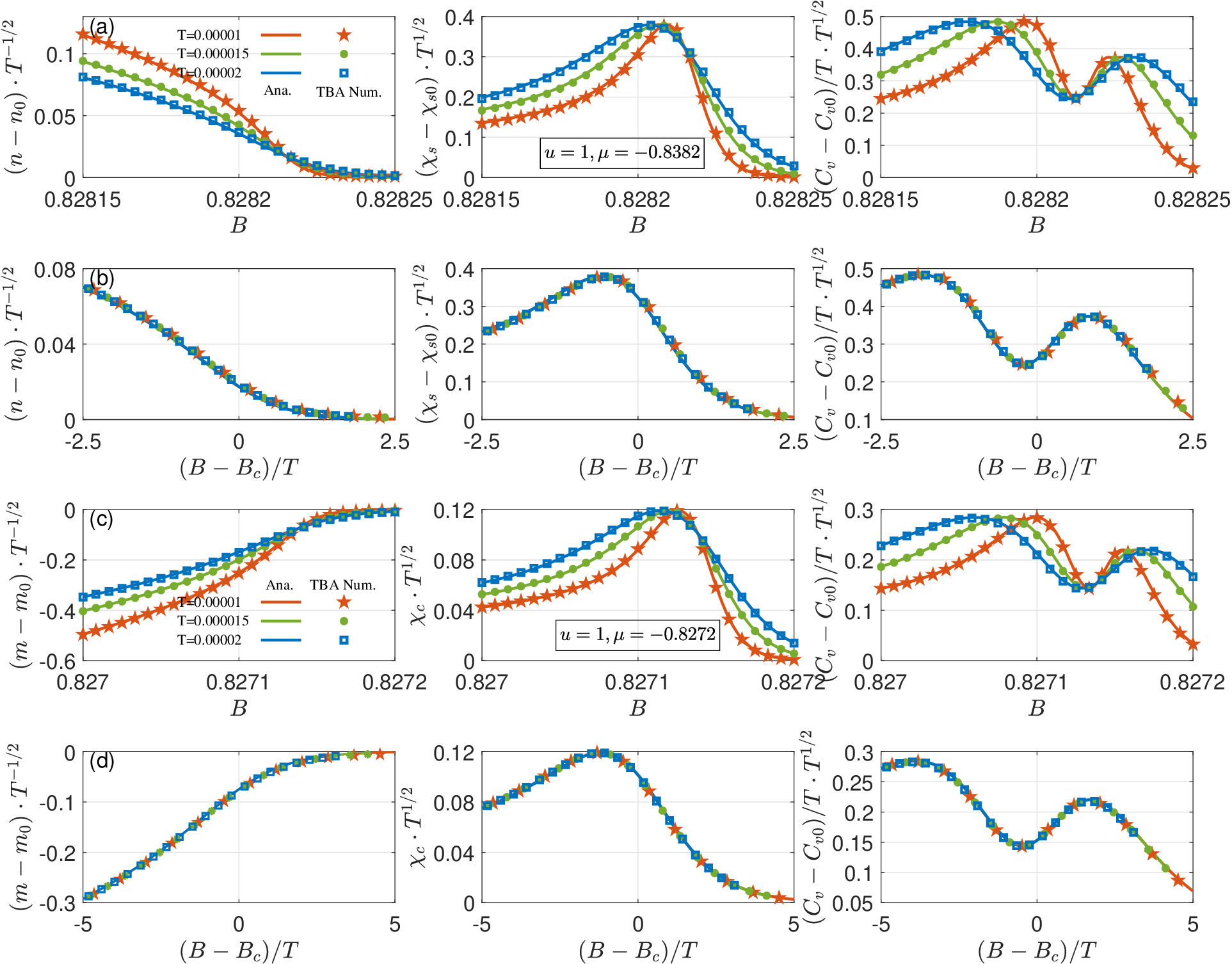}  	
	\caption{Row (a) ((b))  shows universal scaling behaviour of electron density, spin  susceptibility  as well as specific heat  v.s. magnetic field $B$ (the argument $(B-B_c)/T$)  for phase transitions  II-IV. The analytical results (solid lines)  (\ref{QC-II-IV-1})-(\ref{QC-II-IV-5}) agree well with the numerical solutions of the TBA equations. The parameter setting is $u=1,\,\mu=-0.8382$.
		Row (c) ((d))  shows universal scaling behaviour of magnetization, charge   susceptibility  as well as specific heat  v.s. magnetic field $B$ (the argument $(B-B_c)/T$)  for phase transitions  IV-V. The analytical results (solid lines)   (\ref{QC-IV-V-n})-(\ref{QC-IV-V-heat})  agree well with the numerical solution of the TBA equations. The parameter setting is $u=1,\,\mu=-0.8272$.  }      
	\label{fig-45}   
\end{figure}

 In figure~\ref{fig-45} (a) (b) and figure~\ref{fig-45} (c) (d), we plot universal scaling behaviour of electron density and magnetization, spin and charge susceptibilities as well as specific heat  for phase transitions II-IV and IV-V, respectively. 
They show that  analytic expressions of their scaling functions (\ref{QC-II-IV-1})-(\ref{QC-II-IV-5}) and (\ref{QC-IV-V-n})-(\ref{QC-IV-V-heat}) are in good agreement with numerical results obtained from the TBA equations. 
We note that in figure~\ref{fig-45} (a), (c) all lines at different temperatures  intersect at QCP, while figure~\ref{fig-45} (b), (d) show the scaling function  invariant in terms of $\frac{\Delta B}{T}$. 
It is essential to note that all these scaling functions read off the dynamical exponent $z=2$ and correlation critical exponent $\nu=1/2$, for example, the susceptibility (\ref{QC-chi}), also see  \cite{sachdev2011quantum}. 
 In the critical region, the polylog function represents the free fermion type of generating function associated with dynamic critical exponent $z=2$ and correlation critical exponent $v=\frac{1}{2}$ \cite{sachdev2011quantum,continentino2017quantum}.
  In the zero temperature limit, the singular part of the susceptibility can be expressed as $\chi^s_s\propto T^{-\frac{1}{2}}(\Delta B/T)^{-\frac{1}{2}}=(\Delta B)^{-\frac{1}{2}}$. 
Regarding the definition of the critical exponent  $\gamma$ with respect to the general form $\chi\propto (g-g_c)^{-\gamma}$,  here $g$ is the driving parameter, we find   $\gamma=\frac{1}{2}$ for the second order derivatives of the free energy. 
On the other hand, the correlation length can be expressed as $\xi \propto T^{-\frac{1}{z}}$. 
In summary, at the critical point $\Delta B=0$, we also find that thermodynamical properties $\frac{C_v}{T},\, \chi_s,\, \xi$ satisfy the following scaling laws 
\begin{eqnarray}
\frac{C_v}{T}\propto T^{\frac{d-z}{z}}, \quad  \chi_s\propto T^{-\frac{\gamma}{vz}},\quad   \xi\propto T^{-\frac{1}{z}}, 
\end{eqnarray}
which signify the non-Fermi liquid behaviour at QCP\cite{continentino1989critical,continentino2005quantum,continentino2017quantum}.

\subsection{III.6 Universal scaling functions at quantum criticality}
In the previous subsection, we presented some analytic results for each phase transition. 
We observe that the coefficients $a_0$ in (\ref{f-24}) and $b_0$ in (\ref{f-45}) of free energies solely rely on the root densities (\ref{rho-Q}) and (\ref{sigma-A}) with $a_0\approx\sigma(0),b_0\approx2\pi\rho(\pi)$. %
Significantly, it is found that the free energies are related to densities and dressed energies in compact forms for different phase transitions
\begin{eqnarray}
\text{I-II: }\, f&=&u+T^{\frac{3}{2}}\pi^{\frac{1}{2}}\rho(0)\left(\frac{\kappa^{''}(0)}{2}\right)^{-\frac{1}{2}}\mathrm{Li}_{\frac{3}{2}}\left(-\mathrm{e}^{-\frac{\kappa(0)}{T}}\right), \label{1-2}\\
\text{II-III: }\, f&=&f_0+T^{\frac{3}{2}}\pi^{\frac{1}{2}}\rho(\pi)\left(\frac{-\kappa^{''}(\pi)}{2}\right)^{-\frac{1}{2}}\mathrm{Li}_{\frac{3}{2}}\left(-\mathrm{e}^{\frac{\kappa(\pi)}{T}}\right), \label{2-3}\\
\text{V-III: }\, f&=&f_0+T^{\frac{3}{2}}\pi^{\frac{1}{2}}\sigma_1(0)\left(\frac{\varepsilon^{''}_1(0)}{2}\right)^{-\frac{1}{2}}\mathrm{Li}_{\frac{3}{2}}\left(-\mathrm{e}^{-\frac{\varepsilon_1(0)}{T}}\right), \label{3-5}\\
\text{II-IV: }\, f&=&f_0-\frac{\pi T^2}{6v_c}+T^{\frac{3}{2}}\pi^{\frac{1}{2}}\sigma_1(0)\left(\frac{\varepsilon^{''}_1(0)}{2}\right)^{-\frac{1}{2}}\mathrm{Li}_{\frac{3}{2}}\left(-\mathrm{e}^{-\frac{\varepsilon_1(0)}{T}}\right), \label{2-4}\\
\text{V-IV: }\, f&=&f_0-\frac{\pi T^2}{6v_s}+T^{\frac{3}{2}}\pi^{\frac{1}{2}}\rho(\pi)\left(\frac{-\kappa^{''}(\pi)}{2}\right)^{-\frac{1}{2}}\mathrm{Li}_{\frac{3}{2}}\left(-\mathrm{e}^{\frac{\kappa(\pi)}{T}}\right), \label{4-5}
\end{eqnarray}
where $f_0$ comes from the ground state, the terms with $T^2$ reflect the contributions from the background parts.
$\sigma_1(0)$ denotes the density of length-$1$ spin strings  at $\Lambda=0$,  the second derivative $\varepsilon^{''}_1(0) \equiv \left. \frac{d^2\varepsilon_1}{d\Lambda^2} \right|_{\Lambda=0}$,  and $\rho(0),\, \rho(\pi)$ denotes the charge density at $k=0,\, \pi$, respectively. 
Similarly,  for the charge dressed energy,  $\kappa^{''}(0) \equiv \left. \frac{d^2\kappa}{dk^2} \right|_{k=0},\kappa^{''}(\pi) \equiv \left. \frac{d^2\kappa}{dk^2} \right|_{k=\pi}$.
 The polylog function $\mathrm{Li}_{\frac{3}{2}}$ represents the generating function of free fermion criticality. 
 The above scaling functions of the free energy for different quantum phase transitions are valid for arbitrary interaction strengths and fillings, revealing a microscopic origin of the quantum phase transitions  associated with the dressed energies. 
 The functions $-\varepsilon_1(0),\, -\kappa(0),\, \kappa(\pi)$ serve as criticality and
 depend on the energy gaps  away from the QCPs, i.e.  
\begin{eqnarray}
-\varepsilon_1(0), \, -\kappa(0), \, \kappa(\pi)&\approx&\alpha_B\Delta B+\alpha_\mu\Delta \mu+\alpha_u\Delta u.
\end{eqnarray}
The factors $\alpha_{(B,\mu,u)}$ represent the different transition paths in the vicinities of QCPs driven by external fields.
These expressions (\ref{1-2})-(\ref{4-5}) display concise and elegant configurations independent of specific details for arbitrary filling and interaction strength, and can apply to other models with second order phase transitions associated with the dynamical critical experiment $z=2$ and correlation length exponent $\nu=1/2$. 
The derivations for these formulas are given in \cite{LPG:2022}.

\section{IV. Spin incoherent liquid}

Although spin incoherent Luttinger liquid  has been studied in literature \cite{cheianov2004nonunitary,fiete2004green,fiete2007colloquium,fiete2005theory,fiete2005transport}, almost all those works are based on  the framework of bosonization.
 There still lacks a study of such novel phenomenon from the Bethe ansatz perspective.
 In Section II, we used the variations of $\eta$-pair and spin magnetizations $(\Delta\eta ^z, \Delta S^z)$ to characterize the fractional charge and spinon excitations. Such fractionalized quasi-particles reveal fermionic nature of quasiparticles, forming the Luttinger liquid. 
 Meanwhile we found  the only possible fractional spin excitations which can lead to the spin incoherent liquid at low temperatures. 
 Besides the fractional excitations, here  we present rigorous results of the SILL in terms of specific heat, criticality and correlation function. 
  
 \subsection{IV.1 Thermodynamics in SILL} 
In the previous analysis given in section \ref{sec_cri},  we  observe  that a crossover region fanning out from the critical point  does show the existence of the SILL above the phase boundary of the TLL and up to a critical temperature. 
This phenomenon can be revealed through thermodynamic quantities of the model, such as   specific heat. 
We now  identify different energy scaling of the SILL from the TLL and  quantum criticality.  
We first analyze  the variables $\varepsilon_1(0)$ emerged in the polylog functions of II-IV transition in (\ref{2-4}), 
\begin{equation}
\varepsilon_1(0)=2B-\int_{-\pi}^{\pi}\d k \cos ka_1(\sin k-\Lambda)T\ln(1+e^{-\frac{\kappa(k)}{T}})|_{\Lambda=0}, 
\end{equation}
 where $\varepsilon_1(0)=0,<0,>0$ at exact critical point, at phase IV and phase II, representing the distance away from the phase boundary of II-IV. In this case the Fermi point of the charge $\kappa(Q)=0$ gives   $2\cos Q=-\mu-2u-B$, leading to $\kappa(k)=-2\cos k+2\cos Q$, here  $Q$ is the  Fermi point in charge sector. Thus  $\varepsilon_1(0)$ at zero temperature can be simplified by 
\begin{equation}
\varepsilon_1(0)=2B-\frac{4u}{\pi}\int_{0}^{Q}\mathrm{d}k\frac{\cos k}{u^2+\sin^2k}(\cos k-\cos Q)+O(T^2)\label{e0}
\end{equation}
with the temperature term omitted. By further performing Taylor expansion  around the critical point $B_c$ for fixed $\mu$ and $ u$, thus we have 
\begin{eqnarray}
 Q\approx Q_c+\frac{1}{2\sin Q_c}\Delta B
\end{eqnarray}
 with $Q_c=\arccos\left(-\frac{1}{2}(\mu_c+2u_c+B_c)\right)$.  
 After some algebra, the quantity $\varepsilon_1(0)$ in the vicinity of the critical point of phase transition II-IV can be found to be given by 
\begin{eqnarray}
\varepsilon_1(0)&\approx& 2B-\frac{4u}{\pi}\int_{0}^{Q_c}\mathrm{d}k\frac{\cos k}{u^2+\sin^2k}(\cos k-\cos Q_c+\sin Q_c(Q-Q_c))\nonumber\\
&\approx&2\Delta B-\frac{2u}{\pi}\int_{0}^{Q_c}\mathrm{d}k\frac{\cos k}{u^2+\sin^2k}\Delta B\nonumber\\
&\approx&2\Delta B\left[1-\frac{1}{\pi}\arctan\left(\frac{\sin Q_c}{u}\right)\right],\label{e0-dB}
\end{eqnarray}
where the second line brings in the condition $\varepsilon_1(0)|_{B,Q_c}=2\Delta B$ with limitation $\varepsilon_1(0)|_{B_c,Q_c}=0$. Therefore $\alpha_B=-2\left[1-\arctan\left(\sin Q_c/u\right)/\pi\right]$ can be obtained in terms of the definition $-\varepsilon_1(0)\equiv:\alpha_B \Delta B$.

\begin{figure}[H] 
	\centering    
	\includegraphics[scale=0.39]{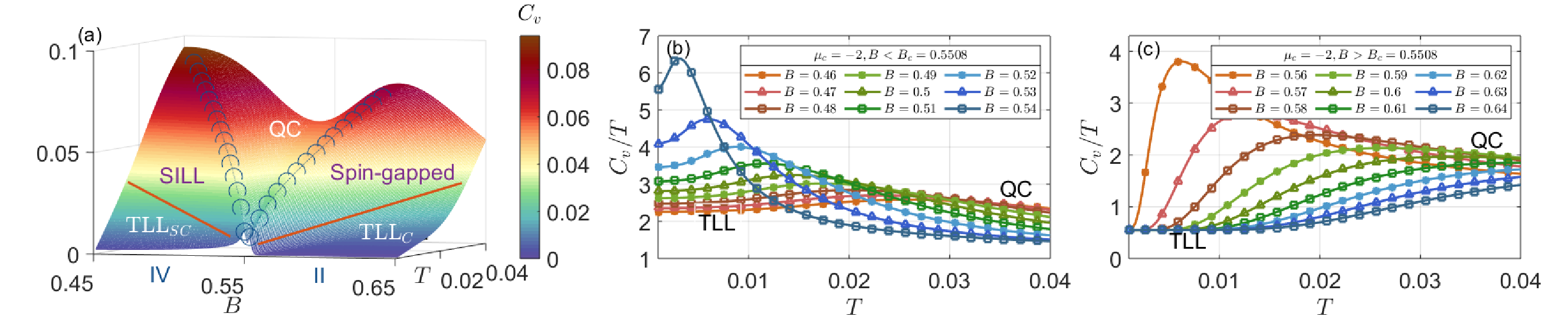}  	
	\caption{(a) 3D plot of specific heat in the $T-B-C_v$ coordinate at $\mu=-2,u=1$ for IV-II phase transition. The blue circled symbols present the critical temperatures determined by the  maximum values of the specific heat (equation (\ref{B-cv})), consistent with numerical simulation. The red lines mark  the TLL phase boundaries below which the  specific heat show a linear temperature dependent.  Crossover regimes between the blue circled symbols and the red lines  denote the SILL phase (on the left) and the spin-gapped phase (on the right). (b) and (c) show the specfic heat v.s. temperature near the phase transition  IV-II for the magnetic field is  less than (greater than) critical magnetic field $B_c$, respectively.  The temperature-independent region and the temperature-square-dependent region in the ratio $C_v/T $  can be visible, charactering the thermodynamics of the TLL and SILL, respectively. }            
	\label{COR}    
\end{figure}

 On the other hand, in the quantum critical region, heat capacity satisfies the universal scaling form 
\begin{equation}
C_v/T =  c_0+c_1T^{-1/2} \left[ \frac{3}{4} \Li_{\frac{3}{2}}\left(-\e^x\right)- x\Li_{\frac{1}{2}}\left(-\e^x\right) + x^2\Li_{-\frac{1}{2}}\left(-\e^x\right)\right]+O((\Delta B/T)^{5/2}), \label{B-cv}
\end{equation}
where $c_0$ is the zero temperature background, $c_1$ a coefficient depending on the  transition point, explicitly see (\ref{QC-II-IV-5}), and $x=-\varepsilon_1(0)/T=\alpha_B \Delta B/T$. 
A brief discussion about interaction-driven case is given in \cite{LPG:2022}.
One characteristic of heat capacity is that it displays bimodal structures around QCP. Thus it is efficient to mark the QC boundaries in terms of the maxima points given by heat capacity.
The local maxima can be determined by $\partial C_v/\partial B=0$, i.e. 
\begin{equation}
\frac{1}{4} \mathrm{Li}_{\frac{1}{2}}\left(-\mathrm{e}^x\right)- x\mathrm{Li}_{-\frac{1}{2}}\left(-\mathrm{e}^x\right) - x^2\mathrm{Li}_{-\frac{3}{2}}\left(-\mathrm{e}^x\right)=0,
\end{equation}
which gives two solutions $x_1=-1.5629, \, x_2=3.6205$. 
Figure \ref{COR} (a) shows an overall behavior of quantum criticality in the vicinity of the IV-II phase transition in the $T-B-C_v$ coordinate. 
The blue circled symbols denote the maxima of specific heat from analytical results (\ref{B-cv}), showing a good agreement between numerical TBA equations. 
 In this critical region, $T\gg \Delta B= B-B_c$, other thermodynamic properties  also show universal scaling behaviour given by equations (\ref{QC-II-IV-1})-(\ref{QC-II-IV-5}).

As temperature decreases gradually from the QC part to a certain extent, TLL regions appear, see the areas below the red lines in figure \ref{COR} (a).
Around the critical point $B_c=0.55$, there emerges two TLLs. In the region with $B<B_c$, denoted as TLL$_{SC}$, the system lies in phase IV with spin and charge degrees of freedom coexisting. More than $B_c$, the system lies in phase II with only charge degree, denoted as TLL$_{C}$.
In the TLL region, the  specific heat $C_v$ is linearly dependent on $T$, also see discussion in Section III. 
In the crossover region between QC  and the TLL$_C$ phase, the spin sector is gapped. 
By utilizing the asymptotic behaviour of polylog function \cite{mohankumar2007two} and expanding (\ref{QC-II-IV-5}), the specific heat is given by
\begin{equation}
C_v\approx \frac{\pi T}{3v_c}+\frac{3\pi^{\frac{1}{2}}a_0}{4} \left(\eta_1\right)^{-\frac{1}{2}}T^{\frac{1}{2}}\mathrm{e}^{\alpha_B\Delta B/T}+O\left(\Delta B\mathrm{e}^{\alpha_B\Delta B/T}\right). \label{QC-II-IV-cv-1}
\end{equation}

By comparison, in the crossover region between QC and the TLL$_{SC}$ phase, the SILL lies in the temperature range $E_s\sim k_Fv_s\ll  k_BT\ll E_c\sim k_Fv_c$. 
This area corresponds to SILL (spin-incoherent Luttinger liquid) \cite{cheianov2004nonunitary,fiete2004green,fiete2007colloquium,fiete2005theory,fiete2005transport} with specific heat
\begin{equation}
C_v\approx \frac{\pi T}{3v_c}+\frac{\pi^2a_0\left(\eta_1\right)^{-\frac{1}{2}}(-\varepsilon_1(0))^{-\frac{1}{2}}T}{3}\left[1+\frac{21\pi^2}{40}(-\varepsilon_1(0))^{-2}T^2\right]+O(T^4),
\end{equation}
which manifests a gas-liquid co-existence. 
 We note that the coefficient before the square bracket should be $\pi T/(3v_s)$. The proof is given in  following. 
 The spin dressed energy around QCP is written as $\varepsilon_1(\Lambda)=D_1\Lambda^2+D_2$. Thus for the SILL region, the spin Fermi point $A$ is related with $\varepsilon_1(A)=D_1A^2+D_2=0$, resulting in $A=(-D_2/D_1)^{1/2}$. On the other hand, by the definition of $D_{1,2}$ given in (\ref{D1})  and (\ref{D2}), we have $D_1=\varepsilon^{''}_1(0)/2=\eta_1,D_2=\varepsilon_1(0)$.
 Therefore the Fermi point $A$ is expressed as $A=(-\varepsilon_1(0)/\eta_1)^{1/2}$. Using the definition of $a_0$, see (\ref{f-24}), 
 we have $a_0=\sigma_1(0)\approx\sigma_1(A)$ and $\varepsilon_1^{'}(A)=2\eta_1A$, which renders 
\begin{eqnarray}
&&\frac{\pi^2a_0\left(\eta_1\right)^{-\frac{1}{2}}(-\varepsilon_1(0))^{-\frac{1}{2}}T}{3}\nonumber\\&=&\frac{\pi^2\sigma_1(A)T\left(\eta_1\right)^{\frac{1}{2}}}{3}\frac{1}{\eta_1(-\varepsilon_1(0))^{\frac{1}{2}}}\nonumber\\&=&\frac{\pi^2\sigma_1(A)T}{3\eta_1A}\nonumber=\frac{\pi\cdot2\pi\sigma_1(A)T}{3\varepsilon_1^{'}(A)}\nonumber\\&=&\frac{\pi T}{3v_s}. 
\end{eqnarray}
This immediately gives an universal thermodynamic relation of the SILL
\begin{eqnarray}
C_v\approx \frac{\pi T}{3}\left( \frac{1}{v_c} + \frac{1}{v_s} \right) +\frac{7\pi^3T^3}{40 v_s(-\varepsilon_1(0))^2} +O(T^4). \label{QC-II-IV-cv-2}
\end{eqnarray}

Furthermore, in figure \ref{COR} (b) and (c), we plot the specific heat below and above the QCP for different values of magnetic fields. 
It is obvious to see the region of the linear-temperature dependent specific heat, a crossover region of the SILL with both the linear- and cubic- temperature-dependent specific heat. 
The latter marks the crossover region  $E_s\sim k_Fv_s\ll  k_BT\ll E_c\sim k_Fv_c$, showing the thermodynamic behaviour of the SILL.

\subsection{IV.2 Correlation functions in SILL}

The SILL regime with $k_Fv_s<k_BT\ll k_Fv_c$ lies between  the boundaries of TLL and QC.
In this region, the spin degree of freedom behaves like hot spins, whereas the  charge still behaves as a collective motion of bosons.
As a result, the SILL largely behaves like a spin and charge decohered liquid, i.e., possessing solely a propagating charge mode but not a spin mode, see \cite{cheianov2004nonunitary,fiete2004green,fiete2007colloquium,fiete2005theory,fiete2005transport}. 
In the spin sector, the magnetic exchange energy is lower than the Fermi energy, resulting in spin thermally excited with equal probability \cite{fiete2005transport}. 
In the strong coupling regime, the spin degrees of freedom shows a spin dynamics of  Heisenberg chain with nondispersive spinon band due to small effective exchange coupling $J=4t^2/U$.
 Whereas the charge acts as noninteracting fermions with dispersive spectrum $\kappa(k)=-2t\cos(k)$. 
 This SILL theory can also be captured in the excitation spectrum in the low density regime in figure \ref{Cph-Shh} (b).
 The concept of SILL is helpful to explain the appearance of conductance plateau in a quantum wire\cite{hew2008spin}. 
 When  interaction increases or in the vicinity of QCP,  the spin velocity progressively dwindles to zero. This means that the spin sector loses dynamics, 
  indicating the emergence of the SILL.

In this regime $E_s\ll T\ll E_c$, the spin sector is totally  thermal averaged and equally excited, whereas the  charge remains at relevant low-energy performance, rendering the correlation functions independent of temperature \cite{fiete2007colloquium}.
We note that the TLL theory has its own applicable condition $T\ll E_c,\, E_s$. 
Whereas for the region $E_s\ll T\ll E_c$, the energy scales of charge and spin degrees of freedom can be dealt with separately \cite{cheianov2004nonunitary}.
Under such a circumstance, the  finite-temperature correlation functions in terms of conformal field theory \cite{schulz1990correlation,frahm1990critical,frahm1991correlation} still remain valid for the charge and spin  degrees of freedom operating with different limits,  i.e. 
\begin{eqnarray}
|x\pm \mathrm{i}v_ct|\ll v_c/T,\qquad  |x\pm \mathrm{i}v_st|\gg v_s/T. \label{SILL-condition}
\end{eqnarray}
This  is essential to capture the asymptotic behaviour of the SILL. 
Here we would like to mention that the remaining temperature term in the spin sector can be replaced by the typical energy scale of spin  $T\sim E_s\sim J\sim \left(k_{F\uparrow}+k_{F\downarrow}\right)/2\cdot v_s\equiv k_Fv_s$. 
With the help of the finite temperature asymptotics of correlation functions  under the condition (\ref{SILL-condition}), the two-point correlation functions of prime fields can be obtained as 
\begin{eqnarray}
&&\left\langle \phi(x,t)\phi(0,0)\right\rangle=\sum A(D_c,D_s,N^{\pm}_c,N^{\pm}_s)\text{exp}(-2\mathrm{i}D_ck_{F,\uparrow}x)\text{exp}(-2\mathrm{i}(D_c+D_s)k_{F,\downarrow}x)\nonumber\\&&\times\frac{1}{(x-\mathrm{i}v_ct)^{2\Delta^{+}_c}(x+\mathrm{i}v_ct)^{2\Delta^{-}_c}}\left(2\pi \alpha k_F\right)^{2\Delta^{+}_s+2\Delta^{-}_s}\e^{-\pi\alpha\left(2\Delta^{+}_s+2\Delta^{-}_s\right) k_Fx},
\label{G-SILL}
\end{eqnarray}
where the conformal dimensions in gapless phases read
\begin{eqnarray}
2\Delta^{\pm}_c(\Delta \bm{N},\bm{D})&=&\left(\mathrm{Z}_{cc}D_c+\mathrm{Z}_{sc}D_s\pm \frac{\mathrm{Z}_{ss}\Delta N_c-\mathrm{Z}_{cs}\Delta N_s}{2\mathrm{det}\mathrm{Z}}\right)^2+2N^{\pm}_c, \nonumber\\
2\Delta^{\pm}_s(\Delta \bm{N},\bm{D})&=&\left(\mathrm{Z}_{cs}D_c+\mathrm{Z}_{ss}D_s\pm \frac{\mathrm{Z}_{cc}\Delta N_s-\mathrm{Z}_{sc}\Delta N_c}{2\mathrm{det}\mathrm{Z}}\right)^2+2N^{\pm}_s.
\end{eqnarray}
Here  $\mathrm{Z}$ is the dressed charge matrix given in (\ref{Z}), while $N_\alpha^\pm,\Delta
\vec N,\vec D$ are related to the three types of excitations: adding particle from left and right Fermi points, changing total particle number, moving of the particle center displacement, also see  \cite{essler2005one}. 
From equation (\ref{G-SILL}), the spin-spin correlation in the spin sector displays exponential decay as a function of distance, while the correlation in the charge sector shows a power-law decay. 
 Explicitly, based on the correlation functions obtained from the finite temperature CFT, we may calculate various two-point correlations of field operators close to critical field $B_c$ in SILL regime, including the  single particle Green's function
\begin{enumerate}[(1)]
	\item $G^{\uparrow}$
	\begin{equation}
	G^{\uparrow }_{B\rightarrow B_c}\sim\frac{\text{exp}(-\mathrm{i}k_{F,\uparrow}x)}{(x-\mathrm{i}v_ct)^{1-\frac{2}{\pi}\sqrt{1-\frac{B}{B_c}}}}\left(2\pi \alpha k_F\right)^{\frac{1}{2}-\frac{1}{\pi}\sqrt{1-\frac{B}{B_c}}}\e^{-\pi \alpha \left(\frac{1}{2}-\frac{1}{\pi}\sqrt{1-\frac{B}{B_c}}\right) k_Fx}+h.c..
	\end{equation}
This  result was derived for the  region close to the critical field $B_c$, corresponding to the phase transition II-IV. Similarly, for other correlations 
		
	\item $G^{\downarrow}$
	\begin{eqnarray}
	G^{\downarrow }_{B\rightarrow B_c}&\sim&\frac{\text{exp}(-\mathrm{i}k_{F,\downarrow}x)}{(x-\mathrm{i}v_ct)^{\frac{1}{4}+\frac{1}{\pi}\sqrt{1-\frac{B}{B_c}}}(x+\mathrm{i}v_ct)^{\frac{1}{4}-\frac{1}{\pi}\sqrt{1-\frac{B}{B_c}}}}\nonumber\\
	&&\times \left(2\pi \alpha k_F\right)^{1-\frac{2}{\pi}\sqrt{1-\frac{B}{B_c}}}\e^{-\pi \alpha \left(1-\frac{2}{\pi}\sqrt{1-\frac{B}{B_c}}\right) k_Fx}+h.c..
	\end{eqnarray}
	
	\item $G^{n}$
	\begin{eqnarray}
	 G^{n}_{B\rightarrow B_c}&\sim&n^2 +\left(\text{exp}(2\mathrm{i}k_{F,\uparrow}x)+\text{exp}(-2\mathrm{i}k_{F,\uparrow}x)\right)\frac{1}{|x-\mathrm{i}v_ct|^{2-\frac{8}{\pi}\sqrt{1-\frac{B}{B_c}}}}\nonumber\\
	 &&\times \left(2\pi \alpha k_F\right)^{2-\frac{4}{\pi}\sqrt{1-\frac{B}{B_c}}}\e^{-\pi \alpha \left(2-\frac{4}{\pi}\sqrt{1-\frac{B}{B_c}}\right) k_Fx}\nonumber\\
	&&+\left(\text{exp}(2\mathrm{i}k_{F,\downarrow}x)+\text{exp}(-2\mathrm{i}k_{F,\downarrow}x)\right)\left(2\pi \alpha k_F\right)^{2-\frac{4}{\pi}\sqrt{1-\frac{B}{B_c}}}\e^{-\pi \alpha \left(2-\frac{4}{\pi}\sqrt{1-\frac{B}{B_c}}\right) k_Fx}\nonumber\\
	&&+\left(\text{exp}(2\mathrm{i}\left(k_{F,\uparrow}+k_{F,\downarrow}\right)x)+\text{exp}(-2\mathrm{i}\left(k_{F,\uparrow}+k_{F,\downarrow}\right)x)\right)\frac{1}{|x-\mathrm{i}v_ct|^2}.
	\end{eqnarray}
	
	\item $G^{\perp}$
	\begin{eqnarray}
	G^{\perp }_{B\rightarrow B_c}&\sim&\left(\text{exp}(\mathrm{i}\left(k_{F,\uparrow}+k_{F,\downarrow}\right)x)+\text{exp}(-\mathrm{i}\left(k_{F,\uparrow}+k_{F,\downarrow}\right)x)\right)\frac{1}{|x-\mathrm{i}v_ct|^{\frac{1}{2}}}\nonumber\\
	&&\times \left(2\pi \alpha k_F\right)^{\frac{1}{2}+\frac{1}{\pi}\sqrt{1-\frac{B}{B_c}}}\e^{-\pi \alpha \left(\frac{1}{2}+\frac{1}{\pi}\sqrt{1-\frac{B}{B_c}}\right) k_Fx}.
	\end{eqnarray}
		
	\item $G^{p}$
	\begin{eqnarray}
	G^{p }_{B\rightarrow B_c}&\sim&\text{exp}(-\mathrm{i}\left(k_{F,\uparrow}+k_{F,\downarrow}\right)x)\frac{1}{(x-\mathrm{i}v_ct)^{\frac{9}{4}}(x+\mathrm{i}v_ct)^{\frac{1}{4}}}\nonumber\\
	&&\times \left(2\pi \alpha k_F\right)^{\frac{1}{2}-\frac{3}{\pi}\sqrt{1-\frac{B}{B_c}}}\e^{-\pi \alpha \left(\frac{1}{2}-\frac{3}{\pi}\sqrt{1-\frac{B}{B_c}}\right) k_Fx}+h.c..
	\end{eqnarray}
	
\end{enumerate}
It should be noticed that although these results are obtained in a very rough approximation, they capture the essential features of the SILL  \cite{fiete2007colloquium}, i.e.,
spin-spin correlation decays exponentially while the correlation in the charge degree of freedom behaves like spinless noninteracting fermions.
 The field theory approach was given in  \cite{cheianov2004nonunitary,fiete2004green,fiete2007colloquium}.
  The reason why the crossover in the vicinity of phase transition  between IV and V can not develop such a similar concept as charge incoherent Luttinger liquid is that near a half-filled lattice, the charge velocity monotonically and exponentially tends to zero \cite{giamarchi2003quantum}.
  This means subtlety occurring  near the Mott phase transition due to rapid vanishing charge collective mode. 
   This limitation is called as the holon confinement.
   In the next section, we will develop a new concept, Contact susceptibility, to study the Mott phase transition.

\section{V. Contact and Contact susceptibility}

In this section, we focus on an experimentally measurable quantity double occupancy \cite{campo2012thermal,van2018competing}, as well as the associated  Contact, revealing the competition between thermal fluctuations and quantum fluctuations from different sources, i.e., external fields and interaction.
The double occupancy serves as an efficient instrument to demarcate phase transition, especially the Mott phase transition in the extended Hubbard model with long range interaction \cite{glocke2007half,de2011thermodynamics}.
On the other hand, the partial wave Contact, which was first proposed in ultracold Fermi gas \cite{tan2008energetics,ZhangSZ:2009}, has become 
an important theme in the study of ultracold atoms \cite{Braaten:2009,Stewart:2010,Yu:2015,Yoshida:2015,Cui:2016,chen2014critical}. 
Here we will introduce the concept of  Contact susceptibilities with respect to the temperature, magnetic field and chemical potential, and investigate applications of these Contact susceptibilities.
We will show that the susceptibilities build up a general connection between interaction-driven quantum criticality and the phase transitions induced by external fields. 
Using these relations, we will also obtain caloric effect in interaction-driven quantum refrigeration and scaling laws at quantum criticality. 

\begin{figure}[t] 
	\centering    
	\includegraphics[scale=0.36]{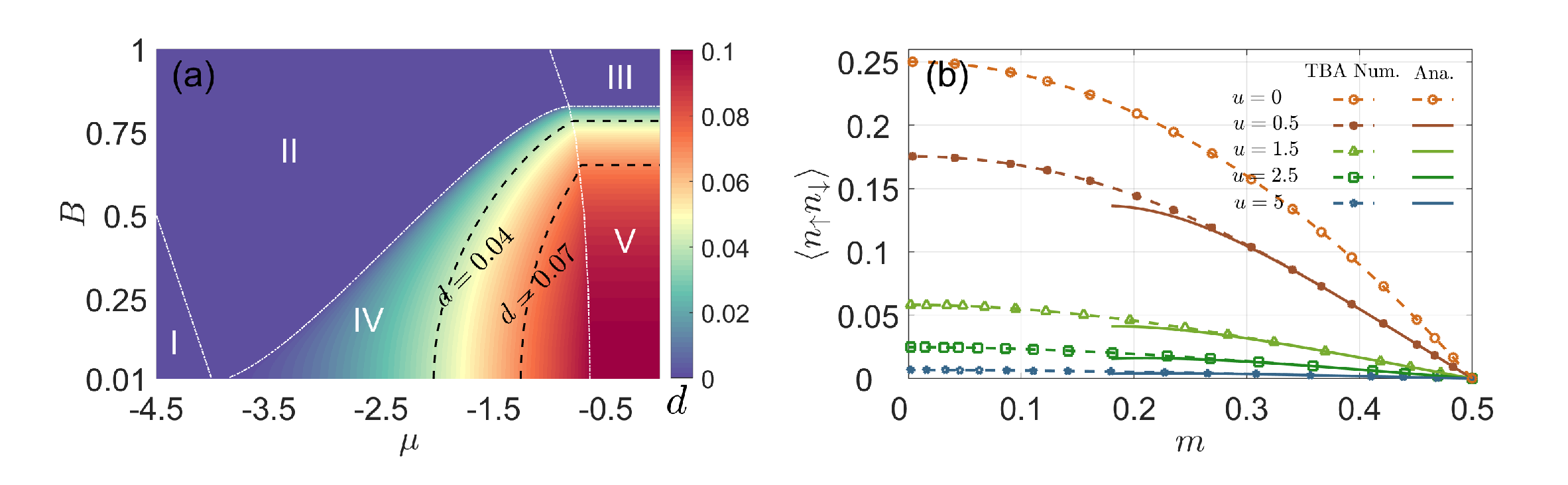}  	
	\caption{(a): Double occupancy maps out the phase diagram of the 1D Hubbard model plotted. The white dot-dashed lines show the zero temperature phase boundaries. While the thick dashed lines show the contour lines of $d=0.04$ and $d=0.07$, respectively. At the critical points, the double occupancy suddenly changes.  Parameter setting for numerics is  the same as figure (\ref{fig-R}) (a), i.e. $T=0.005$ and $u=1$. 
		(b): Double occupancy versus magnatization  in Mott phase for  various coupling strengths. The numerics from the TBA equations confirms the accuracy of the analytical result (\ref{DO-d}) with the Contact given by (\ref{Contact-ana}).}  
	\label{DO-C}    
\end{figure}

\subsection{V.1 Double occupancy and Mott phase}
In the Hubbard model, the lattice filling parameter, interaction strength and external fields can drive different phase transitions \cite{giamarchi1991umklapp,giamarchi1997mott}.
In contrast to the dilute limit case, the Mott insulator phase induced by interaction is much less understood. 
When the interaction strength increases up to a critical value $u_c $\cite{glocke2007half,de2011thermodynamics,giamarchi2003quantum} in the canonical ensemble, the system reaches a Mott insulator state in the extended Hubbard model. 
The half-filled phase can be delineated by the Luther-Emery liquid with one gapless and one gapped sectors \cite{luther1974backward}, in which the excitations  comply with bosonic and fermionic statistics, respectively.
 In order to explore the magnetic order and detect the Mott phase transition, one can introduce  the double occupancy $d=\frac{1}{N}\sum_i\left\langle n_{i,\uparrow}n_{i,\downarrow} \right\rangle$, which  depicts the probability of two electrons with opposite spin occupying a single lattice site with potential energy $E_{\mathrm{pot}}=u d$. 
The double occupancy has been invested extensively in theoretical and experimental research on the Hubbard model at half filling.
It appears to show discontinuity as the interaction approaches critical coupling in interaction-induced Mott transition (note that $u_c=0$ for  this model in the canonical ensemble in which this phenomenon is unconspicuous) and exhibits nonmonotonic behavior in the half-filled state as temperature varies, similar to the Pomeranchuk effect (the melting pressure of liquid Helium-3 shows a trend of first decreasing and then increasing with temperature) \cite{glocke2007half,de2011thermodynamics,wietek2021mott}.
It also reveals the competing physics from the charge and spin fluctuations. 
For doping-induced Mott transition, $d$ is significantly suppressed in the area of $\mu<u$ and generates a pronounced signal when density exceeds unity \cite{de2011thermodynamics}.
It is useful to locate the Mott transition point through the detection of double occupancy. 

In the canonical ensemble, the double occupancy can be obtained from the free energy $f$ by 
\begin{eqnarray}
 d&=&\frac{1}{4} \frac{\partial f}{\partial u}-\frac{1}{4}+\frac{n_c}{2},\label{DO-d}
 \end{eqnarray}
  in which the last two  terms $-1/4+n_c/2$ stem from the extra terms $-2uN+uL$ in the Hamiltonian.
We can define $C=\partial f/\partial u$ to be the lattice version of Contact $C$,
\begin{eqnarray}
C=\partial f/ \partial u=4 d -2n_c+1,\label{Contact}
\end{eqnarray}
 which is analogous to the Tan's Contact \cite{tan2008energetics,chen2014critical} in the continuous systems of ultracold atoms.
 Regarding the role that the double occupancy can reflect the phase information, we expect that $d$  distinguishes different phases with and without internal degrees of freedom, see figure \ref{DO-C} (a). 
 $d$ always vanishes for phases I, II, and III and has no demarcation lines  between these phases. 
 The boundaries of II and IV or III and V are obvious, while for IV and V the inflection point of the contour line marks the phase transition. 
 By the fact that the double occupancy essentially reflects charge fluctuations at Mott insulator accompanying with the existence of the antiferromagnetic order, we demonstrate the roles of magnetic field $B$ and interaction strength $u$ on $d$ in figure (\ref{DO-C}) (b). 
 It is  clear to observe that repulsive interaction always lowers the possibility of double occupancy which is quite intuitive. 
 When the  interaction increases, the electrons with different spins are repel each other more strongly and the system becomes more incompressible.
 The infinity coupling $u\to \infty$ makes the particles fully localized. 
 The decrease caused by magnetization originates from the tendency that the magnetic field tends to align the spins of electrons in the field direction. 
 Through the derivative of the background term in (\ref{f-45}) with respect to $u$, the exact analytic expression of the  Contact $C_0$ at Mott phase V of ground state can be obtained as
\begin{equation}
C_0=-1-\frac{4}{3}\frac{\partial}{\partial u}\left(\lambda_1\eta_1\right)A^3-4\lambda_1\eta_1A^2\frac{\partial A}{\partial u}-\frac{2}{15}\frac{\partial}{\partial u}\left(\lambda_2\eta_1\right)A^5-\frac{2}{3}\lambda_2\eta_1A^4\frac{\partial A}{\partial u}, \label{Contact-ana}
\end{equation}
where $\partial A/\partial u$ is given in (\ref{A-u}) below.
We plotted the double occupancy through  (\ref{Contact-ana}) (solid lines) in figure \ref{DO-C} (b), showing a good agreement between this analytical result and the numerical calculation from the TBA equations. 
Based on the relation from (\ref{Contact}), it is more essential to investigate the variations of the Contact with respect to the  changes of the external fields and temperature. 

\begin{figure}[H] 
	\centering    
	\includegraphics[scale=0.45]{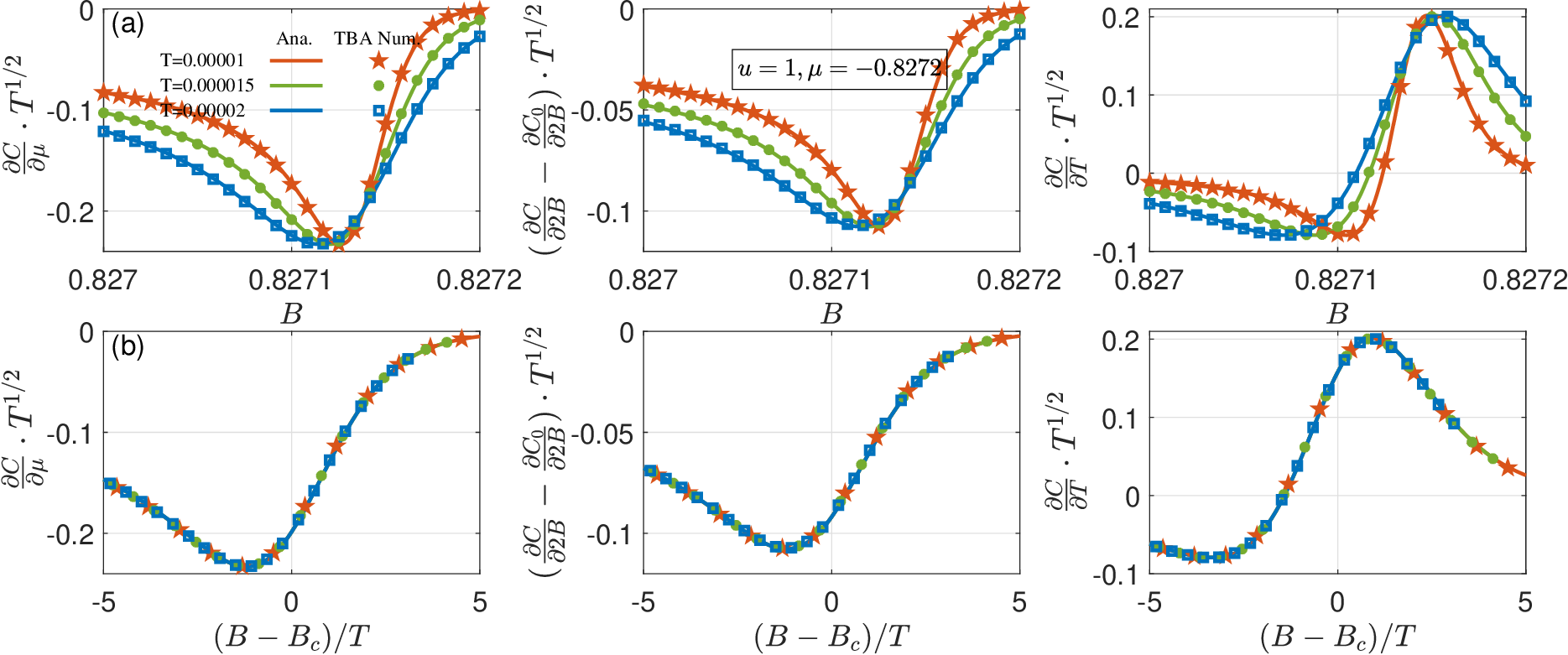}  	
	\caption{Universal Scaling behaviours of susceptibilities: (a) Contact susceptibilities with respect to chemical potential, magnetic field and temperature v.s. magnetic field  at the quantum criticality of phase transition IV-V. 
		(b) Contact  susceptibilities v.s. the argument $(B-B_c)/T$ show  universal  scaling forms at quantum criticality. Parameter setting reads $u=1,$ and $\mu=-0.8272$. Numerical results obtained from the TBA equations (symbols) confirm the scaling functions of Contact susceptibilities (\ref{CS-C})-(\ref{CS-C-T})  (solid lines). }       
	\label{C45}      
\end{figure}

\subsection{V.2 Contact susceptibilities and Mott phase transition}
In comparison to the double occupancy $d$, the Contact $C$ is more essential to capture many-body effects induced by interaction strength. 
Using the thermal potential $f=e-\mu n_c-2Bm-Ts-uC$ and the Maxwell relations in its derivatives with respect to the temperature, chemical potential and magnetic fields, we may build up general relations between Contact susceptibilities and interaction-driven variations of density, magnetization and entropy:
\begin{eqnarray}
\frac{\partial n_c}{\partial u}&=&-\frac{\partial C}{\partial \mu}, \label{umu} \\
\frac{\partial m}{\partial u}&=&-\frac{\partial C}{\partial (2B)},\\
\frac{\partial s}{\partial u}&=&-\frac{\partial C}{\partial T} \label{uT}. \label{uB}
\end{eqnarray} 
We  prove that these Contact susceptibilities will provide striking features of interaction effect in the thermodynamics of the model. 

Building on the Contact susceptibilities, we further show that the  Contact susceptibilities exhibit universal scaling behaviour in the quantum critical region \cite{chen2014critical}. 
The Mott insulator phase with average one electron occupying one site is of particular interest in many-body physics.
Here using the free energy (\ref{f-45}) for the transition from IV to V, we obtain the Contact susceptibilities
\begin{eqnarray}
C&=&-1-\frac{4}{3}f_{11}A^3-4\lambda_1\eta_1A^2\frac{\partial A}{\partial u}-\frac{2}{15}f_{21}A^5-\frac{2}{3}\lambda_2\eta_1A^4\frac{\partial A}{\partial u}+\frac{T^{\frac{1}{2}}b_0b_1}{2\pi^{\frac{1}{2}}}f_{\frac{1}{2}}, \label{CS-C}\\
\frac{\partial C}{\partial \mu}&=&-\frac{b_0b_1}{2\pi^{\frac{1}{2}} T^{\frac{1}{2}}}f_{-\frac{1}{2}}, \label{CS-C-Mu}\\
\frac{\partial C}{\partial (2B)}&=&\frac{\partial C_0}{\partial 2B} -\frac{b_0b_1}{4\pi^{\frac{1}{2}}T^{\frac{1}{2}}}\left(1-\frac{4A_c}{\pi u+A_c}\right)f_{-\frac{1}{2}},  \label{CS-C-B}\\
\frac{\partial C}{\partial T}&=&\frac{b_0b_1}{4\pi^{\frac{1}{2}}T^{\frac{1}{2}}}f_{\frac{1}{2}}+\frac{b_0b_1}{2\pi^{\frac{1}{2}}T^{\frac{3}{2}}}\left[\Delta \mu+\left(1-\frac{4A_c}{\pi u+A_c}\right)\Delta B\right]f_{-\frac{1}{2}}, \label{CS-C-T}
\end{eqnarray}
where $C_0$ denotes the background terms of ground state, that is equation (\ref{Contact-ana}). We have denoted the background of the magnetic susceptibility and other functions as 
\begin{eqnarray}
\frac{\partial C_0}{\partial 2B}&=&-4A^2\left(f_{11}+\frac{1}{6}f_{21}A^2\right)\frac{\partial A}{\partial( 2B)}-8\eta_1A\left(\lambda_1+\frac{1}{3}\lambda_2 A^2\right)\frac{\partial A}{\partial (2B)}\frac{\partial A}{\partial u}, \nonumber\\
&&-4\eta_1A^2\left(\lambda_1+\frac{1}{6}\lambda_2A^2\right)\frac{\partial^2 A}{\partial (2B) \partial u},  \\
f_{11}&=&\frac{\partial}{\partial u}\left(\lambda_1\eta_1\right),f_{21}=\frac{\partial}{\partial u}\left(\lambda_2\eta_1\right),\qquad \frac{\partial A}{\partial B}=-\frac{1}{\eta_1 A\left(1+\frac{A}{\pi u}\right)},\nonumber \\
b_1&=&\frac{\partial (2+C_2)}{\partial u}=-2-\frac{8}{\pi u\left(1+u^2\right)^{\frac{3}{2}}}A^2\frac{\partial A}{\partial u}.\nonumber
\end{eqnarray}
 Solving (\ref{A-c}), we have 
\begin{eqnarray}
\frac{\partial A}{\partial u}&=&\left(1+u^2\right)\frac{u-\left(1+u^2\right)^{\frac{1}{2}}}{A\left(1+\frac{A}{\pi u}\right)}+\frac{3u}{2\left(1+u^2\right)}\frac{A\left[1+\frac{2}{9\pi u}\left(4+\frac{1}{u^2}\right)A\right]}{1+\frac{A}{\pi u}}\label{A-u},\\
\frac{\partial^2 A}{\partial B \partial u}&=& \frac{1+\frac{2A}{\pi u}}{\eta_1 A^2 \left(1+\frac{A}{\pi u}\right)^2}\frac{\partial A}{\partial u}
-\frac{1}{\eta_1\pi u^2\left(1+\frac{A}{\pi u}\right)^2}.
\end{eqnarray}
In figure \ref{C45}, we demonstrate scaling behaviour of Contact susceptibilities, confirming the analytic expressions (\ref{CS-C})-(\ref{CS-C-T})  (solid lines) with the numerical data from TBA (symbols).

\subsection{V.3 Interaction-driven quantum cooling}

It is remarkable to observe that equation (\ref{uT}) essentially relates entropy, temperature and interaction strength. 
This relation is of great importance for interaction driven quantum cooling since the interaction  can be tuned in cold atom experiments via Feshbach resonance, and temperature plays a vital role in controlling the entropy.
Therefore an adiabatic cycle process can be used in realization of quantum refrigeration in real physical systems. 
Inspired by this, here we consider an isentropic process by ramping up or down the interaction strength. 
Focusing on the $(T,u)$ coordinates for a fixed magnetic field, we perform total derivatives on entropy $s$, i.e., 
\begin{equation}\d s=\frac{\partial s}{\partial u}\d u+\frac{\partial s}{\partial T}\d T=0.
\end{equation}
Using equation (\ref{uT}) and the relation $\partial s/\partial T=C_v/T$, the points on the isentropic line in the $(u,T)$ coordinates satisfy the relation:
 \begin{equation}
 \frac{C_v}{T} \frac{\partial T}{\partial u}=\frac{\partial C}{\partial T}.
\end{equation} 

\begin{figure}[t] 
	\centering    
	\includegraphics[scale=0.4]{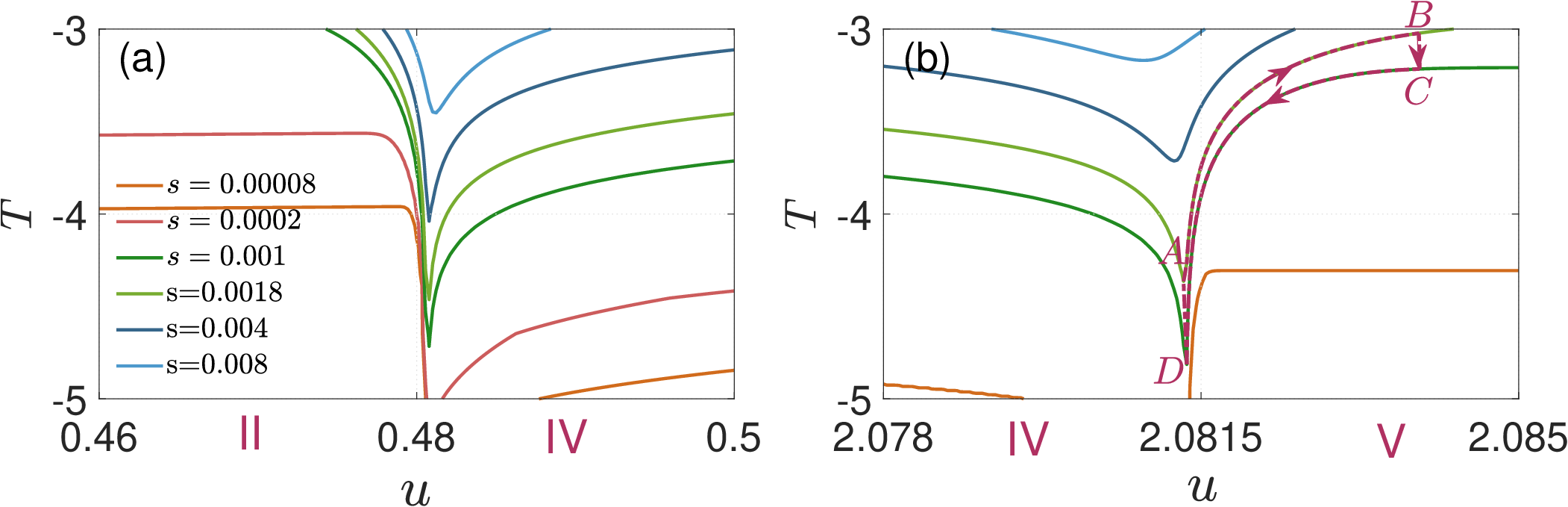}  
	\caption{Plot of isentropic lines in $T-u$  plane for  the interaction-driven  phase transitions II-IV (a)  and IV-V (b) at $B=0.15$ and $\mu=-2.5$, of which the temperature on the vertical axis is logarithmic.  The purple dotted lines in (b) denote the interaction driven Otto  cycle, where the stages A and B (or  stages  C, D) lie on the  isentropic lines, see the main text.}       
	\label{s-II-IV-IV-V}     
\end{figure}

Thus the interaction driven Gr{\"u}neisen ratio $\Gamma_{\mathrm{int}}$ is related to $\partial C/\partial T$ via the relation $\Gamma_{\mathrm{int}}=\partial C/\partial T\cdot u/{C_v}$. 
Note that $\partial C/\partial T$ dramatically changes near quantum phase transition  (see equation (\ref{CS-C-T}) and figure~\ref{C45}), where $C_v$ has a  minimum (see figure~\ref{fig-45} ) and the entropy has a maximum (see figure~\ref{s-II-IV-IV-V}). 
A good cooling effect is observed  when the interaction drives the system approaching a critical point.
 The lowest temperature point can be estimated by $\partial C/\partial T=0$. 
 To characterise the refrigerating efficiency, we give a brief description on the lowest reachable temperature during an adiabatic cooling cycle. 
 From the results (\ref{CS-C-T}) of $\partial C/\partial T$, the condition for determining the lowest temperature reads
\begin{equation}
 \frac{1}{2}\Li_{\frac{1}{2}}\left(-\e^{x}\right)-x\Li_{-\frac{1}{2}}\left(-\e^{x}\right)=0, \label{relation-cooling}
\end{equation} 
where $x=\alpha_u\Delta u/T$. 
It is found that, to an good approximation, $x\approx1.3117$. 
Thus the extreme point can be determined. 
In figure~\ref{s-II-IV-IV-V}, we plot the isentropic lines near the phase boundaries of II-IV (a) and IV-V (b), respectively.
 It is obvious that entropy shows a minimum in  the quantum critical regions. Away from QCP with temperature $T\ll |u-u_c|$, entropy linearly depends on temperature, see the left part of figure~\ref{s-II-IV-IV-V} (a) and right part of figure~\ref{s-II-IV-IV-V} (b). 
 In figure~\ref{s-II-IV-IV-V} (b) we draw an  Otto  cycle for the  interaction-driven refrigeration process.
 The stages A,\, D lie around QCP, whereas the stages B,C are located in the TLL area. 
 There contains four steps to cool the target material. For A $\rightarrow$ B, the working substance is  adiabatically ramped up from the target  temperature $T_{\rm target}$ to the nonthermal  higher   temperature stage B.  
 Then through a hot isochore process B $\rightarrow$ C, the working substance comes into contact with  the ambient, transferring heat to the  high temperature  source. 
 While the temperature of working substance reduces to the one at  the  thermal state C. 
Next, for the isentrope process C $\rightarrow$ D, the working substance is  adiabatically ramped down to the low temperature stage D.
This is an opposite process  contrast to the  A $\rightarrow$ B.
Finally, for the isochore process D $\rightarrow$ A, the working substance contacts with the target object, absorbing heat from the target material and reaching the thermal state A. 
 Consequently, the target object is cooled down by this cycle.  

Now let us determine the lowest temperature which can be reached  through an isentropic process indicated in the figure~\ref{s-II-IV-IV-V}. 
From equations (\ref{1-2})-(\ref{4-5}),  the phase II (V)  contains one charge (spin) degrees of freedom.
Consequently,  their entropy $s_{\mathrm{L1}}$ and $ s_{\mathrm{L2}}$ are given by (\ref{f-TLLc}) and (\ref{f-TLLs}), respectively, namely,
\begin{eqnarray}
s_{\mathrm{L1}}&\approx& \frac{\pi T_{\mathrm{L1}}}{3v_c},
\\s_{\mathrm{L2}}&\approx& \frac{\pi T_{\mathrm{L2}}}{3v_s}.
\end{eqnarray}
Comparing isentropic lines with the same entropy  for phase II, IV and V, for example  $s=0.00008$ in  figure~\ref{s-II-IV-IV-V},  the temperature of TLL$_{C}$ (phase II) is higher than that of TLL$_{S} $ (phase V) since charge velocity $v_c$ changes  faster than spin velocity $v_s$ when the interaction is changed around the critical point, i.e.,
\begin{equation}
T_{\mathrm{L1}}>T_{\mathrm{L2}}.\label{TT}
\end{equation}
On the other hand, when it approaches the QCP, i.e., at the extreme low temperature for each isentropic line, entropy have explicit expressions for the transition II-IV and IV-V
\begin{eqnarray}
s_{\mathrm{s1}}&\approx&\lambda_3 \pi^{1/2}\sigma_1(0)(\varepsilon^{''}_1(0)/2)^{-1/2}T_{c1}^{1/2},\\
s_{\mathrm{s2}}&\approx&\lambda_3 \pi^{1/2}\rho(\pi)(-\kappa^{''}(\pi)/2)^{-1/2}T_{c2}^{1/2},
\end{eqnarray}
respectively, where  $\lambda_3=x\Li_{1/2}\left(-\e^{x}\right)-3/2\Li_{3/2}\left(-\e^{x}\right)\approx 1.3467$.

With the above analysis, we observe that the entropy shows a square root dependence on the temperature at extreme point, it is proportional to temperature in the Luttinger liquid. 
 Therefore, considering an isentropic cooling  process  through the ramping up or down in the $T-u$  plane around critical phase transitions from II to IV or from V to IV, see figure~\ref{s-II-IV-IV-V}, 
 the minimum temperatures can be reached 
\begin{eqnarray}
	\text{II-IV:  } \frac{T_{c1}^{1/2}}{ T_{L1}}&=&\frac{\pi^{1/2}(\varepsilon^{''}_1(0)/2)^{1/2}}{3\lambda_3v_c\sigma_1(0)},\\
	\text{V-IV:  } \frac{T_{c2}^{1/2}}{ T_{L2}}&=&\frac{\pi^{1/2}(-\kappa^{''}(\pi)/2)^{1/2}}{3\lambda_3v_s\rho(\pi)},
	\end{eqnarray}
respectively. A brief discussion about interaction-driven quantum cooling is given in \cite{LPG:2022}. Based on previous results (\ref{kappa-Q})- (\ref{sigma-A}), the leading contributions to minimum temperature near quadruple critical point are 
\begin{eqnarray}
	\text{II-IV:  } \frac{T_{c1}^{1/2}}{ T_{L1}}&\approx&\frac{\pi^{1/2}\eta_1^{1/2}}{6\lambda_1\lambda_3\delta}, \\
	\text{V-IV:  } \frac{T_{c2}^{1/2}}{ T_{L2}}&\approx&\frac{2\lambda_1\pi^{5/2}}{3\lambda_3\eta_1 A}.
\end{eqnarray}
These suggest that  the lowest temperature can be reached around QCPs, which can be also reachable through  adiabatic demagnetizaion cooling, see the study of the Gr{\"u}neisen parameters in \cite{yu2020gruneisen}. 
Moreover, we further  note that the location of the  minimum temperature for each contour entropy line  in the $T$-$u$ plane is governed by  the relation $\alpha_u \Delta u/T\approx1.3117$ in (\ref{relation-cooling}), where $\alpha_u$ needs to be determined.

\subsection{V.4 Calculation of $\alpha_u$ for  phase transition IV-V}

\begin{figure}[t] 
	\centering    
	\includegraphics[scale=0.36]{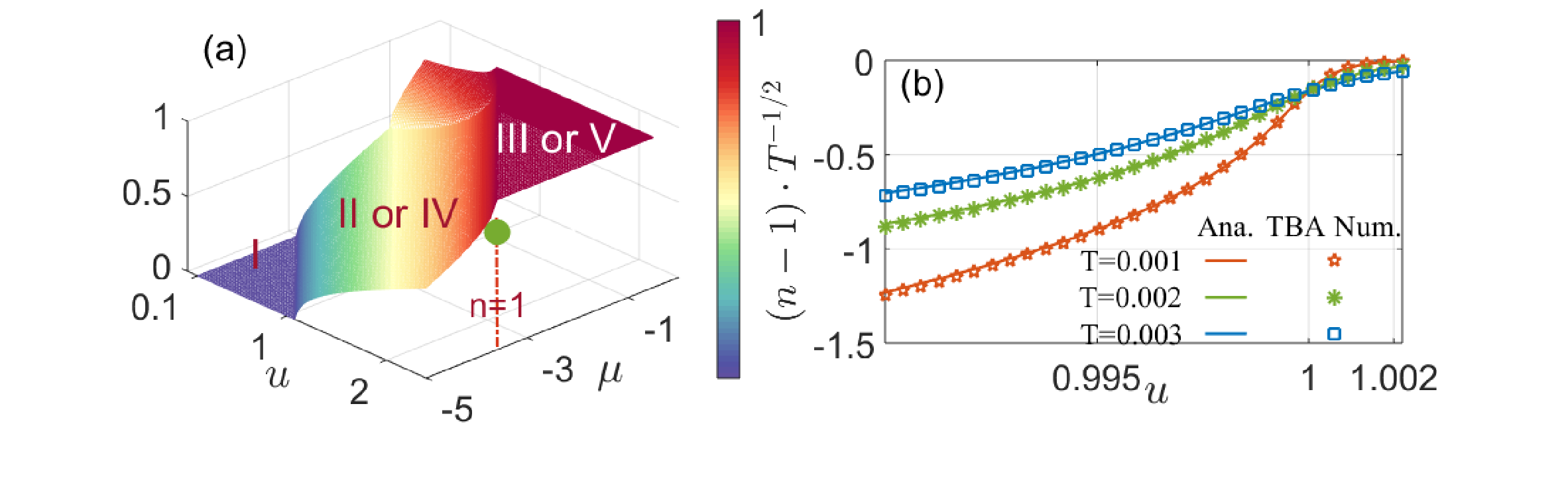}  	
	\caption{(a) 3D plot of density $n$ in $u-\mu-n$ coordinate with fixed $B=0.82714$: showing quantum phases I, II (or IV) and III (or V). The red dash-dotted line in the $u-\mu$ plane represents the projection of 3D when the density firstly reaches unity from phase II or IV. And the green circle on the $n=1$ line  denotes small areas around $u_c=1$. (b) Quantum scaling behaviour of density near phase transition from  IV to V driven  by interaction. Numerical result from the TBA equations and analytical result are in good agreement. }       
	\label{45-n}     
\end{figure}

One remaining problem from the discussion in the previous subsection is that the underlying  coefficient $\alpha_u$ is not given. 
It not only determines the scaling factor of the critical temperature at quantum criticality driven by the interaction, but also determines the extreme point associated with the lowest temperature in an isotropic process in the refrigeration cycle. 
We know that $\alpha_u$ emerged in polylog function is related to the phase transition line via the formula $\kappa(\pi)=\alpha_B\Delta B+\alpha_{\mu}\Delta \mu+\alpha_u\Delta u$. 
For transition IV-V, the phase boundary between phase IV and V is marked by $\kappa(\pi)=0$, where $\kappa(\pi)$ is expressed via $\varepsilon_1(\Lambda)$ by 
\begin{eqnarray}
\kappa(\pi)&=&2-\mu-2 u-B+\int_{-A}^{A}\d \Lambda  a_1(\Lambda) \varepsilon_1(\Lambda),\label{b45-1} \\
\varepsilon_1(\Lambda)&=&2 B-4 \textmd{Re} \sqrt{1-(\Lambda-\mathrm{i} u)^{2}}+4u
-\int_{-A}^{A} \d \Lambda^{\prime} a_{2}\left(\Lambda-\Lambda^{\prime}\right)\varepsilon_1(\Lambda^{\prime}).\label{b45-2}
\end{eqnarray}
However for the transition IV-V, the corresponding equations (\ref{b45-1}) and (\ref{b45-2}) are not in polynomial forms in terms of interaction $u$.
 In this part, we present a numerical scheme with equation (\ref{umu}) based on the quality that phase transition from IV to V occurs due to the emergence of Mott insulator with constant density. 
 In addition to the magnetic field $B$ and chemical potential $\mu$ that can drive the phase transition, the interaction also drives the system from one phase to others.
  In figure~\ref{45-n} (a), we plot the phase diagram in the ($u-\mu-n_c$) coordinate at fixed magnetic field $B=0.82714$. 
  The vacuum phase I has  $n_c=0$. 
  The  density is $0<n_c<1$ corresponding to phase II or IV and $n_c=1$ for the phase III or V. 
   The boundary line for IV and V (or II and III) is marked with a constant density $n_c=1$. On this transition line, we perform total derivatives of $n_c$, yielding
\begin{equation}
\d n_c=\frac{\partial n_c}{\partial u}\d u+\frac{\partial n_c}{\partial \mu}\d\mu=0.\label{dn0}
\end{equation}
Substituting (\ref{umu}) for $\partial n_c/\partial u=-\partial C/\partial \mu$ with (\ref{CS-C-Mu}) and $\partial n_c/\partial \mu=\chi_c$ with (\ref{QC-IV-V-chic}) into (\ref{dn0}), near the boundary line $n_c=1$ in the $u-\mu$ plane, we have
\begin{equation}
\alpha_u=-\alpha_{\mu}\frac{\partial \mu}{\partial u}.\label{45-umu}
\end{equation}
where we use $b_1=\alpha_u,-1=\alpha_{\mu}$. Near the phase transition IV to V, we see $\alpha_{\mu}=-1$ since the chemical potential $\mu$ only appears in the charge leading term, no contribution to the driving terms in spin string, see (\ref{b45-1}) and (\ref{b45-2}). 
At this point, we relate the value of $\alpha_u$ to the slope at QCP in the $u_c-\mu_c$ plane. 
Therefore  equation (\ref{45-umu}) provides us with a way to obtain the unknown  coefficient $\alpha_u$.
 Moreover, $\partial \mu/\partial u$ can also be obtained directly from (\ref{b45-1}) and (\ref{b45-2})
\begin{eqnarray}
2\frac{\partial \mu}{\partial u}&=&-2+\int_{-A}^{A}\d \Lambda  \left[\frac{\partial a_1(\Lambda)}{\partial u} \varepsilon_1(\Lambda)+a_1(\Lambda)\frac{\partial \varepsilon_1(\Lambda)}{\partial u}\right],\label{pp} \\
\frac{\partial \varepsilon_1(\Lambda)}{\partial u}&=&-4\frac{\partial}{\partial u}[ \textmd{Re} \sqrt{1-(\Lambda-\mathrm{i} u)^{2}}-u]
-\int_{-A}^{A} \d \Lambda^{\prime} \frac{\partial}{\partial u}[a_{2}\left(\Lambda-\Lambda^{\prime}\right)\varepsilon_1(\Lambda^{\prime})],
\end{eqnarray}
where the factor $2$ emerged at the leftmost end of (\ref{pp}) arises from the derivative of $\kappa(\pi)$ and $\mu$. Then $\partial \mu/\partial u$ can be obtained by iteratively solving the above two equations. 
Here, we use color map to get $\alpha_u$ instead, which can apply to arbitrary systems without knowing explicit analytic formulas. In figure~\ref{45-n} (a), we plot density phase  diagram containing QCP $(u_c, \mu_c)$ of interest.
In order to evaluate $\alpha_u(u_c,\mu_c)$ numerically for the interaction-driven  phase transitions, compared to figure~\ref{fig-45} (c) (d) driven by external fields, we choose two adjacent points $(u_1,\mu_1),(u_2,\mu_2)$ near QCP $u_c=1$, see the green areas in figure~\ref{45-n} (a). 
It is evaluated from (\ref{45-umu}) that $\alpha_u(u_c=1)\approx(\mu_1-\mu_2)/(u_1-u_2)\approx-1.9627$. 
Using this value as  the argument factor of the scaling function for the phase transition IV-V with the other part already known in equations (\ref{f-45}) and (\ref{4-5}), we plot the density scaling law in figure~\ref{45-n} (b) in terms of the variation of interaction strength. 
One can see that the obtained $\alpha_u$ captures well the thermodynamic scaling law.

\subsection{V.5 Calculation of $\alpha_u$ for phase transition II-IV}
Similar to the analysis of the parameter $\alpha_u$ for  the transition of IV to V, the relevant coefficients $\alpha_B,\alpha_{\mu},\alpha_u$ for transition II to IV are related to each other. 
Contrast to the Mott-insulator transition, this transition arises from the introduction of spin-down electrons, i.e., the phase transition points are determined by zero  spin-down particle density  $n_{\downarrow}=0$.
 Considering a magnetic field $B$ and coupling strength $u$ driven phase transition, we perform total derivative of $n_{\downarrow}$, i.e.,
\begin{equation}
\d n_{\downarrow}=\frac{\partial n_{\downarrow}}{\partial u}\d u+\frac{\partial n_{\downarrow}}{\partial (2B)}\d(2B)=0. \label{dnd0}
\end{equation}
Substituting $\partial m/\partial B=\chi_s$ with (\ref{QC-II-IV-4}) and (\ref{uB}) into the (\ref{dnd0}), with the help of $n_{\downarrow}=n_c/2-m$ we obtain 
\begin{equation}
\alpha_u=-\alpha_B\frac{\partial B}{\partial u}.\label{24-uB}
\end{equation}

Recall that (\ref{e0-dB}) gives us the value of $\alpha_B=-2\left[1-\arctan\left(\sin Q_c/u\right)/\pi\right]$.
We also note that the quantity $\partial B/\partial u$ can be obtained from (\ref{e0}), namely, near the critical line, we have the following relation 
\begin{equation}
B=\frac{2u}{\pi}\int_{0}^{Q}\mathrm{d}k\frac{\cos k}{u^2+\sin^2k}(\cos k-\cos Q).
\end{equation}
Therefore, $\partial B/\partial u$ is easily obtained through the derivative on both sides of this equation with respect to $u$. 
From equation (\ref{24-uB}), we finally get the following analytic expression
\begin{eqnarray}
 \alpha_u&=&\frac{4}{\pi}\left[\int_{0}^{Q_c}\mathrm{d}k\frac{\cos^2 k}{u_c^2+\sin^2 k}-2u_c^2\int_{0}^{Q_c}\mathrm{d}k\frac{\cos^2 k}{(u_c^2+\sin^2 k)^2}
\right.\nonumber\\
&&\left. +\arctan\left(\frac{\sin Q_c}{u_c}\right)+\frac{\sin(2Q_c)}{2(u_c^2+\sin Q_c^2)}\right].
\end{eqnarray}

In summary, there exist simple relations between any pair of these scaling factors $\alpha_B,\, \alpha_{\mu}$ and $\alpha_u$. 
In the previous two subsections, via Contact susceptibilities expressions (\ref{umu}) and (\ref{uB}), we build up relationships between $\alpha_u$ and $\alpha_{\mu}$ as well as between $\alpha_u$ and $\alpha_B$, see (\ref{45-umu}) and (\ref{24-uB}). In fact, in terms of thermodynamic potential, Maxwell relations can also build a connection between the charge  susceptibility $\partial n_c/\partial (2B)$ and magnetization susceptibility $\partial m/\partial \mu$. Moreover, (\ref{45-umu}) and (\ref{24-uB}) remain valid for  other transitions, like I-II, II-III and III-V.  Based on this feature, we can conclude that the following relations
\begin{eqnarray}
\frac{\alpha_u}{\alpha_{\mu}}&=&-\frac{\partial \mu}{\partial u}\label{oumu},\\
\frac{\alpha_u}{\alpha_{B}}&=&-\frac{\partial B}{\partial u}\label{ouB},\\
\frac{\alpha_B}{\alpha_{\mu}}&=&-\frac{\partial \mu}{\partial B}\label{omuB}
\end{eqnarray} 
hold true in general, i.e., for arbitrary interaction strength and density.
In the above,  we have discussed the  exact values of $\alpha_B$ for the phase transition II-IV and $\alpha_{\mu}$ for the phase transition IV-V. The remaining  coefficients can be derived through equations (\ref{oumu}), (\ref{ouB}), (\ref{omuB}) in a straightforward way.
Here, we give the values of $\alpha_{\mu}$ for II-IV and $\alpha_B$ for IV-V
\begin{eqnarray}
	\text{II-IV: } \alpha_{\mu}&=&\frac{2}{\pi}\arctan\left(\frac{\sin Q_c}{u}\right),\\
	\text{IV-V: } \alpha_{B}&=&-\frac{1}{2}+\int_{0}^{A}\d \Lambda  a_1(\Lambda)\frac{\partial \varepsilon_1(\Lambda)}{\partial B}.
\end{eqnarray}
Together with equations (\ref{1-2})-(\ref{4-5}), the results obtained here offer more general description of quantum criticality in terms of full internal and external potentials in the 1D Hubbard model in arbitrary experimentally controllable parameters. 
These hold true for the second order phase transition in higher dimensional quantum systems too.

\section{VI. Conclusion and remarks}

We have presented analytical results of thermal and magnetic properties of 1D Hubbard model, ranging from elementary spin and change  excitations, to the  spin incoherent Luttinger liquid, universal thermodynamics, quantum criticality and interaction-driven refrigeration. 
A summary of our new results is the following:\\
1)	We have studied elementary  excitations involving  fractional spin and charge excitations, gapped charge $k-\Lambda$ strings and spinon bound state excitations as well as some combinations of them in section II. These are complementary to that studied in \cite{essler2005one}.
Based on  the study of such excitations, together with the dimensionless Wilson ratio, universal scalings of thermodynamics of the model, we have a rigorous investigation of the spin incoherent Luttinger liquid, which was previously studied only in the framework of effective theory via bosonization.
From the conformal field theory point of view, we have given rigorously the characteristics of various correlation functions near the phase transition from phase II to phase IV, showing an existence of collective mode in charge degrees of freedom rather than the spin degrees of freedom.  \\  
2) We have presented general  analytical results of thermodynamics, independent of microscopic details of the model.
Explicitly, we have determined the additivity rules of charge and spin susceptibilities in grand canonical ensemble equations (\ref{chi-gc}) (\ref{chi-gs}) and  canonical ensemble equations (\ref{chi-c}) (\ref{chi-s}).
Away from the critical point and in  the low-energy regime, in general,   coherent spin and charge degrees of freedom give rise to the well known phenomenon of spin-charge  separation, indicating the nature of the TLL.
The crossover regime between the TLL and the quantum critical region belongs to the SILL,  indicating a coexistence of liquid and gas, see equation (\ref{QC-II-IV-cv-2}). 
Besides the thermodynamics, the quantum  criticality and universal scaling laws induced by the variation of  magnetic field and chemical potential have been obtained analytically and confirmed numerically, see general results of criticality equations (\ref{1-2})-(\ref{4-5}). 
In addition, the interaction-driven quantum critical behaviour has been studied too. \\ 
3) We have  introduced the lattice version of the Contact and  Contact susceptibilities with respect to the external fields and temperature. 
In particular, we have  investigated applications of Contact susceptibilities, that build up a general connection between interaction-driven quantum criticality and the phase transitions induced by external fields.
By virtue of the Contact susceptibilities, we have  discussed Mott transition, quantum refrigeration and interaction-induced quantum transitions, see section V.

In view of the rapid advances in trapping and controlling ultracold atoms in experiment, the results obtained here will provide direct guidance to explore experimentally various many-body phenomena in the 1D Hubbard model, such as quantum criticality, spin coherent and incoherent Luttinger liquids, generalized hydrodynamics and transport properties. 
These relations reveal deep insights into quantum criticality  of the Hubbard model in  higher dimensions. 
Furthermore, applications of our method to quantum metrology and other quantum technologies are also highly desirable. 

\section*{Acknowledgements}
We thank the Integrable Theory Group  at APM for their helpful discussions. X.W.G. acknowledges kind hospitalities  of Rice University and Institute for Advanced Study, Tsinghua University during his visits  in 2018 and 2021, respectively. 

J.J.L. performed the analytical and numerical study of the model. X.W.G.  supervised J.J.L. for conducting the results reported in this paper. Both X.W.G. and H.P. initiated this study. 

J.J.L. and X.W.G. is supported by the NSFC key grant No. 12134015, the NSFC grant  No. 11874393 and No. 12121004.
X.W.G. is also partially supported by the Innovation Program for Quantum Science and Technology 2021ZD0302000. 
H.P. acknowledges support from the US NSF (PHY-2207283) and the Welch Foundation (Grant No. C-1669).

\begin{appendix}
	
	\section{APPENDIX A: Universal thermodynamics}
	\setcounter{equation}{0}
	\renewcommand\theequation{A\arabic{equation}}
	
	Universal low energy physics can be derived from the TBA equations under the conditions of $B/T\gg1$ and $|\mu|/T\gg1$.
	In such low energy limits,  the TBA equations (\ref{eq-TBA-1}) and (\ref{eq-TBA-2}) reduce to two coupled  equations  in terms of charge quasimomentum  $k$ and spin  length-1 $\Lambda$ string, see  (\ref{ku}) and (\ref{eu}). 
	Analysing the structure of the simplified TBA equations (\ref{ku}) and (\ref{eu}), we find that it is efficient  to take Sommerfeld expansion of the kernel functions in terms of the Fermi points. 
	Let us denote the fermi point  $Q,\,A$ for charge and spin degrees of freedom satisfying $\kappa(Q)=0,\varepsilon_1(A)=0$.
	For our convenience, we denote  $\frac{\partial{\bar{a}_1(\sin k,\Lambda)}}{\partial{\Lambda}}=a_1(\sin k-\Lambda)$, $\frac{\partial{\bar{a}(\sin k,\Lambda)}}{\partial{\Lambda}}=a_1(\sin k-\Lambda)+a_1(\sin k+\Lambda)$.
	Now we focus on  the second term in the (\ref{ku})
	\begin{eqnarray}
	&&-\int_{-\infty}^{\infty}\mathrm{d} \Lambda  a_1(\sin{k}-\Lambda) T\ln(1+\mathrm{e}^{-\frac{\varepsilon_1(\Lambda)}{T}})\nonumber \\
	&=&-\bar{a}_1(\sin k,\Lambda) T\ln(1+\mathrm{e}^{-\frac{\varepsilon_1(\Lambda)}{T}})|_{-\infty}^{\infty}-\int_{0}^{\infty}\mathrm{d} \Lambda  \bar{a}(\sin{k},\Lambda)\frac{1}{1+\mathrm{e}^{\frac{\varepsilon_1(\Lambda)}{T}}}\frac{\partial\varepsilon_1(\Lambda)}{\partial{\Lambda}}\nonumber\\
	&=&-\int_{0}^{\infty}\mathrm{d} \Lambda  \bar{a}(\sin{k},\Lambda)\frac{1}{1+\mathrm{e}^{\frac{\varepsilon_1(\Lambda)}{T}}}\frac{\partial\varepsilon_1(\Lambda)}{\partial{\Lambda}}\nonumber\\
	&=&-\int_{\varepsilon_1(0)}^{\varepsilon_1(\infty)}\mathrm{d} \varepsilon_1 \bar{a}(\sin{k},\Lambda(\varepsilon_1))\frac{1}{1+\mathrm{e}^{\frac{\varepsilon_1}{T}}}\nonumber \\
	&=&-\int_{\frac{\varepsilon_1(0)}{T}}^{\frac{\varepsilon_1(A)}{T}}\mathrm{d}x T \bar{a}(\sin{k},\Lambda(Tx))\frac{1}{1+\mathrm{e}^{x}}-\int_{\frac{\varepsilon_1(A)}{T}}^{\frac{\varepsilon_1(\infty)}{T}}\mathrm{d}x T \bar{a}(\sin{k},\Lambda(Tx))\frac{1}{1+\mathrm{e}^{x}}\nonumber \\
	&=&-\int_{\frac{\varepsilon_1(0)}{T}}^{\frac{\varepsilon_1(A)}{T}}\mathrm{d}x T \bar{a}(\sin{k},\Lambda(Tx))\left(1-\frac{1}{1+\mathrm{e}^{-x}}\right)-\int_{\frac{\varepsilon_1(A)}{T}}^{\frac{\varepsilon_1(\infty)}{T}}\mathrm{d}x T \bar{a}(\sin{k},\Lambda(Tx))\frac{1}{1+\mathrm{e}^{x}}\nonumber \\
	&=& -\int_{\frac{\varepsilon_1(0)}{T}}^{0}\mathrm{d}x T \bar{a}(\sin{k},\Lambda(Tx))+\int_{0}^{-\frac{\varepsilon_1(0)}{T}}\mathrm{d}x T \bar{a}(\sin{k},\Lambda(-Tx))\frac{1}{1+\mathrm{e}^{x}}\nonumber\\
	&&-\int_{0}^{\frac{\varepsilon_1(\infty)}{T}}\mathrm{d}x T \bar{a}(\sin{k},\Lambda(Tx))\frac{1}{1+\mathrm{e}^{x}}\nonumber 
	\\
	&=&-\int_{\varepsilon_1(0)}^{0}\mathrm{d} \varepsilon_1 \bar{a}(\sin{k},\Lambda(\varepsilon_1))+\int_{0}^{\infty}\mathrm{d}x T \frac{\bar{a}(\sin{k},\Lambda(-Tx))-\bar{a}(\sin{k},\Lambda(Tx))}{1+\mathrm{e}^{x}}\nonumber \\
	&=&\int_{\varepsilon_1(0)}^{0}\mathrm{d} \varepsilon_1 \frac{\partial{\Lambda}}{\partial{\varepsilon_1}} \varepsilon_1 (a_1(\sin{k}-\Lambda)+a_1(\sin{k}+\Lambda))\nonumber \\
	&&-\int_{0}^{\infty}\mathrm{d}x T \frac{(a_1(\sin{k}-\Lambda(0))+a_1(\sin{k}+\Lambda(0)))\Lambda^{'}(0)2Tx}{1+\mathrm{e}^{x}}\nonumber \\
	&=&\int_{0}^{A}\mathrm{d} \Lambda \varepsilon_1(\Lambda) (a_1(\sin{k}-\Lambda)+a_1(\sin{k}+\Lambda))-2T^2(a_1(\sin{k}-A)+a_1(\sin{k}+A))\Lambda^{'}(0)\frac{\pi^2}{12}\nonumber \\
	&=&\int_{0}^{A}\mathrm{d} \Lambda \varepsilon_1(\Lambda) (a_1(\sin{k}-\Lambda)+a_1(\sin{k}+\Lambda))\nonumber\\
	&&-\frac{\pi^2T^2}{6}\frac{1}{\varepsilon_1^{'}(A)}(a_1(\sin{k}-A)+a_1(\sin{k}+A)).
	\end{eqnarray}
	In the above calculations, the following equations are used:
	\begin{eqnarray}
	&& \frac{\bar{a}(\sin{k},\Lambda(Tx))-\bar{a}(\sin{k},\Lambda(-Tx))}{2Tx}=\frac{\partial{\bar{a}}}{\partial{\Lambda}}|_{\Lambda=\Lambda(0)}\Lambda^{'}(0),\nonumber \\
	&& \Lambda^{'}(0)=\frac{1}{\frac{d\varepsilon_1}{d\Lambda}|_{\varepsilon_1=0}}=\frac{1}{\varepsilon^{'}_1(A)},\qquad 
	\int_{0}^{\infty}\mathrm{d} x \frac{x}{1+\mathrm{e}^{x}}=\frac{\pi^2}{12}.
	\end{eqnarray}
	
	Conducting  similar calculation  in the integrals of $\varepsilon_1(\Lambda)$, the dressed energies can be converted to
	\begin{eqnarray}
	\kappa(k)&=&-2 \cos{k}-\mu-2 u-B\nonumber+\int_{-A}^{A}\mathrm{d} \Lambda \varepsilon_1(\Lambda) a_1(\sin{k}-\Lambda)\\&&-\frac{\pi^2T^2}{6}\frac{1}{\varepsilon_1^{'}(A)}(a_1(\sin{k}-A)+a_1(\sin{k}+A))\label{kl0-m}, \\
	\varepsilon_1(\Lambda)&=&2 B- \frac{\pi^2T^2}{6\kappa^{'}(Q)}\cos Q (a_1(\sin{Q}-\Lambda)+a_1(\sin{Q}+\Lambda))\nonumber\\
	&& +\frac{\pi^2T^2}{6\varepsilon_1^{'}(A)} (a_2(\Lambda-A)+a_2(\Lambda+A))\nonumber\\&&+\int_{-Q}^{Q}\mathrm{d} k \kappa(k) \cos k a_1(\sin{k}-\Lambda)\nonumber\\
	&& -\int_{0}^{A}\mathrm{d} \Lambda^{'} \varepsilon_1(\Lambda^{'}) (a_2(\Lambda-\Lambda^{'})+a_2(\Lambda+\Lambda^{'})). \label{el0-m}
	\end{eqnarray}
	
	Compare above equations with root densities (\ref{rho}) and (\ref{sigma}), we gain insight that there is an implicit connection between the two sets of equations which are highly symmetric. In order to get a compact and closed form of the free energy, combing  equations (\ref{rho}) and (\ref{sigma}) with (\ref{kl0-m}) and (\ref{el0-m}) and using Sommerfeld expansion, we obtain  the free energy 
	\begin{eqnarray}
	f&=&-\int_{-\pi}^{\pi} \frac{\mathrm{d} k}{2\pi} T\ln(1+\mathrm{e}^{-\frac{\kappa(k)}{T}})+u\nonumber\\
	&=&-\frac{\pi^2T^2}{6\kappa^{'}(Q)}+\int_{0}^{Q} \frac{\mathrm{d} k}{\pi}\kappa(k)+u\nonumber\\
	&=& -\frac{\pi^2T^2}{6\kappa^{'}(Q)} +u \nonumber\\
	&& +\int_{-Q}^{Q} \mathrm{d}k\left[-2 \cos{k}-\mu-2 u-B-\frac{\pi^2T^2}{6}\frac{1}{\varepsilon_1^{'}(A)}(a_1(\sin{k}-A)+a_1(\sin{k}+A))\right]\rho(k)\nonumber\\
	&&+\int_{-A}^{A} \mathrm{d}\Lambda \left[2 B- \frac{\pi^2T^2}{6\kappa^{'}(Q)}\cos Q (a_1(\sin{Q}-\Lambda)+a_1(\sin{Q}+\Lambda)) \right.\nonumber \\
	&&\left. +\frac{\pi^2T^2}{6\varepsilon_1^{'}(A)} (a_2(\Lambda-A)+a_2(\Lambda+A))\right]\sigma_{1}(\Lambda)\nonumber\\
	&=&-\frac{\pi^2T^2}{6\kappa^{'}(Q)} +u+\left[\int_{-Q}^{Q} \d k\left(-2 \cos{k}-\mu-2 u-B\right)\rho(k)+2B\int_{-A}^{A} \d\Lambda \sigma_1(\Lambda)\right] \nonumber\\&&-\frac{\pi^2T^2}{6\varepsilon_1^{'}(A)}\left[\int_{-Q}^{Q} \d k(a_1(\sin{k}-A)+a_1(\sin{k}+A))\rho(k)-(a_2(\Lambda-A)+a_2(\Lambda+A))\sigma_{1}(\Lambda)\right] \nonumber\\&&-\frac{\pi^2T^2}{6\kappa^{'}(Q)}\left[\int_{-A}^{A} \mathrm{d}\Lambda\cos Q (a_1(\sin{Q}-\Lambda)+a_1(\sin{Q}+\Lambda))\right]\nonumber\\	
	&=&-\frac{\pi^2T^2}{6\kappa^{'}(Q)}+f_0-\frac{\pi^2T^2}{6\varepsilon_1^{'}(A)}\left[\sigma_{1}(A)+\sigma_{1}(-A)+\int_{-A}^{A} \mathrm{d}\Lambda(a_2(\Lambda-A)+a_2(\Lambda+A))\sigma_{1}(\Lambda)\right]\nonumber\\
	&&-\frac{\pi^2T^2}{6\kappa^{'}(Q)}\left[\rho(Q)+\rho(-Q)-\frac{1}{\pi}\right]+\frac{\pi^2T^2}{6\varepsilon_1^{'}(A)}\int_{-A}^{A} \mathrm{d}\Lambda(a_2(\Lambda-A)+a_2(\Lambda+A))\sigma_{1}(\Lambda)\nonumber\\
	&=&f_0-\frac{\pi^2T^2\rho(Q)}{3\kappa^{'}(Q)}-\frac{\pi^2T^2\sigma_{1}(A)}{3\varepsilon_1^{'}(A)}\nonumber\\&=&f_0-\frac{\pi T^2}{6}\left(\frac{1}{v_c}+\frac{1}{v_s}\right),
	\end{eqnarray}
	where $f_0=\int_{-Q}^{Q} \d k\left(-2 \cos{k}-\mu-2 u-B\right)\rho(k)+2B\int_{-A}^{A} \d\Lambda \sigma_1(\Lambda)$ is the background contribution from zero temperature. The form of the free energy serves as universal thermodynamics of the TLL in terms of spin and charge degrees of freedom. 
	Notice that in the above calculation process, following integral identity  was  used 
	\begin{equation}\left\{\begin{array}{l}
	f_1=f^{(0)}_1+K_{12}*f_2\\f_2=f^{(0)}_2+K_{21}*f_1+K_{22}*f_2\\
	g_1=g^{(0)}_1+K_{21}*g_2\\g_2=g^{(0)}_2+K_{12}*g_1+K_{22}*g_2
	\end{array}\rightarrow \int f_1g^{(0)}_1+\int f_2g^{(0)}_2=\int g_1f^{(0)}_1+\int g_2f^{(0)}_2.
	\right.\end{equation}
	
	In spin polarized band II, spin degree vanishes. The free energy in low temperature is simplified as $f=f_0-\frac{\pi T^2}{6}\frac{1}{v_c}$, specific heat $\frac{C_v}{T}=\frac{\pi }{3}\frac{1}{v_c}$ accordingly. While in phase V with charge half filled, the free energy is denoted as $f=f_0-\frac{\pi T^2}{6}\frac{1}{v_s}$ and specific heat $\frac{C_v}{T}=\frac{\pi }{3}\frac{1}{v_s}$. 
	
	\section{APPENDIX B: Free energy of phase III}
	\setcounter{equation}{0}
	\renewcommand\theequation{B\arabic{equation}}
	
	In previous consideration, the effects from $k-\Lambda$ strings are  ignored at low-temperature limit $T\ll B,\, \mu$. 
	Meanwhile, one can infer from equations (\ref{eq-TBA-2}) and (\ref{eq-TBA-3}) that the leading orders of $\varepsilon_n^{\prime}$ and high $\Lambda$ string with length $n$ contributing to free energy are $\e^{\frac{2n\mu}{T}}$ and $\e^{-\frac{2nB}{T}}$,  respectively, rendering these gapped terms omitted reasonably.
	In order to estimate  those energy scales, in this appendix, we  rigorously calculate the contributions to the  thermodynamics from the lowest gapped $k-\Lambda$ strings $\varepsilon_1^{\prime}$ and from the real spin rapidity   $\varepsilon_1$.
	To make calculations less difficult, we choose area that  $\varepsilon_1$ is  strictly positive and  make the two driving terms $\e^{\frac{2\mu}{T}},\, \e^{-\frac{2B}{T}}$ commensurable. 
	Here we treat the problem in the weak chemical potential region  and  high particle density regime, i.e. close to the quadruple critical point. 
	Consequently, we are intend to discuss the case in phase III.
	On the basis of above limitations,  for the phase III, the TBA equations  in terms of length-1 $\varepsilon_1$ and $\varepsilon_1^{\prime}$ are  rewritten as 
	\begin{eqnarray}
	\varepsilon_1(\Lambda)&=&2B-4\operatorname{Re}\sqrt{1-(\Lambda-iu)^2}+4u+\sum_{m=1}^{\infty} A_{1 m} * T\ln \left(1+e^{-\frac{\varepsilon_m}{T}}\right)\nonumber\\&&-2J_1a_1(\Lambda)+2J_2\left[\frac{\pi}{u}a^2_1(\Lambda)-\frac{4\pi^2}{u^2}\Lambda^2a^3_1(\Lambda)\right], \label{APP-B-TBA-1}\\
	\varepsilon^{'}_1(\Lambda)&=&-2\mu-2J_1a_1(\Lambda)+2J_2\left[\frac{\pi}{u}a^2_1(\Lambda)-\frac{4\pi^2}{u^2}\Lambda^2a^3_1(\Lambda)\right],  \label{APP-B-TBA-2}\\
	\kappa(k)&=&-2 \cos{k}-C_12\sin^2k+C_2, \label{APP-B-TBA-3}
	\end{eqnarray}
	in which $C_1,C_2$ are the integral functions related to the gapped dressed energies $\varepsilon_1$ and $\varepsilon_1^{\prime}$
	\begin{eqnarray}
	C_1&=&\int_{0}^{\infty}\d \Lambda \left[\frac{\pi}{u}a^2_1(\Lambda)-\frac{4\pi^2}{u^2}\Lambda^2a^3_1(\Lambda)\right] \left[T\ln(1+\e^{-\frac{\varepsilon^{'}_1(\Lambda)}{T}})-T\ln(1+\e^{-\frac{\varepsilon_1(\Lambda)}{T}})\right],\label{A-C1} \\
	C_2&=&-\mu-2 u-B+\int_{0}^{\infty}\d \Lambda  2a_1(\Lambda) \left[T\ln(1+\e^{-\frac{\varepsilon^{'}_1(\Lambda)}{T}})-T\ln(1+\e^{-\frac{\varepsilon_1(\Lambda)}{T}})\right],\label{A-C2}
	\end{eqnarray} 
	and $J_1,J_2$ come from charge degree 
	\begin{eqnarray}
	J_1&=&\frac{T^{\frac{3}{2}}}{\sqrt{1+2C_1}}\Gamma(\frac{3}{2})\Li_{\frac{3}{2}}\left(-\e^{\frac{2+C_2}{T}}\right)-\frac{T^{\frac{5}{2}}}{8(1+2C_1)^{\frac{5}{2}}}\Gamma(\frac{5}{2})\Li_{\frac{5}{2}}\left(-\e^{\frac{2+C_2}{T}}\right),\\
	J_2&=&\frac{T^{\frac{5}{2}}}{3(1+2C_1)^{\frac{3}{2}}}\Gamma(\frac{5}{2})\Li_{\frac{5}{2}}\left(-\e^{\frac{2+C_2}{T}}\right).
	\end{eqnarray} 
	Above expressions are similar to previous equations (\ref{QC-kappa})-(\ref{QC-J2}), except that the contribution from $k-\Lambda$ string is taken into account.
	
	In terms of the above formations of TBA equations within high density region, the Gibbs free energy per site is given by 
	\begin{eqnarray}
	f&=&-\mu-u-B-T \int_{-\infty}^{\infty} \frac{\d \Lambda}{\pi} \operatorname{Re} \frac{1}{\sqrt{1-(\Lambda-\mathrm{i} u)^{2}}}\ln \left(1+\e^{-\frac{\varepsilon_1(\Lambda)}{T}}\right)\nonumber\\&&-T\ln \left(1+\e^{\frac{\kappa(0)}{T}}\right)+\frac{1}{\pi}\frac{T^{\frac{3}{2}}}{\sqrt{1+2C_1}}\Gamma(\frac{3}{2})\Li_{\frac{3}{2}}\left(-\e^{\frac{2+C_2}{T}}\right)\nonumber\\
	&&+\frac{1}{24\pi}\frac{T^{\frac{5}{2}}(1+8C_1)}{(1+2C_1)^{\frac{5}{2}}}\Gamma(\frac{5}{2})\Li_{\frac{5}{2}}\left(-\e^{\frac{2+C_2}{T}}\right).\label{Append-free-energy}
	\end{eqnarray}
	
	In order to get $C_{1,2}$ or $J_{1,2}$,  we should solve  the coupled equations (\ref{APP-B-TBA-1})- (\ref{APP-B-TBA-3}). 
	We may  ignore the convolution terms whose leading order is $1+\e^{-\frac{2B}{T}}$.
	After carefully analysing  the terms related to the parameter $\Lambda$, the integral term  $\ln(1+\e^{-\frac{\varepsilon_1(\Lambda)}{T}})$ is approximately equal  to $\e^{-\frac{\varepsilon_1(\Lambda)}{T}}$, and with expansions around $\Lambda=0$ we arrive at, 
	\begin{eqnarray}
	\e^{-\frac{\varepsilon_1(\Lambda)}{T}}&\approx&
	\e^{-\frac{2B+\eta_0}{T}}\e^{-\frac{\eta_1\Lambda^2}{T}}\e^{\frac{g_{\Lambda}}{T}},
	\end{eqnarray}
	where we denote  $g_{\Lambda} =2J_1a_1(\Lambda)-\frac{2\pi}{u}J_2a^2_1(\Lambda)+\frac{8\pi^2}{u^2}J_2\Lambda^2a^3_1(\Lambda)$. 
	Conducting similar manipulation to charge bound states and using (\ref{A-C1}) and (\ref{A-C2}) related with the integral $\varepsilon_1$ and $\varepsilon_1^{\prime}$, we get the coefficients $C_1,\, C_2$
	\begin{eqnarray}
	C_1&=&T\e^{\frac{2\mu}{T}}\left[\frac{\pi}{u}t_2-\frac{4\pi^2}{u^2}t_3\right]-T\e^{-\frac{2B+\eta_0}{T}}\left[\frac{\pi}{u}s_2-\frac{4\pi^2}{u^2}s_3\right], \\
	C_2&=&-\mu-2u-B+2Tt_1\e^{\frac{2\mu}{T}}-2Ts_1\e^{-\frac{2B+\eta_0}{T}}, 
	\end{eqnarray}
	in which 
	\begin{eqnarray}
	s_1(s_2,s_3)&=&\int_{0}^{\infty}\d \Lambda  a_1(\Lambda)(a^2_1(\Lambda),\Lambda^2 a^3_1(\Lambda))\e^{-\frac{\eta_1\Lambda^2}{T}}\e^{\frac{g_{\Lambda}}{T}},\\
	t_1(t_2,t_3)&=&\int_{0}^{\infty}\d \Lambda  a_1(\Lambda)(a^2_1(\Lambda),\Lambda^2 a^3_1(\Lambda))\e^{\frac{g_{\Lambda}}{T}},
	\end{eqnarray}
	resulting from the subleading terms of $\varepsilon_1,\, \varepsilon_1^{\prime}$.
	To this point, we note that (\ref{APP-B-TBA-1}) and (\ref{APP-B-TBA-2}) depend on the value of $J_1,\,J_2$, while (\ref{APP-B-TBA-3}) relies on $C_1,C_2$.
	Meanwhile, 
	we observe that $C_1,\, C_2$ are  given  by $J_1,\, J_2$, in turn $J_1$ and $J_2$ also contain $C_1$ and $C_2$. Therefore, the remaining problem is to get the expressions for $C_1,\, C_2$ and the entire story can be solved.
	By complicated and lengthy iterative procedures, the $J_1$ and $J_2$ can be obtained explicitly 
	\begin{eqnarray}
	J_1&=&\frac{1}{2}\pi^{\frac{1}{2}}T^{\frac{3}{2}}\tau_1+\frac{1}{2}\pi^{\frac{1}{2}}T^{\frac{5}{2}}f_{\frac{3}{2}}\e^{-\frac{2B+\eta_0}{T}}\left(\frac{\pi}{u}s^{(0)}_2-\frac{4\pi^2}{u^2}s^{(0)}_3\right)-\frac{3}{32}\pi^{\frac{1}{2}}T^{\frac{5}{2}}\tau_2,\\
	J_2&=&\frac{\pi^{\frac{1}{2}}T^{\frac{5}{2}}}{4}\tau_2,
	\end{eqnarray}
	where $f_n=\Li_n\left(-\e^{\frac{2-\mu-2u-B}{T}}\right),\, s^{(0)}_2=\frac{1}{2\pi^{3/2}u}\left(a^{-\frac{1}{2}}-a^{-\frac{3}{2}}\right),\, s^{(0)}_3=\frac{\pi^{-5/2}a^{-3/2}}{4},a=\frac{\eta_1u^2}{T}$, \, erfc is complementary error function, and $\tau_1,\, \tau_2$ are defined by 
	\begin{eqnarray}
	\tau_1&=&f_{\frac{3}{2}}+\left(1+\frac{T^{\frac{1}{2}}f_{\frac{3}{2}}}{2\pi^{\frac{1}{2}} u}+\frac{3Tf^2_{\frac{3}{2}}}{16\pi u^2}\right)\e^{\frac{2\mu}{T}}f_{\frac{1}{2}}-\left[\e^a \operatorname{erfc}(\sqrt{a})+\right.  \nonumber\\
	&&\left. \left(\frac{T^{\frac{1}{2}}f_{\frac{3}{2}}}{\pi^{\frac{3}{2}} u}+\frac{Tf^2_{\frac{3}{2}}}{2\pi^2 u^2}\right)\frac{\pi^{\frac{1}{2}}}{a^{\frac{1}{2}}}-\left(\frac{2T^{\frac{1}{2}}f_{\frac{3}{2}}}{\pi^{\frac{3}{2}} u}+\frac{3Tf^2_{\frac{3}{2}}}{2\pi^2 u^2}\right)\frac{\pi^{\frac{1}{2}}}{2a^{\frac{3}{2}}}\right]\e^{-\frac{2B+\eta_0}{T}}f_{\frac{1}{2}}, \nonumber\\
	\tau_2&=&f_{\frac{5}{2}}+\e^{\frac{2\mu}{T}}f_{\frac{3}{2}}-\e^a \operatorname{erfc}(\sqrt{a})\e^{-\frac{2B+\eta_0}{T}}f_{\frac{3}{2}}.
	\end{eqnarray}
	Substituting these  results obtained here  into the expression of free energy (\ref{Append-free-energy}),  then the final solution (\ref{f}) is finally obtained. 
	
\end{appendix}



\end{document}